%% file: main.tex
\documentclass[preprint,11pt,3p,times]{elsarticle}

\usepackage[group-separator={,},exponent-product = \cdot]{siunitx}
\usepackage{amsmath}
\usepackage[mathscr]{eucal}
\usepackage{graphicx}
\usepackage{svg}
\usepackage{caption}
\usepackage{subcaption}
\usepackage{hyphenat}
\usepackage{bm}
\usepackage{upgreek}
\usepackage{adjustbox}
\usepackage{dsfont}

\usepackage{multirow}

\usepackage{url}
\usepackage{color}
\definecolor{colUniBwOr}{rgb}{0.929,0.431,0.0} % Unibw Orange
\definecolor{colUniBwGr}{RGB}{113,112,114} % Unibw Gray
\definecolor{matlabBlue}{rgb}{0 0.4470 0.7410} % MATLAB blue
\definecolor{matlabOrange}{rgb}{0.8500 0.3250 0.0980} % MATLAB orange
\definecolor{matlabYellow}{rgb}{0.9290 0.6940 0.1250} % MATLAB yellow
\definecolor{matlabGreen}{rgb}{0.4660 0.6740 0.1880} % MATLAB green

\usepackage{tikz}
\usepackage{tikz-3dplot}
\usepackage{pgfplots}
\usetikzlibrary{shapes,arrows,calc,fit,backgrounds,shapes.multipart}

\usepackage{pgfplots}
\pgfplotsset{compat = newest}

\usepackage{amsthm}
\theoremstyle{definition}

\newtheoremstyle{remarkstyle} % https://latex.org/forum/viewtopic.php?t=18631
  {}{}{}{}{\bfseries}{.}{.5em}{{\thmname{#1 }}{\thmnumber{#2}}{\thmnote{ (#3)}}}
\theoremstyle{remarkstyle}

\graphicspath{{figures/}}

\include{defines}

\usepackage[colorlinks=true,
            linkcolor=black,
            citecolor=black,
            urlcolor=black,
            bookmarksnumbered=true,
            pdfstartpage=1,
            pdfpagemode=UseOutlines]{hyperref}

\usepackage[acronym]{glossaries}
\setacronymstyle{long-short}
\glsdisablehyper % Don't treat acronyms as hyperlinks

\makeatletter
\def\ps@pprintTitle{%
  \let\@oddhead\@empty
  \let\@evenhead\@empty
  \def\@oddfoot{\reset@font\hfil\thepage\hfil}
  \let\@evenfoot\@oddfoot
}
\makeatother

\begin{document}

\graphicspath{{figures/}{../figures/}}

\begin{frontmatter}

\title{Modified augmented Lagrangian preconditioning for {\mixeddimensional} {\beamsolid} coupling}

\author[imcs]{Max Firmbach\corref{cor1}}\ead{max.firmbach@unibw.de}
\author[imcs]{Ivo Steinbrecher}\ead{ivo.steinbrecher@unibw.de}
\author[imcs]{Alexander Popp}\ead{alexander.popp@unibw.de}
\author[imcs,dsc]{Matthias Mayr}\ead{matthias.mayr@unibw.de}
\address[imcs]{Institute for Mathematics and Computer-Based Simulation, Universit\"{a}t der Bundeswehr M\"{u}nchen,\\Werner-Heisenberg-Weg 39, D-85577 Neubiberg, Germany}
\address[dsc]{Data Science \& Computing Lab, Universit\"{a}t der Bundeswehr M\"{u}nchen,\\Werner-Heisenberg-Weg 39, D-85577 Neubiberg, Germany}
%\address[dlr]{Institute for the Protection of Terrestrial Infrastructures, Deutsches Zentrum f\"{u}r Luft- und Raumfahrt, North Rhine-Westphalia, Germany}
\cortext[cor1]{corresponding author}

\begin{abstract}
This paper presents modified augmented Lagrangian block preconditioners for the {\mixeddimensional} coupling of three-dimensional
solid bodies with embedded one-dimensional {\torsionfree} {\KirchhoffLove} beams using Lagrange multipliers for constraint enforcement.
The finite element discretization of this mixed formulation leads to an indefinite {\saddlepoint} system.
An augmented Lagrangian formulation is employed to regularize the linear system while maintaining exact enforcement
of the coupling constraints. Starting from the corresponding ideal augmented Lagrangian  block preconditioner, more practical block-triangular
variants are derived in which the solid, beam, and Schur complement blocks can be treated independently. In addition, different variants of Schur
complement approximations are introduced. Numerical experiments investigate the dependence on model parameters and mesh refinement,
as well as weak and strong scalability for short-fiber composite representative volume elements. For the tested parameter regime, a suitable choice of the
Schur complement approximation yields favorable results, indicating the suitability of the proposed approach for large-scale simulations
of {\mixeddimensional} models in solid and structural mechanics, as illustrated by an engineering example involving a composite sandwich plate.
\end{abstract}

\end{frontmatter}

\section{Introduction}
\label{sec:introduction}

In both nature and technical systems,
fibers embedded into solids improve the functional properties of the coupled system,
for example by serving as reinforcements to improve load-bearing capacities under tensile loading.
Examples can be found in many fields:
In human biology, collagen fibers serve as the ubiquitous load-bearing and reinforcing element on the nanometer scale
and, thus, form an important structural basis in both healthy and diseased tissue,
for example in arterial walls~\cite{Holzapfel2008a} or for the design of scaffolds in regenerative medicine~\cite{Cunniffe2011a}.
In civil engineering, fiber-reinforced concrete, {\ie} concrete with randomly dispersed short steel fibers,
comes with many advantages such as improved structural strength (especially in the tensile regime),
crack width reduction, improved abrasion- and impact-resistance, and even post-cracking tensile strength, {\cf} \cite{Aitcin1998a,Fehling2014a,Naaman2011a} among others.
%Similarly, steel-reinforced concrete relies on embedded steel bars to improve its tensile strength.
To analyze and design such systems,
numerical modeling and simulation nowadays plays a key role.
Naturally, accurate but efficient models are required to obtain useful and predictive simulation capabilities.

\begin{figure}[h]
	\centering
	\begin{subfigure}{0.32\textwidth}
		\centering
		\includegraphics[width=\textwidth]{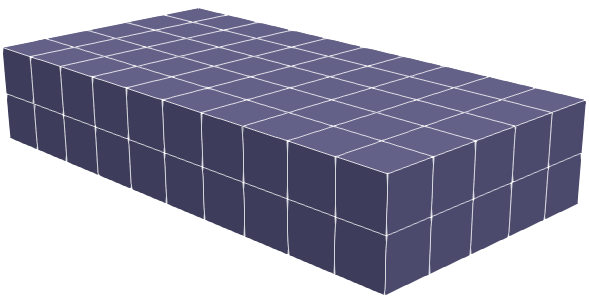}
		\caption{Homogenization}
		\label{fig:ModelingHomogenization}
	\end{subfigure}
	\hfill
	\begin{subfigure}{0.32\textwidth}
		\centering
		\includegraphics[width=\textwidth]{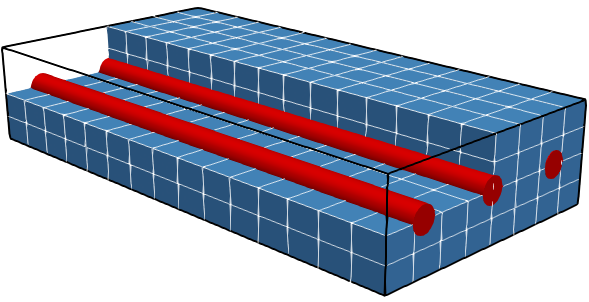}
		\caption{Embedded fibers in continuum}
		\label{fig:ModelingBeamToVolume}
	\end{subfigure}
	\hfill
	\begin{subfigure}{0.32\textwidth}
		\centering
		\includegraphics[width=\textwidth]{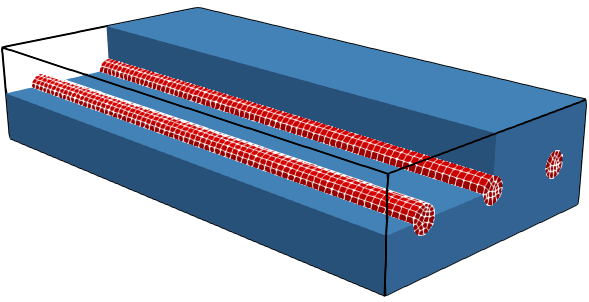}
		\caption{Fully resolved 3D model}
		\label{fig:ModelingFullyResolved}
	\end{subfigure}
	\caption{Spectrum of modeling techniques for fibers embedded into three-dimensional solids~\cite{Steinbrecher2020}.}
	\label{fig:DifferentModelingApproaches}
	\end{figure}

Different modeling approaches are available for the embedding of slender fibers into solid continua (see \figref{fig:DifferentModelingApproaches}):
On the one hand,
homogenization techniques incorporate all fiber information into the bulk constitutive law,
usually leading to anisotropic formulations with preferential directions along the fiber orientation~\cite{Agarwal2017a}.
On the other hand,
both the embedding solid as well as the fibers can be modeled as distinct 3D solid bodies with coupling conditions along the surfaces of the fibers.
While this fully resolved approach delivers the highest modeling accuracy,
its computational effort renders it infeasible for practical application scenarios with many fibers.
As a remedy, {\mixeddimensional} modeling techniques exploit the slenderness of the embedded fibers to represent them by reduced-dimensional
structural models such as trusses or beams, which are then embedded into three-dimensional solid bodies.
A first Gauss-point-to-segment approach using spring-like ``junction elements'' has been outlined in~\cite{Durville2007a}.
A hybrid approach, where fibers are modeled as fully resolved three-dimensional (3D) continua embedded into the
solid via the extended finite element method in the zone of interest and are represented by simple truss elements elsewhere,
has been proposed in~\cite{Le2017a}.
A variationally consistent overlapping domain decomposition approach for {\mixeddimensional} {\beamsolid} coupling has been described in~\cite{Khristenko2021a}.
Recently, we have proposed a {\mixeddimensional} approach to {\fibersolid} coupling using a mortar-type method for the
constraint discretization combined with a penalty regularization for the solution process~\cite{Steinbrecher2025a,Steinbrecher2020,Steinbrecher2022a}.
In addition, we have developed an approximate block factorization preconditioner to tackle the specific challenges of the penalty regularization~\cite{Firmbach2024a}.
Furthermore, {\mixeddimensional} coupling is successfully used for fibers immersed in fluid domains~\cite{Hagmeyer2024a,Hagmeyer2022a,Lespagnol2024}.

Constraint enforcement based on penalty regularization, as used in our prior work~\cite{Firmbach2024a,Steinbrecher2020,Steinbrecher2022a}, results in reduced system sizes,
a straightforward and robust nonlinear solution procedure, and ease of implementation as well as are readily applicable for many engineering systems.
However, their resulting formulation comes with several well-known drawbacks
including variational inconsistency, ill-conditioning of the linear system due to the penalty parameter, and inexact constraint enforcement.
In contrast, the direct use of a Lagrange multiplier field gives rise to a linear system of equations exhibiting {\saddlepoint} structure~\cite[Equation 38]{Steinbrecher2020}.
Yet, it enables exact constraint enforcement
as well as alleviates the user from choosing a penalty parameter, that potentially harms the conditioning and solvability of the arising linear system of equations.
Still, the {\saddlepoint} structure of the discretized linear system poses a challenge,
as preconditioners need to be tailored to the system properties in order to yield an effective iterative solution process.
For indefinite linear systems arising in computational physics, a variety of factorization-based preconditioning strategies is available~\cite{Murphy2000}.
An extensive summary and classification of preconditioning strategies for {\saddlepoint} systems can be found in~\cite{Benzi2005a}.

So far, most of the research is
concerned with the efficient preconditioning of problems related to fluid mechanics, where the discretization of the Navier-Stokes equation with finite elements results
in a {\saddlepoint} system~\cite{Benzi2006a,Benzi2011b,Elman1999,Farrell2019a}. In solid mechanics, the need for {\saddlepoint} block preconditioners
arises when additional constraints in the form of Lagrange multipliers are added to the system as in contact mechanics~\cite{Adams2004a,Franceschini2022a,Wiesner2021}
or fracture mechanics~\cite{Franceschini2019} for example. Furthermore, the coupling of domains with different dimensionality has been realized with Lagrange multipliers. The resulting
{\mixeddimensional} formulation is also of {\saddlepoint} structure and, thus, needs special treatment~\cite{Dimola2024,Kuchta2019,Kuchta2016}.
Another popular class of preconditioners for {\saddlepoint} systems is based on augmented Lagrangian methods~\cite{Benzi2005a}. 
Recently, such methods have been successfully applied to fictitious domain formulations and the associated interface problems~\cite{BenziPreprint,Benzi2026a}.
Except for a diagonal preconditioner based on the augmented Lagrangian being outlined in~\cite{Budisa2020, Budisa2021},
their application to {\mixeddimensional} formulations remains mostly unexplored so far. 
Furthermore, existing work  on specialized preconditioners for {\mixeddimensional} modeling is limited to transport phenomena in the one-dimensional (1D) domain.
To this end, scalable block preconditioners for {\mixeddimensional} {\beamsolid} coupling problems governed by vector-valued elasticity formulations
and using Lagrange multipliers for constraint enforcement remain largely unexplored.

In this contribution, we develop modified augmented Lagrangian block preconditioners for {\mixeddimensional} {\beamsolid} coupling problems with Lagrange multiplier constraint enforcement.
The resulting finite element discretization gives rise to an indefinite {\saddlepoint} system whose efficient solution requires suitable Schur complement approximations and is further complicated
by the presence of a singular beam sub-problem.
The singular beam sub-problem arises because the considered fully embedded beams are a pure Neumann problem, only the coupling to the surrounding solid constrains the beam rigid body modes.
Our main idea is to reformulate the original Lagrange multiplier system in an augmented Lagrangian framework. This reformulation is motivated
not only by the preconditioner construction, but also by the pure Neumann character of the beam sub-problem, which gets regularized by the augmented term.
Starting from the corresponding ideal augmented Lagrangian block preconditioner,
we derive practical block-triangular preconditioners that permit independent treatment of the solid, beam, and Schur complement blocks.
In this context, we investigate several Schur complement approximations,
including variants based on sparse approximate inverses, spectral equivalence properties of the augmentation term,
and separate augmentation parameters for the solid and beam contributions.
Further, we analyze the influence of physical, geometric, discretization, and solver parameters on the resulting methods and
investigate the convergence behavior under mesh-refinement as well as weak and strong scalability on parallel computing clusters.
Finally, we use the proposed approach for large-scale {\mixeddimensional} simulations in solid and structural mechanics.

The remainder of this manuscript is structured as follows. \Secref{sec:equations} presents the relevant equations describing the coupling approach and introduces
the finite element discretization used for the {\mixeddimensional} {\beamsolid} problem, which leads to a block system with {\saddlepoint}
structure. \Secref{sec:analysis} discusses important features of the arising block system, the relevant modeling parameters, and positions the
modeling scheme compared to existing approaches and the current state of literature. \Secref{sec:block_preconditioner} introduces block
preconditioners tailored to the linear system, discusses different variants and Schur complement approximations for practical computations
and compares the preconditioning approach to existing methods in literature.
\Secref{sec:numerical_examples} provides numerical experiments to validate and show the findings related to the proposed preconditioner.
The results include insights into the choice of governing hyper-parameters, modeling parameter influence, strong and weak scaling behavior
as well as the applicability to real world problems. \Secref{sec:conclusion_and_outlook}
summarizes our findings and hints at possible future research directions.

\section{Modeling approach of the {\mixeddimensional} {\beamsolid} problem}
\label{sec:equations}

\begin{figure}
\centering
\includegraphics[scale=1.0]{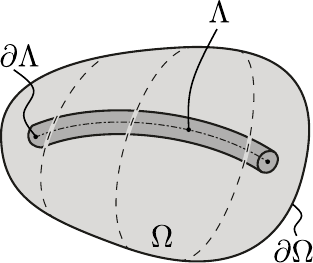}
\caption{Geometrical setup and the corresponding notation showing the three-dimensional solid domain $\dom$,
its boundary $\partial\dom$ and the one-dimensional beam centerline $\Lambda$ with its boundary $\partial\Lambda$.}
\label{fig:domain_visualization}
\end{figure}

We now introduce the governing equations and discretization used to model the coupling of a three-dimensional solid body
with one-dimensional fibers. First, we introduce each physical field individually before combining them to form the overall
coupled problem. An illustration of the geometrical setting is given in \figref{fig:domain_visualization} showing the solid
domain $\Omega$, its boundary $\partial\Omega$ and the beam centerline $\Lambda$ with its boundary $\partial\Lambda$. As the main focus of this publication is the construction
of robust and efficient preconditioning techniques, we mainly give a brief overview of the
mechanical theory. The interested reader is referred to \cite{Steinbrecher2020} for more details.

Before dealing with equations, we introduce
some notation that is used throughout the remainder of this manuscript: All quantities denoted with $(\cdot)^{\indexSolid}$ are
associated with the solid continuum, while quantities with superscript $(\cdot)^{\indexBeam}$ are related to beam contributions.
In addition, let $\mathcal{D}$ be a bounded domain in $\mathbb{R}^{\ndim}$ with $\ndim=1,2,3$. Then, $H^{s}(\mathcal{D})$ for $s \in \mathbb{N}_0$ are the
classical Sobolev spaces. %and with a slight abuse of notation, we consider $H^{-s}(\mathcal{D})$ to be the dual of $H^{s}(\mathcal{D})$.
As we solely treat vector-valued unknowns with $\ndim$ components, we also introduce the corresponding product
Sobolev spaces $[H^{s}(\mathcal{D})]^\ndim$. To distinguish continuous quantities from their discrete counterparts, we use the subscript $(\cdot)_h$
to denote the latter. In addition, $\variation$~indicates virtual, but kinematically admissible quantities in line with the concept of
virtual work. Tensors of second order and higher are denoted in bold font.

\subsection{Pure solid problem}

The solid domain $\dom \subset \REalSp{3}$ is governed by the equations of quasi-static nonlinear elasticity,
stated in the material configuration as
\begin{alignat}{2}
\matDiv \; \tensor{P} + \hat{b} &= 0  &&\quad\quad \text{in} \; \dom \nonumber,\\
\indexedSolid{u} &= \indexedSolid{u}_D &&\quad\quad \text{on} \; \Gamma_D, \label{eq:solid_strong_form} \\
\tensor{P} \cdot n &= \hat{t} &&\quad\quad \text{on} \; \Gamma_N \nonumber,
\end{alignat}
where $\Gamma_D$ and $\Gamma_N$ denote the Dirichlet and Neumann parts of the boundary
$\partial\dom = \Gamma_D \cup \Gamma_N$ with $\Gamma_D \cap \Gamma_N = \emptyset$, respectively. Hereby, $\tensor{P}$ represents the first {\PiolaKirchhoff} stress
tensor, while $\indexedSolid{u}$ defines the displacement, $\hat{b}$ the body forces acting on the solid continuum, $\hat{t}$ the tractions at the Neumann boundary
and $n$ the outward-pointing normal vector in the material configuration. We choose
$\indexedSolid{\mathscr{V}}(\Omega) := [H^1(\Omega)]^3$ as the underlying Sobolev space for test and basis functions. Further, we transform \eqref{eq:solid_strong_form}
into the variational formulation: \\ Find the displacement vector $\indexedSolid{\unknown} = (\indexedSolid{\unknown}_x, \indexedSolid{\unknown}_y, \indexedSolid{\unknown}_z)
\in \indexedSolid{\mathscr{V}}_D(\Omega):=\{\indexedSolid{\unknown}\in\indexedSolid{\mathscr{V}}(\Omega)\;|\;\text{trace}_{\Gamma_D}\;\indexedSolid{u}=\indexedSolid{u}_D\}$ such that
\begin{equation}
\indexedSolid{\weakForm} = 0 \quad\quad \forall \; \indexedSolid{\variation\unknown} \in \indexedSolid{\mathscr{V}}_{0}(\Omega):=\{\indexedSolid{\variation\unknown}\in\indexedSolid{\mathscr{V}}(\Omega)\;|\;\text{trace}_{\Gamma_D} \;\indexedSolid{\variation\unknown}=0\}.
\label{eq:solid_weak_form}
\end{equation}
Therein, we define the overall virtual work $\indexedSolid{\weakForm}$ as difference between the variation of the potential energy~$\indexedSolid{\Pi}$ of the solid and
the respective virtual work of the external load reading
\begin{equation*}
\indexedSolid{\weakForm} = \indexedSolid{\variation\Pi}_{\text{int}} - \indexedSolid{\weakForm}_{\text{ext}}
= \int_{\Omega} \tensor{S}:\variation \tensor{E} \; \mathrm{d}V
- \int_{\Omega} \hat{b} \cdot \indexedSolid{\variation\unknown} \; \mathrm{d}V
- \int_{\Gamma_N} \hat{t} \cdot \indexedSolid{\variation\unknown} \; \mathrm{d}A,
\end{equation*}
with $\tensor{S} := \inv{\tensor{F}}\cdot\tensor{P}$ and $\tensor{E}$ representing the second {\PiolaKirchhoff} stress tensor and the {\GreenLagrange} strain tensor, respectively.
Here, $\tensor{F}$ describes the deformation gradient. In addition, we define strain tensor~$\tensor{E}$ and stress tensor~$\tensor{S}$ as
\begin{equation*}
\tensor{E}:=\frac{1}{2}(\trans{\tensor{F}}\cdot\tensor{F}-\tensor{I}) \quad \text{and} \quad
\tensor{S} = \frac{\partial\Psi}{\partial\tensor{E}},
\end{equation*}
with strain energy density $\Psi$. For a {\StVenantKirchhoff} material, $\Psi$ follows as
\begin{equation*}
\Psi := \frac{\indexedSolid{\youngs}\poisson}{2(1+\nu)(1-2\poisson)}(\text{trace} \; \tensor{E})^2 + \frac{\indexedSolid{\youngs}}{2(1+\poisson)} \tensor{E} : \tensor{E},
\end{equation*}
where $\indexedSolid{\youngs}$ represents the solid Young's modulus and $\poisson$ the Poisson's ratio of the solid body.
Without loss of generality, we restrict ourselves to a {\StVenantKirchhoff} material model in the scope of this publication.
By discretizing \eqref{eq:solid_weak_form} with continuous linear {\Lagrangian} finite elements,
we obtain a linear system to be solved stated as
\begin{equation*}
\indexedSolid{K}\indexedSolid{\unknown}_h = \indexedSolid{f},
\end{equation*}
with stiffness matrix $\indexedSolid{K}\in\mathbb{R}^{ \indexedSolid{n}_{\text{DOF}}\times \indexedSolid{n}_{\text{DOF}}}$, load vector $\indexedSolid{f}\in\mathbb{R}^{ \indexedSolid{n}_{\text{DOF}}}$
and $\indexedSolid{\unknown_h}\in\mathbb{R}^{ \indexedSolid{n}_{\text{DOF}}}$ being the vector of the discrete displacement degrees of freedom.

\subsection{Pure beam problem}

We consider {\torsionfree} beams based on the {\KirchhoffLove} formulation \cite{Kirchhoff1859a, Love1944a},
assuming vanishing shear deformations and no torsion contribution to the internal elastic energy,
which results in a purely displacement-based formulation. These assumptions are valid for fibers with high slenderness ratios,
a double symmetric {\crosssection} and a straight centerline in the reference configuration.
The resulting formulation represents a simplification, as rotational degrees of freedom are not considered explicitly. This modeling choice
is made in view of the primary focus of the present work, namely the investigation of the proposed block preconditioning strategy rather than
the treatment of advanced beam kinematics. At the same time, the concepts developed in this work are not restricted to the present setting and
could, in principle, also be extended to more general geometrically exact beam formulations that account for torsion and shear effects,
like the {\SimoReissner} beam theory \cite{Reissner1972, Simo1985}.
We employ beams  with a constant circular
{\crosssection} along the beam centerline $\Lambda$. For a more detailed introduction on geometrically
exact beam formulations, we refer to \cite{Meier2015a, Meier2019a}.
We define $\indexedBeam{\mathscr{V}}(\Lambda):=[H^2(\Lambda)]^3$ and consider the following variational formulation: \\
Find $\indexedBeam{\unknown} = (\indexedBeam{\unknown}_x, \indexedBeam{\unknown}_y, \indexedBeam{\unknown}_z)
\in \indexedBeam{\mathscr{V}}_{D}(\Lambda):=\{\indexedBeam{\unknown}\in\indexedBeam{\mathscr{V}}(\Lambda)\;|\;\text{trace}_{\partial\Lambda_{D}}\;\indexedBeam{u}=\indexedBeam{u_D}\}$ such that
\begin{equation*}
\indexedBeam{\weakForm} = 0 \quad\quad \forall \; \delta \indexedBeam{\unknown} \in \indexedBeam{\mathscr{V}}_{0}(\Lambda):=\{\indexedBeam{\variation\unknown}\in\indexedBeam{\mathscr{V}}(\Lambda)\;|\;\text{trace}_{\partial\Lambda_{D}}\;\indexedBeam{\variation\unknown}=0\}.
\end{equation*}
The global contribution of the beams to the overall virtual
work is again given as the difference of the variation of the potential energy and the variation of the external work
\begin{equation*}
\indexedBeam{\weakForm} = \indexedBeam{\variation\Pi}_{\text{int}} - \indexedBeam{\weakForm}_{\text{ext}} \quad \text{with} \quad \indexedBeam{\Pi_{\text{int}}} = \frac{1}{2} \int_{\Lambda} \indexedBeam{\youngs}A\varepsilon^2 + \indexedBeam{\youngs}I\kappa^2 \; \mathrm{d}s,
\end{equation*}
where $\indexedBeam{\youngs}$ denotes the Young’s modulus, $A$ the {\crosssection} area, $I$ the second moment of area, $\varepsilon$ the axial tension
and $\kappa$ the scalar curvature of the beams, respectively. Due to the representation of the curvature~$\kappa$ by the Frenet--Serret vector, which
solely depends on the beam centerline, the finite element discretization of the beam requires a $C^1$ continuous interpolation of the centerline along~$\Lambda$.
Therefore, we consider continuous finite elements based on cubic Hermite polynomials for the discretization \cite{Meier2015a}, resulting in
\begin{equation*}
\indexedBeam{K}\indexedBeam{\unknown}_h = \indexedBeam{f}.
\end{equation*}
Hereby, $\indexedBeam{K}\in\mathbb{R}^{ \indexedBeam{n}_{\text{DOF}}\times \indexedBeam{n}_{\text{DOF}}}$ represents the stiffness matrix, $\indexedBeam{f}\in\mathbb{R}^{ \indexedBeam{n}_{\text{DOF}}}$
the load vector and $\indexedBeam{\unknown}_h\in\mathbb{R}^{ \indexedBeam{n}_{\text{DOF}}}$ the discretized beam displacement vector.

\subsection{Coupled {\beamsolid} system}
\label{subsec:coupled_problem}

In a final step, we consider the coupled system between the solid and the beam field.
For the mixed-dimensional coupling of the 3D solid domain~$\Omega$ and the 1D beam domain~$\Lambda$, we introduce a Lagrange multiplier
field~${\lambda = (\lambda_x, \lambda_y, \lambda_z)}$ to enforce
\begin{equation}
\gap := \indexedSolid{\unknown} - \indexedBeam{\unknown} = 0 \quad\quad \text{on} \; \Lambda,
\label{eq:positional_coupling_condition}
\end{equation}
with the vector-valued gap function. We note that \eqref{eq:positional_coupling_condition} is of Dirichlet type. The constraints given
in \eqref{eq:positional_coupling_condition} are referred to as positional coupling,
since they enforce the position of the beam {\crosssection} centroid to be coupled to the underlying solid.
We remark that due to the {\torsionfree} {\KirchhoffLove} beam formulation a pure coupling of positions is sufficient.
Considering more complex geometrically exact beam theories makes the coupling of rotational degrees of freedom necessary,
as e.g. presented in \cite{Steinbrecher2022}, yet we do not consider this approach in the scope of this publication.
Due to the introduction of the Lagrange multiplier field, the extended overall
virtual work of the constraint {\mixeddimensional} problem is defined as
\begin{equation*}
\weakForm = \indexedSolid{\weakForm} + \indexedBeam{\weakForm} + \variation\Pi^{\lambda} = 0,
\end{equation*}
with the addition of the variation of the Lagrange multiplier potential $\variation\Pi^{\lambda}$.
The Lagrange multiplier potential~$\Pi^{\lambda}$ and its variation~$\variation\Pi^{\lambda}$ are given as
\begin{equation*}
\Pi^{\lambda} = \int_{\Lambda} \lambda \; \gap \; \mathrm{d}s
\qquad \text{and} \qquad
\variation \Pi^{\lambda} = \int_{\Lambda} \variation\lambda \; \gap \; \mathrm{d}s + \int_{\Lambda} \lambda \; \variation\gap \; \mathrm{d}s
.
\end{equation*}

We define the product space for our displacement and Lagrange multiplier solution as $\indexedSolid{\mathscr{V}}(\Omega) \times \indexedBeam{\mathscr{V}}(\Lambda) \times \mathscr{Q}(\Lambda)$
where $\mathscr{Q}(\Lambda)$ denotes the Lagrange multiplier space.
Thus, the final variational formulation of the coupled problem reads:
Find $\unknown = (\indexedSolid{\unknown}, \indexedBeam{\unknown}, \lm) \in \indexedSolid{\mathscr{V}}_{D}(\Omega) \times \indexedBeam{\mathscr{V}}_{D}(\Lambda) \times \mathscr{Q}(\Lambda)$ such that
\begin{equation}
\label{eq:coupled_weak_form}
\weakForm = 0 \quad \forall \; \delta \indexedSolid{\unknown} \in \indexedSolid{\mathscr{V}}_{0}(\Omega),\; \forall \; \delta \indexedBeam{\unknown} \in \indexedBeam{\mathscr{V}}_{0}(\Lambda), \; \forall \; \delta \lambda \in \mathscr{Q}(\Lambda)
.
\end{equation}
For the spatial discretization of~\eqref{eq:coupled_weak_form}, we consider a mortar-type approach for the discretization of the Lagrange multiplier field.
As the continuous Lagrange multiplier field is defined along the beam centerline,
the interpolation is defined along the one-dimensional beam elements. In the nomenclature
of classical contact mechanics, the beam would be considered the source side, and the
solid the target side \cite{Popp2009, Steinbrecher2020}. After discretization with continuous finite elements,
we obtain the linear system of equations of the coupled problem
\begin{equation}
\linearOperator \unknown_h = \rhs, \;\; \text{with} \;\;
\linearOperator :=
\begin{pmatrix}
\indexedSolid{K} & ~ & ~ -\trans{\matSolidLm} \\
~ & \indexedBeam{K} & \trans{\matBeamLm} \\
-\matSolidLm & \matBeamLm & ~ \\
\end{pmatrix}
\;\; \text{and} \;\;
\rhs := 
\begin{pmatrix}
\indexedSolid{\residualNonlinear} \\
\indexedBeam{\residualNonlinear} \\
\gap \\
\end{pmatrix},
\label{eq:discretized_coupled_system}
\end{equation}
with the discrete solution vector $\unknown_h$ sought in the corresponding product of solid, beam, and Lagrange multiplier finite element spaces
and solved for in every iteration of the nonlinear Newton-type solver. By construction, we consider a non-conforming discretization arising
from non-matching (and even mixed-dimensional) meshes across subdomain interfaces $\Omega \cap \Lambda$.
This gives rise to the off-diagonal coupling operators~$\matSolidLm\in\mathbb{R}^{n^{\lambda}_{\text{DOF}}\times \indexedSolid{n}_{\text{DOF}}}$
and~$\matBeamLm\in\mathbb{R}^{n^{\lambda}_{\text{DOF}}\times \indexedBeam{n}_{\text{DOF}}}$.
Again, the interested reader is referred to our previous work \cite{Steinbrecher2020} for detailed derivations.
In the absence of explicit Dirichlet boundary conditions on the embedded beams, the respective displacement unknowns are constrained only through the Lagrange multiplier field.
Consequently, the discrete beam operator itself is singular, as it inherits the rigid-body nullspace of the corresponding pure Neumann beam problem. Nevertheless, the positional
coupling eliminates this nullspace within the coupled system, so the singularity of the isolated beam operator does not compromise the well-posedness of the coupled formulation.

\section{Analysis of the coupled problem}
\label{sec:analysis}

Finding a solution of the block system given in \eqref{eq:discretized_coupled_system} in a fast and efficient manner
for increasing problem sizes is only feasible with an iterative linear solver. These methods usually converge slowly without a properly
constructed preconditioner $\mathcal{P}$. The aim of this publication is to find an appropriate block preconditioner
specifically tailored to the underlying physical problem. For the sake of completeness, the preconditioned system reads
\begin{equation*}
\inv{\mathcal{P}}\linearOperator \unknown_h = \inv{\mathcal{P}}\rhs.
\end{equation*}
In the scope of this publication, we consider block triangular preconditioners of the generic form
\begin{equation}
\mathcal{P}
:=
\begin{pmatrix}
\indexedSolid{K} & ~ & ~ \\
~ & \indexedBeam{K} & ~ \\
-\matSolidLm & \matBeamLm & S \\ 
\end{pmatrix}
\label{eq:naive_block_preconditioner}
\end{equation}
with the Schur complement $S:=-\matSolidLm \inv{(\indexedSolid{K})} \trans{\matSolidLm} - \matBeamLm \inv{(\indexedBeam{K})} \trans{\matBeamLm}$.
In the following, we see that this naive preconditioning approach is not feasible for the range of applications we consider.
Nevertheless we propose to use \eqref{eq:naive_block_preconditioner} as a starting point for more sophisticated versions of $\mathcal{P}$
later on. Before constructing these block preconditioners though, we analyze the linear system given in \eqref{eq:discretized_coupled_system} first.
We consider the influence of the relevant modeling parameters and highlight important features of the block matrix structure. In a next
step, we compare the introduced Lagrange multiplier coupling approach to methods relying on a penalty-based regularization to enforce the coupling
constraints. In a last step, we discuss the solvability of the beam sub-problem.

Before starting the analysis of the linear system, we introduce the scaling matrix $W\in\mathbb{R}^{n^{\lambda}_{\text{DOF}}\times n^{\lambda}_{\text{DOF}}}$
as it will be used throughout the remainder of this manuscript. We postpone the discussion on how to choose $W$ to later. For now, we state that $W$ is derived
from the findings given in \citep{Steinbrecher2020} being represented by a diagonal matrix of mass-matrix-type at the coupling interface $\Lambda$.

\subsection{Influence of modeling parameters and modeling assumptions}
\label{subsec:modeling_parameters}

The {\mixeddimensional} coupling between a solid continuum and fibers is dominated by a few modeling parameters, which highly influence the
conditioning of the linear system given in \eqref{eq:discretized_coupled_system}. Those parameters can be related to the material models, the geometry,
and the discretization of the coupled problem. We shortly want to introduce these parameters and discuss their importance for our range of applications.
We note that our choice regarding the discretization and geometric relations is based on the original {\mixeddimensional} model~\cite{Steinbrecher2020},
in particular the assumption of~$\indexedSolid{h} \geq 2\indexedBeam{\beamRadius}$, {\ie} that the solid mesh size~$\indexedSolid{h}$ 
needs to be larger than the beam {\crosssection} dimensions.
Up until that point, the {\mixeddimensional} model is valid and produces similar results to a fully resolved coupling (i.e., coupling on the beam surface).
If $\indexedSolid{h}$ becomes smaller than the beam radius~$\indexedBeam{\beamRadius}$ and violates the modeling assumptions of~\cite{Steinbrecher2020},
the {\mixeddimensional} approach reproduces a singular solution artificially introduced by the 1D-3D modelling assumptions, which is unphysical.
Thus, such cases are not considered in this publication.

We further note that the mathematical analysis of the well-posedness of the continuous 1D-3D coupling entails additional difficulties that are not addressed in the present work.
In particular, for the standard solid space~$[H^1(\Omega)]^3$, the trace of the solid displacement on the 1D beam centerline~$\Lambda$ is not well defined.
Conversely, a Lagrange multiplier distributed along~$\Lambda$ acts as a line load on the three-dimensional continuum and does not belong to the dual of the natural solid energy space.
A rigorous treatment of the continuous 1D-3D problem therefore requires, for instance, weighted Sobolev spaces or suitably defined fractional/operator-based norms, see~\cite{Cattaneo2015,DAngelo2012}
and related analyses in~\cite{Cerroni2019,Dimola2024,Kuchta2021}. For the cases considered in this work, the mortar-type discretization and mass-type scaling introduced in Section~\ref{subsec:coupled_problem}
are discrete devices that mask the underlying continuous ill-posedness problem. This mathematical limitation is consistent with the modeling restriction $\indexedSolid{h} \geq 2\indexedBeam{\beamRadius}$ introduced
in~\cite{Steinbrecher2020} based on mechanical considerations. The present formulation is intended for a regime in which the beam cross-section remains unresolved by the solid discretization,
rather than as an approximation of the singular continuous 1D-3D problem under arbitrary mesh refinement.

We use \cite{Lauff2025b, Lauff2025a, Schneider2022} as guidelines to match the modeling parameters for practical application cases.
An overview of all relevant parameters is given in \tabref{tab:quantities}.
We first discuss the parameter related to the material, which is given by the stiffness ratio $\mathcal{E} := \indexedBeam{\youngs}/\indexedSolid{\youngs}$
of the Young's modulus of the beam and the solid, respectively. We assume $\indexedBeam{\youngs} > \indexedSolid{\youngs}$, which can be interpreted
as the beams acting as a reinforcement of the solid continuum. Usually, the ratio varies between $\mathcal{E} \approx 2$ for metal-based composites and
can go up to $\mathcal{E} > 1000$ for niche applications. We mostly consider $\mathcal{E} \in [10, 1000]$, as one would find for example in steel-reinforced
concrete, short fiber polymer matrix applications or in carbon reinforced composites. Increasing values of the stiffness ratio result in a scaling mismatch and
therefore result in a worse conditioning of \eqref{eq:discretized_coupled_system}, thus complicating the solvability with an iterative linear solver.

The second group of parameters is related to geometric and discretization-based quantities given by the beam {\crosssection} radius $\indexedBeam{\beamRadius}$,
the characteristic beam element length $\indexedBeam{h}$ and the {\beamtosolid} volume ratio $\mathcal{\volume}:=\indexedBeam{\volume}/\indexedSolid{\volume}$.
In order not to violate the geometrically exact beam finite element formulation, we assume beam slenderness ratios $\indexedBeam{\xi} := \indexedBeam{h}/(2\indexedBeam{R}) \gg 1$.
A decreasing beam {\crosssection} radius results in an increasing condition number of the beam sub-matrix in \eqref{eq:discretized_coupled_system}.
In addition, we assume the solid mesh size $\indexedSolid{h}$ to be greater or equal the diameter of the beam element $\indexedSolid{h} \geq (2\indexedBeam{\beamRadius})$
and expect the solid mesh size to be smaller than the characteristic beam element length $\indexedSolid{h} \leq \indexedBeam{h}$ \cite{Steinbrecher2020}.
We denote the ratio between characteristic beam element length and solid mesh size by $\mathcal{H}:=\indexedBeam{h}/\indexedSolid{h}$. In engineering
applications, the amount of beams inside a solid is usually defined by the {\beamtosolid} volume ratio $\mathcal{\volume}$, which is indirectly related to the
geometric quantities introduced before. We consider the {\beamtosolid} volume ratio to be at maximum $\mathcal{\volume}\approx10\%$. We argue that higher values
can be modeled more efficiently by a homogenization of the material properties, without explicitly modeling the beams. We consider application cases like they
would appear for steel-reinforced concrete, which usually feature $\mathcal{\volume}\approx 1\%-3\%$. The {\beamtosolid} volume ratio for short fiber polymer
matrix composites can be  $\mathcal{\volume}\approx 10\%-30\%$. High ratios of $\mathcal{\volume}$ potentially increase the bandwidth of the coupling matrices $D$ and $M$ and therefore
also of the Schur complement, leaving preconditioning and solving of \eqref{eq:discretized_coupled_system} an even more difficult task. 

\begin{table}
\centering
\caption{Overview of the relevant modeling parameters with their notations and units.}
\begin{tabular}{l l l}
\hline
Notation & Parameter & Unit \\
\hline
$\indexedSolid{\youngs}$ & Young's modulus of the solid domain & $\qty{}{\newton/\square\meter}$ \\
$\indexedSolid{h}$ & Solid element size & $\qty{}{\meter}$ \\
$\indexedSolid{\volume}$ & Volume of the solid domain & $\qty{}{\cubic\meter}$ \\
\hline
$\indexedBeam{\youngs}$ & Young's modulus of the beam domain & $\qty{}{\newton/\square\meter}$ \\
$\indexedBeam{h}$ & Beam element length & $\qty{}{\meter}$ \\
$\indexedBeam{\beamRadius}$ & Beam {\crosssection} radius & $\qty{}{\meter}$ \\
$\indexedBeam{\volume}$ & Volume of the beam domain & $\qty{}{\cubic\meter}$ \\
\hline
$\mathcal{\volume} = \indexedBeam{\volume}/\indexedSolid{\volume}$ & Beam-to-solid volume ratio & - \\
$\mathcal{E} = \indexedBeam{\youngs}/\indexedSolid{\youngs}$ & Beam-to-solid stiffness ratio & - \\
$\mathcal{H} = \indexedBeam{h}/\indexedSolid{h}$ & Beam-to-solid element size ratio & - \\
\hline
\end{tabular}
\label{tab:quantities}
\end{table}

%\begin{remark}[Modeling assumptions regarding $\indexedSolid{h}$ and $\indexedBeam{\beamRadius}$]
%\label{re:modeling_assumptions}
%As discussed and shown in the original work~\cite{Steinbrecher2020},
%the {\mixeddimensional} coupling approach described in~\secref{sec:equations} is only valid in a distinct range of parameters.
%Specifically, we assume $\indexedSolid{h} \geq 2\indexedBeam{\beamRadius}$, such that the beam {\crosssection}
%dimensions are smaller than the characteristic solid finite element mesh size~$\indexedSolid{h}$. If $\indexedSolid{h}$ becomes smaller,
%the {\mixeddimensional} approach does not achieve spatial mesh convergence in the fine mesh limit~\cite{Steinbrecher2020}
%and, thus, such cases are not considered in this publication.
%\end{remark}

\subsection{Influence of block matrix structure and sub-solves}
\label{subsec:matrix_structure}

By enforcing the positional Dirichlet coupling $\eqref{eq:positional_coupling_condition}$ by a Lagrange multiplier field,
the resulting linear operator has {\saddlepoint} structure with a zero $(3,3)$-block on the main diagonal. Due to the {\saddlepoint}
structure, the overall block system is indefinite. Both the solid and beam stiffness matrices are stemming from finite element
discretizations of elliptic operators and feature positive definiteness and are in addition also symmetric. As we use a non-symmetric preconditioner
of block triangular structure (as given in \eqref{eq:naive_block_preconditioner}), we consider a generalized minimal residual
method (GMRES) \cite{Saad1986} as linear solver for the remainder of this manuscript. Further, the {\saddlepoint} structure of
the block matrix restricts the type of preconditioners to be applied. Due to the zero block on the main block diagonal,
black-box methods like block Jacobi or block Gauss-Seidel are not applicable, further emphasizing the use of Schur complement
based approximate block factorization preconditioners. As the solid problem is of elliptic nature, standard algebraic
multigrid~(AMG) \cite{Griebel2003a, Vanek1996a} methods present themselves as favorable option for preconditioning
the respective matrix sub-block. For our application case of several short and independent fibers, we consider only interactions
between the fibers and the surrounding solid continuum, but not between the fibers. This results in the matrix $\indexedBeam{K}$ featuring
a block-diagonal sparsity pattern. The size of each diagonal block depends on the amount of beam elements used to discretize a fiber.
Therefore, direct methods and the corresponding factorization are still efficient to a certain degree even for systems with many fibers.
While multigrid methods would in general work too, their application comes with challenges that are still topic of current research,
as highlighted in~\cite[Remark 4.1]{Firmbach2024a}. As we employ different discretizations on the solid, beam and Lagrange multiplier
field, and therefore have non-matching interfaces, the coupling matrices $M$ and $D$ have block structure, but feature no other special
sparsity pattern, for example such as being diagonal like in \cite{Dimola2024}. As the coupling matrices $M$ and $D$ potentially have a large bandwidth,
the Schur complement itself might also feature many {\nonzero} entries. Still, the system size is usually small, which again makes direct
methods feasible.

\subsection{Comparison to a penalty-based formulation}
\label{subsec:penalty_formulation}

Another popular approach in computational mechanics to ease the solvability of \eqref{eq:discretized_coupled_system} is the application
of a penalty regularization. Hereby, the positional coupling \eqref{eq:positional_coupling_condition} is enforced using the relation
\begin{equation}
\lambda = \penaltyParam \inv{W} \gap
\label{eq:penalty_regularization}
\end{equation}
with penalty parameter $\penaltyParam \in \mathbb{R}^{+}$, resulting in a purely displacement-based formulation
\begin{equation}
\linearOperator \unknown_h = \rhs, \;\; \text{with} \;\;
\linearOperator =
\begin{pmatrix}
\indexedSolid{K} + \penaltyParam \trans{\matSolidLm} \inv{W} \matSolidLm & -\penaltyParam \trans{\matSolidLm} \inv{W} \matBeamLm  \\
-\penaltyParam \trans{\matBeamLm} \inv{W} \matSolidLm & \indexedBeam{K} + \penaltyParam \trans{\matBeamLm} \inv{W} \matBeamLm \\
\end{pmatrix}
\;\; \text{and} \;\;
\rhs = 
\begin{pmatrix}
\indexedSolid{\residualNonlinear} - \penaltyParam\trans{M}\inv{W}\gap \\
\indexedBeam{\residualNonlinear} + \penaltyParam\trans{\matBeamLm}\inv{W}\gap \\
\end{pmatrix}.
\label{eq:penalty_discretized_coupled_system}
\end{equation}
Undeniably, one advantage of the penalty-based formulation is the reduced system size due to eliminating the Lagrange multiplier
field as an independent unknown through the penalty regularization \eqref{eq:penalty_regularization}. Yet, we argue that there are
also several disadvantages in using \eqref{eq:penalty_discretized_coupled_system}, which are mainly related to preconditioning and
efficient solvability. The regularization introduces a severe ill-conditioning arising from the penalty parameter
$\penaltyParam$ as well as destroys block-diagonal dominance of $\linearOperator$ in \eqref{eq:penalty_discretized_coupled_system}
as discussed in our prior work~\cite{Firmbach2024a}.
To enforce the constraints \eqref{eq:positional_coupling_condition} properly, the penalty parameter should be
chosen $\penaltyParam \approx \indexedBeam{\youngs}$ \cite{Steinbrecher2020}, which in return results in a high condition number of the
system matrix. Therefore, preconditioners specifically tailored to the problem formulation are necessary. One common approach is the
use of block factorization based preconditioners. The appearing approximated Schur complement in such methods usually results in a sparse matrix 
with comparably high bandwidth due to the introduction of the penalty terms. Therefore, iteratively solving and preconditioning 
\eqref{eq:penalty_discretized_coupled_system} is difficult and computationally demanding, especially leaving traditional
smoothing-based methods as they usually appear in multigrid methods inefficient. While progress has been made recently related
to preconditioning of the penalty-regularized system based on multilevel algorithms \cite{Budisa2024}, its application to
\eqref{eq:penalty_discretized_coupled_system} is not trivial as the coupling is enforced on two fundamentally different partial
differential equations (PDE). On the other hand, block preconditioning
approaches as given in \cite{Firmbach2024a} come with a high computational cost.

\subsection{Pure Neumann problem related to $\indexedBeam{K}$}
\label{subsec:pure_neumann}

As we consider fibers completely embedded into the solid body, usually no explicit Dirichlet conditions are imposed onto the one-dimensional
beam domain. When looking at the entire block system, the rigid body modes of the fibers are fully constrained by the Lagrange multiplier field
that enforces the positional coupling to the solid domain. Hence, the overall system is uniquely solvable. Things change however, when considering
the individual sub-blocks as done inside the block preconditioner and thus the linear solver (given in \eqref{eq:naive_block_preconditioner}).
Therein, the sub-block related to the beam contribution $\indexedBeam{K}$ is singular after discretization due to the missing Dirichlet
conditions on the beam centerline $\Lambda$ and its boundary $\partial\Lambda$. Therefore, the sub-block is not uniquely solvable as solutions
are only determined up to the rigid body modes.

Different approaches to solve pure Neumann problems are presented in literature with the so-called null space playing a crucial role.
For three-dimensional elasticity problems, the null space is defined by the space of rigid body modes consisting of three
translations and three rotations. In our case, the Neumann problem is defined on the domain governed by the equations of a
torsion-free Kirchhoff-Love beam formulation. Due to the fact, that the considered beams cannot represent torsion deformation,
the null space is defined by only five components, which fully describe the space of rigid body modes.

The null space can be used to enforce additional constraints by Lagrange multipliers, which make the system uniquely
solvable \cite{Bochev2005, Kutcha2018}. However, this requires to solve yet another {\saddlepoint} system. Another approach
considers a projection operator to remove the null space from the solution of the linear system~\cite{Bochev2005, Kutcha2018}.
In addition, one can apply a rank-$m$ correction with $m$ being the number of independent null space components to shift the critical eigenvalues
to make the linear operator invertible. However, the resulting projection matrix is usually dense and therefore difficult to treat~\cite{Benzi2024}.
While deflation techniques related to iterative linear solvers can avoid the explicit construction of the projected matrix,
their implementation is invasive, making the use of already available software libraries inconvenient~\cite{Baggio2017}.

In the scope of this publication, we do not further explore null space related approaches, but try to reformulate the
linear system \eqref{eq:discretized_coupled_system} itself to avoid the pure Neumann problem in the first place, as is
shown in \secref{sec:block_preconditioner}.

\section{Block preconditioning of the {\mixeddimensional} problem}
\label{sec:block_preconditioner}

We now discuss the novelty and main idea of this publication: the construction of modified augmented Lagrangian block preconditioners
for the {\mixeddimensional} {\beamsolid} coupled problem. As our naive preconditioning approach \eqref{eq:naive_block_preconditioner} is not
directly applicable to the coupled problem and based on the general findings discussed in \secref{sec:analysis}, we propose to reformulate
\eqref{eq:discretized_coupled_system} into its augmented formulation~\cite{Fortin1983, Hestenes1969, Powell1969} reading
\begin{equation}
\linearOperator \unknown_h = \rhs, \;\; \text{with} \;\;
\linearOperator =
\begin{pmatrix}
\indexedSolid{K} + \penaltyParam \trans{\matSolidLm} \inv{W} \matSolidLm & -\penaltyParam \trans{\matSolidLm} \inv{W} \matBeamLm & -\trans{\matSolidLm} \\
-\penaltyParam \trans{\matBeamLm} \inv{W} \matSolidLm & \indexedBeam{K} + \penaltyParam \trans{\matBeamLm} \inv{W} \matBeamLm & \trans{\matBeamLm} \\
-\matSolidLm & \matBeamLm & ~ \\ 
\end{pmatrix}
\;\; \text{and} \;\;
\rhs =
\begin{pmatrix}
\indexedSolid{\residualNonlinear} - \penaltyParam\trans{M}\inv{W}\gap \\
\indexedBeam{\residualNonlinear} + \penaltyParam\trans{\matBeamLm}\inv{W}\gap \\
\gap
\end{pmatrix}.
\label{eq:augmented_discretized_coupled_system}
\end{equation}
The linear system given in \eqref{eq:augmented_discretized_coupled_system} can be seen as a combination of the pure Lagrange multiplier approach
\eqref{eq:discretized_coupled_system} and the method based on a penalty regularization~\eqref{eq:penalty_discretized_coupled_system}. The augmented
Lagrangian formulation combines the advantages of both approaches by keeping the {\saddlepoint} structure of the original system and therefore solving exactly for the constraints,
but also adding a penalty contribution to stabilize the solution process \cite{Burman2023}. The system given in~\eqref{eq:augmented_discretized_coupled_system}
is therefore less sensitive to changes in parameters and better conditioned than the original one. In addition, the Schur complement is modified, which is
discussed further in \secref{subsec:choice_of_schur_complement}. We opt to use the augmented Lagrangian formulation not only in the preconditioner itself,
but already assemble the full system during the nonlinear solution process, as the relevant terms are already available there. This also avoids the complexity
of handling two system matrices, namely~\eqref{eq:discretized_coupled_system} and~\eqref{eq:augmented_discretized_coupled_system}, at the same time for the
nonlinear iteration, the preconditioner and the respective linear solver.

Considering the model parameters described in \secref{subsec:modeling_parameters}, their influence on the augmented system~\eqref{eq:augmented_discretized_coupled_system}
does not change in comparison to the pure Lagrange multiplier formulation. Similar to the penalty regularization, we introduce one additional parameter to steer the
constraint enforcement, given as the penalty parameter $\penaltyParam \in \mathbb{R}^{+}$. However, one important difference is the choice of $\penaltyParam$: While for the
penalty-regularized system $\penaltyParam \approx \indexedBeam{\youngs}$ should hold true, with the augmented formulation we have greater freedom in choosing the penalty
parameter to our needs, without having an explicit restriction. As we end up with the original system for $\penaltyParam = 0$ and still solve for the original
solution, $\penaltyParam$ can usually be chosen much smaller than $\indexedBeam{\youngs}$. This in turn allows us to control the ill-conditioning of the
related matrix sub-blocks. We later see in \secref{subsec:modified_augmented_lagrangian_block_preconditioner}, that $\penaltyParam$ plays an important role
in the construction of the block preconditioner.

Following the discussion of \secref{subsec:pure_neumann}, the addition of the penalty contribution can be seen as indirect enforcement of the coupling conditions.
Matrix $D$ encodes the coupling constraints between the beam and the Lagrange multiplier field and is injective on the rigid-body subspace of the embedded
{\KirchhoffLove} beams. In particular, it eliminates all five rigid-body modes of the beam, and thus $\operatorname{null}(\indexedBeam{K}) \cap \operatorname{null}(D) = \{0\}$.
This heals the problematic nature of the pure Neumann problem on the beam sub-block even for small values of the penalty parameter. This, therefore, results in an
invertible augmented beam matrix sub-block, thus making direct methods applicable as outlined in \secref{subsec:matrix_structure}. 
Further, considering the augmentation of the solid matrix still yields a symmetric and positive definite operator, which underlines the applicability of AMG.

\begin{figure}
\centering
\begin{subfigure}{0.49\textwidth}
\includegraphics[trim={1.5cm 1cm 1.5cm 0.5cm}, clip, width=\textwidth]{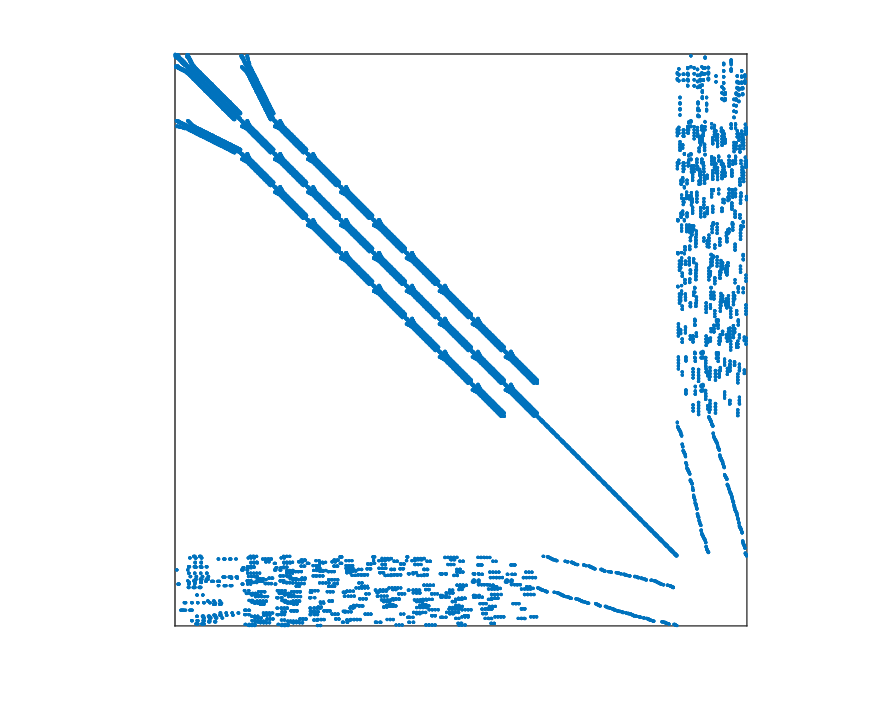}
\caption{Sparsity pattern of the coupled problem \eqref{eq:discretized_coupled_system} showing the {\saddlepoint} structure ($\text{nnz = 302135}$).}
\end{subfigure}
\hfill
\begin{subfigure}{0.49\textwidth}
\includegraphics[trim={1.5cm 1cm 1.5cm 0.5cm}, clip, width=\textwidth]{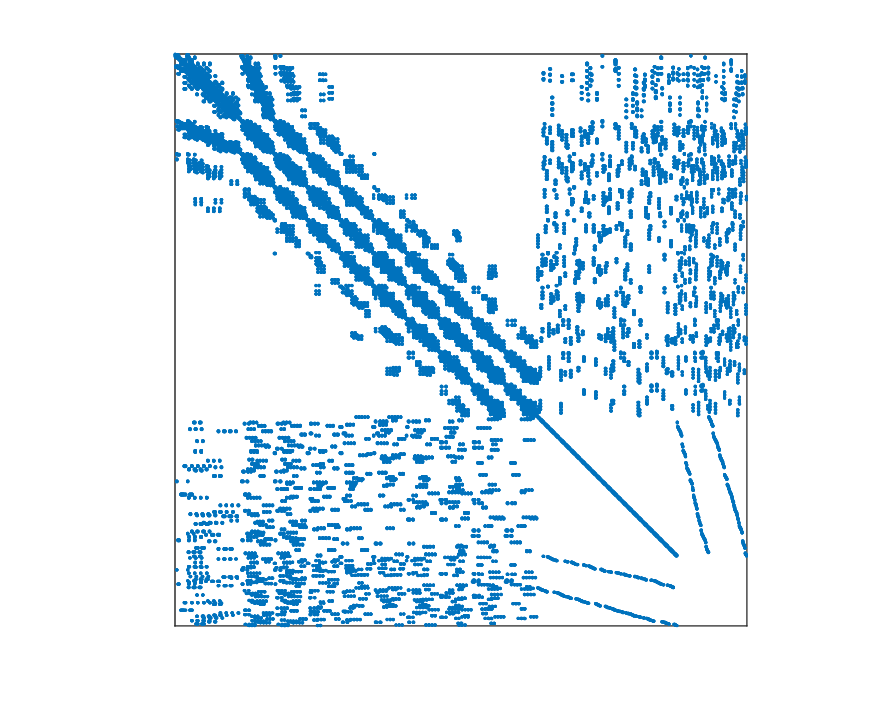}
\caption{Sparsity pattern of the augmented coupled problem \eqref{eq:augmented_discretized_coupled_system} ($\text{nnz = 447651}$).}
\end{subfigure}
\caption{Comparison of the sparsity patterns of the coupled problem and the augmented version, with the latter one showing an increase in the number of
{\nonzeros} by roughly $50\%$.}
\label{fig:sparsity_comparison}
\end{figure}

The introduction of additional terms increases the density of the stiffness matrix, but the bandwidth
can be controlled quite well due to the choice of $W$ being a diagonal matrix~\cite{Benzi2026a, Cerroni2019}.
In addition, coupling matrices directly acting on the solid and beam domain are added to the
off-diagonal blocks, where before plain zero blocks appeared, changing the overall matrix structure and increasing its density. If the beam sub-matrix is block
diagonal and the respective sub-blocks are comparably small, as it is the case for many, but short embedded fibers, the addition of penalty
contributions is negligible, as the matrix bandwidth will not grow outside of the block
diagonal sparsity pattern. All contributions related to the Lagrange multiplier field are not altered. The most influential part is on the solid
stiffness $\indexedSolid{K}$ increasing the bandwidth of the respective matrix block. The overall system matrix becomes denser, yet this increase in entries can
be controlled quite well. An illustration of the sparsity patterns of the {\saddlepoint} system \eqref{eq:discretized_coupled_system} and the augmented version
\eqref{eq:augmented_discretized_coupled_system}  for an exemplary case is shown in \figref{fig:sparsity_comparison}, with the overall amount of {\nonzero}
entries increased by $\approx 50\%$.

For the sake of simplified notation, we consider terms with subscript $(\cdot)_{\penaltyParam}$ to be related to the augmentation.
Thus we can rewrite the augmented system~\eqref{eq:augmented_discretized_coupled_system} resulting in the compact notation
\begin{equation}
\linearOperator_{\penaltyParam} \unknown_h = \rhs_{\penaltyParam}, \;\; \text{with} \;\;
\linearOperator_{\penaltyParam} =
\begin{pmatrix}
\indexedSolid{K}_{\penaltyParam} & -\trans{B_{\penaltyParam}} & -\trans{\matSolidLm} \\
-B_{\penaltyParam} & \indexedBeam{K}_{\penaltyParam} & \trans{\matBeamLm} \\
-\matSolidLm & \matBeamLm & ~ \\ 
\end{pmatrix}
\;\; \text{and} \;\;
\rhs_{\penaltyParam} =
\begin{pmatrix}
\indexedSolid{\residualNonlinear}_{\penaltyParam} \\
\indexedBeam{\residualNonlinear}_{\penaltyParam} \\
\gap
\end{pmatrix},
\label{eq:simple_augmented_discretized_coupled_system}
\end{equation}
with the coupling matrix $B_{\penaltyParam} := \penaltyParam \trans{\matBeamLm} \inv{W} M$. We also denote that $\penaltyParam$ has the unit $\qty{}{\newton/\square\meter}$,
yet with a little abuse of notation, we omit the explicit mentioning of this property in most parts of the discussion.

\subsection{Modified augmented Lagrangian block preconditioning}
\label{subsec:modified_augmented_lagrangian_block_preconditioner}

Based on the augmented linear system given in \eqref{eq:simple_augmented_discretized_coupled_system}, we now construct a
block preconditioner of the form given by $\mathcal{P}$ initially introduced in \secref{sec:analysis}. An almost obvious
preliminary choice is the use of an augmented Lagrangian preconditioner. To the authors' knowledge, this type of preconditioner
has mostly been applied to the Stokes and Oseen equations~\cite{Benzi2006a, Farrell2019a} as well as for fictitious domain problems~\cite{Benzi2026a},
yet its application to problems from computational solid mechanics is mainly limited to contact mechanics~\cite{Franceschini2022a}.

In the following analysis, we assume that $\indexedSolid{K}\in\mathbb{R}^{\indexedSolid{n}_{\text{DOF}}\times \indexedSolid{n}_{\text{DOF}}}$
and $W\in\mathbb{R}^{n^{\lambda}_{\text{DOF}}\times n^{\lambda}_{\text{DOF}}}$ are symmetric positive definite.
Furthermore, for~$M\in\mathbb{R}^{n^{\lambda}_{\text{DOF}}\times \indexedSolid{n}_{\text{DOF}}}$ and~$D\in\mathbb{R}^{n^{\lambda}_{\text{DOF}}\times \indexedBeam{n}_{\text{DOF}}}$,
we assume $\operatorname{rank}(M) = \operatorname{rank}(D) = n^{\lambda}_{\text{DOF}}$.
Thus, $M$ and $D$ have full row rank, whereas full column rank is not required.
To obtain a transparent comparison between the ideal and modified augmented Lagrangian preconditioners, the following auxiliary analysis assumes that the rigid-body nullspace
of~$\indexedBeam{K}\in\mathbb{R}^{\indexedBeam{n}_{\text{DOF}}\times \indexedBeam{n}_{\text{DOF}}}$, has been removed, so that $\indexedBeam{K}$ is also symmetric positive
definite. This makes the respective inverse expressions and their asymptotic limits well defined and allows the changes in the preconditioned operator to be attributed specifically to
the dropping of specific matrix sub-blocks, rather than to the singularity of the isolated beam sub-problem. In practice, the inverse of the singular matrix $\indexedBeam{K}$ is never applied,
as all beam sub-solves involve the augmented block $\indexedBeam{K}_{\penaltyParam}$.

\paragraph{Ideal augmented Lagrangian preconditioner}
The ideal augmented Lagrangian preconditioner based on the augmented {\mixeddimensional} {\beamsolid} problem \eqref{eq:simple_augmented_discretized_coupled_system}
reads
\begin{equation}
\mathcal{P}_{\penaltyParam}
:=
\begin{pmatrix}
\begin{matrix}
\indexedSolid{K}_{\penaltyParam} & -B_{\penaltyParam}^{T}\\
-B_{\penaltyParam} & \indexedBeam{K}_{\penaltyParam}
\end{matrix} &
\begin{matrix}
~ \\[0pt] ~
\end{matrix}
\\[0pt]
\begin{matrix}
-\matSolidLm & \matBeamLm
\end{matrix} &
S_{\penaltyParam}
\end{pmatrix}.
\label{eq:augmented_lagrangian_preconditioner}
\end{equation}
Similar to \cite{Benzi2011b}, the term ``ideal'' refers to using all the augmented sub-blocks inside the preconditioner.
The respective action of $\inv{\mathcal{P}_{\penaltyParam}}$ is given as
\begin{equation}
\inv{\mathcal{P}_{\penaltyParam}}
:=
\begin{pmatrix}
I & ~ & ~ \\
~ & I & ~ \\
~ & ~ & \inv{(S_{\penaltyParam})}
\end{pmatrix}
\begin{pmatrix}
I & ~ & ~ \\
~ & I & ~ \\
\matSolidLm & -\matBeamLm  & I
\end{pmatrix}
\begin{pmatrix}
\left(
\begin{matrix}
\indexedSolid{K}_{\penaltyParam} & -B_{\penaltyParam}^{T}\\
-B_{\penaltyParam} & \indexedBeam{K}_{\penaltyParam}
\end{matrix}
\right)^{\!-1} &
\begin{matrix}
~ \\[0pt] ~
\end{matrix}
\\[0pt]
\begin{matrix}
~ & ~
\end{matrix} &
I
\end{pmatrix}.
\label{eq:augmented_lagrangian_preconditioner_action}
\end{equation}
Hereby, $S_{\penaltyParam}$ is the augmented Schur complement
\begin{align}
S_{\penaltyParam} :&= -
\begin{pmatrix}
-\matSolidLm & \matBeamLm \\
\end{pmatrix}
\begin{pmatrix}
\indexedSolid{K}_{\penaltyParam} & -B_{\penaltyParam}^T \\
-B_{\penaltyParam} & \indexedBeam{K}_{\penaltyParam}
\end{pmatrix}^{-1}
\begin{pmatrix}
-\trans{\matSolidLm} \\
\trans{\matBeamLm} \\
\end{pmatrix} \\
&=
-\begin{pmatrix}
-\matSolidLm & \matBeamLm \\
\end{pmatrix}
\Biggl(
\begin{pmatrix}
\indexedSolid{K} & ~ \\
~ & \indexedBeam{K} \\
\end{pmatrix}
+
\penaltyParam
\begin{pmatrix}
-\trans{\matSolidLm} \\
\trans{\matBeamLm}
\end{pmatrix}
\inv{W}
\begin{pmatrix}
-\matSolidLm & \matBeamLm
\end{pmatrix}
\Biggr)^{-1}
\begin{pmatrix}
-\trans{\matSolidLm} \\
\trans{\matBeamLm} \\
\end{pmatrix}.
\label{eq:full_augmented_schur_complement}
\end{align}
By applying the Sherman-Morrison-Woodbury formula ({\eg} \cite[Theorem 3.2]{Bacuta2006a}), we can rewrite \eqref{eq:full_augmented_schur_complement} into
\begin{equation*}
\inv{S_{\penaltyParam}}=
-
\Biggl(
\begin{pmatrix}
\matSolidLm & -\matBeamLm \\
\end{pmatrix}
\begin{pmatrix}
\indexedSolid{K} & ~ \\
~ & \indexedBeam{K} \\
\end{pmatrix}^{-1}
\begin{pmatrix}
-\trans{\matSolidLm} \\
\trans{\matBeamLm} \\
\end{pmatrix}
\Biggr)^{-1}
-
\penaltyParam\inv{W},
\end{equation*}
which in the end simplifies to the statement
\begin{equation}
\inv{S_{\penaltyParam}} = \inv{S}-\penaltyParam \inv{W}
= \inv{(-\matSolidLm \; \inv{(\indexedSolid{K})} \; \trans{\matSolidLm} - \matBeamLm \; \inv{(\indexedBeam{K})} \; \trans{\matBeamLm})}
- \penaltyParam\inv{W}.
\label{eq:augmented_schur_complement}
\end{equation}
The resulting inverse of the augmented Schur complement is a simple combination of $S$ from the original system \eqref{eq:discretized_coupled_system} and the contribution
from the augmentation. Based on \eqref{eq:augmented_schur_complement}, we can argue that if $\penaltyParam \rightarrow \infty$ for a fixed mesh size $h$,
the augmentation term $\penaltyParam\inv{W}$ dominates $\inv{S}_{\penaltyParam}$ and leads to a spectrally equivalent representation if the scaling matrix $W$
is properly chosen (e.g. \cite[Remark 4.4]{Benzi2006a} and \cite{Golub2003}). This asymptotic observation concerns the exact augmented Schur complement only.
Applying the ideal augmented Lagrangian preconditioner to our system matrix results in
\begin{align*}
	\inv{\mathcal{P}_{\penaltyParam}}\mathcal{A}_{\penaltyParam} &= 
	\begin{pmatrix}
		I & ~ & ~ \\
		~ & I & ~ \\
		~ & ~ & \inv{S}_{\penaltyParam}
	\end{pmatrix}
	\begin{pmatrix}
		I & ~ & ~ \\
		~ & I & ~ \\
		M & -D & I
	\end{pmatrix}
	\begin{pmatrix}
		\left(
		\begin{matrix}
			\indexedSolid{K}_{\penaltyParam} & -B_{\penaltyParam}^{T}\\
			-B_{\penaltyParam} & \indexedBeam{K}_{\penaltyParam}
		\end{matrix}
		\right)^{\!-1} &
		\begin{matrix}
			~ \\[0pt] ~
		\end{matrix}
		\\[0pt]
		\begin{matrix}
			~ & ~
		\end{matrix} &
		I
	\end{pmatrix}
	\begin{pmatrix}
		\indexedSolid{K}_{\penaltyParam} & -\trans{B_{\penaltyParam}} & -\trans{\matSolidLm} \\
		-B_{\penaltyParam} & \indexedBeam{K}_{\penaltyParam} & \trans{\matBeamLm} \\
		-\matSolidLm & \matBeamLm & ~ \\ 
	\end{pmatrix}
	=
	\begin{pmatrix}
		I & ~ & Q_{13} \\
		~ & I & Q_{23} \\
		~ & ~ & Q_{33}
	\end{pmatrix},
\end{align*}
with the individual matrix blocks~$Q_{13}$, $Q_{23}$, and~$Q_{33}$ given as
\begin{align*}
Q_{13} &= \inv{(\indexedSolid{K}_\penaltyParam)}(-\trans{M}+\trans{B}_\penaltyParam Q_{23}), \\
Q_{23} &= \inv{(\indexedBeam{K}_{\penaltyParam} - B_\penaltyParam \inv{(\indexedSolid{K}_\penaltyParam)} \trans{B}_\penaltyParam)} (\trans{D}-B_\penaltyParam \inv{(\indexedSolid{K}_\penaltyParam)}\trans{M}), \\
Q_{33} &= I.
\end{align*}
With these statements, we analyze the asymptotic behavior of the matrix blocks~$Q_{13}$, $Q_{23}$, and~$Q_{33}$ given by the following limits
\begin{alignat*}{2}
\lim_{\penaltyParam\rightarrow0} \norm{Q_{13}} &{}= \norm{-\inv{(\indexedSolid{K})}\trans{M}}, &\qquad \lim_{\penaltyParam\rightarrow\infty} \norm{Q_{13}} &{}= 0, \\
\lim_{\penaltyParam\rightarrow0} \norm{Q_{23}} &{}= \norm{\inv{(\indexedBeam{K})}\trans{D}}, &\qquad \lim_{\penaltyParam\rightarrow\infty} \norm{Q_{23}} &{}= 0, \\
\lim_{\penaltyParam\rightarrow0} \norm{Q_{33}} &{}= 1, &\qquad \lim_{\penaltyParam\rightarrow\infty} \norm{Q_{33}} &{}=1.
\end{alignat*}
For the exact ideal augmented Lagrangian preconditioner and every penalty parameter~$\penaltyParam>0$,
the preconditioned operator is unit upper triangular, and therefore all its eigenvalues are equal to one.
For~$\penaltyParam\rightarrow\infty$, matrix blocks $Q_{13}$ and $Q_{23}$ vanish and, ultimately, the preconditioned operator tends towards identity,
$\inv{\mathcal {P}_{\penaltyParam}}\mathcal{A}_{\penaltyParam}\longrightarrow I$.
Thus, while its spectrum is independent of the penalty parameter, the preconditioned operator itself approaches the identity as the penalty parameter increases.

\paragraph{Modified augmented Lagrangian preconditioner}
As the augmented sub-blocks represent the penalty-based system discussed in \secref{subsec:penalty_formulation} and its preconditioner
is non-trivial, as outlined in \cite{Firmbach2024a}, the application of \eqref{eq:augmented_lagrangian_preconditioner_action}
is not optimal in the context of computationally efficient augmented Lagrangian preconditioning for {\mixeddimensional} {\beamsolid} problems. The core component
of the respective preconditioner would be the same as the one proposed in \cite{Firmbach2024a}, yet the goal of this publication is to
construct a more efficient method. Therefore, we aim to remove block $\trans{B_{\penaltyParam}}$ from $\mathcal{P}_{\penaltyParam}$ to
obtain a block triangular preconditioner resulting in a modified version of the ideal augmented  Lagrangian preconditioner proposed in
\cite{Benzi2011a, Benzi2011b} and expanded in \cite{Benzi2013} stated as
\begin{equation}
\mathcal{P}_{\penaltyParam, \text{mod}}
:=
\begin{pmatrix}
\indexedSolid{K}_{\penaltyParam} & ~ & ~ \\
-B_{\penaltyParam} & \indexedBeam{K}_{\penaltyParam} & ~ \\
-\matSolidLm & \matBeamLm & S_{\penaltyParam} \\ 
\end{pmatrix},
\label{eq:modified_augmented_lagrangian_preconditioner}
\end{equation}
with its action defined by
\begin{equation}
\inv{\mathcal{P}}_{\penaltyParam, \text{mod}}
:=
\begin{pmatrix}
I & ~ & ~ \\
~ & I & ~ \\
~ & ~ & \inv{S}_{\penaltyParam}
\end{pmatrix}
\begin{pmatrix}
I & ~ & ~ \\
~ & I & ~ \\
\matSolidLm & -\matBeamLm  & I
\end{pmatrix}
\begin{pmatrix}
I & ~ & ~ \\
\inv{(\indexedBeam{K}_{\penaltyParam})} B_{\penaltyParam} & I & ~ \\
~ & ~ & I 
\end{pmatrix}
\begin{pmatrix}
\inv{(\indexedSolid{K}_{\penaltyParam})} & ~ & ~ \\
~ &\inv{(\indexedBeam{K}_{\penaltyParam})} & ~ \\
~ & ~ & I \\
\end{pmatrix}.
\end{equation}
The action of the preconditioner defined by $\inv{\mathcal{P}}_{\penaltyParam, \text{mod}}$ enables us to approximate the inverse
of each main diagonal block individually, yet it violates the Schur complement assumption given above \cite{Benzi2011b}, due to
the removal of block $\trans{B_{\penaltyParam}}$. Again, we consider the preconditioned operator, this time applying the modified augmented Lagrangian approach resulting in
\begin{align*}
	\inv{\mathcal{P}_{\penaltyParam,\text{mod}}}\mathcal{A}_{\penaltyParam} &= 
	\begin{pmatrix}
		I & ~ & ~ \\
		~ & I & ~ \\
		~ & ~ & \inv{S}_{\penaltyParam}
	\end{pmatrix}
	\begin{pmatrix}
		I & ~ & ~ \\
		~ & I & ~ \\
		\matSolidLm & -\matBeamLm  & I
	\end{pmatrix}
	\begin{pmatrix}
		I & ~ & ~ \\
		\inv{(\indexedBeam{K}_{\penaltyParam})} B_{\penaltyParam} & I & ~ \\
		~ & ~ & I 
	\end{pmatrix}
	\begin{pmatrix}
		\inv{(\indexedSolid{K}_{\penaltyParam})} & ~ & ~ \\
		~ &\inv{(\indexedBeam{K}_{\penaltyParam})} & ~ \\
		~ & ~ & I \\
	\end{pmatrix}
	\begin{pmatrix}
		\indexedSolid{K}_{\penaltyParam} & -\trans{B_{\penaltyParam}} & -\trans{\matSolidLm} \\
		-B_{\penaltyParam} & \indexedBeam{K}_{\penaltyParam} & \trans{\matBeamLm} \\
		-\matSolidLm & \matBeamLm & ~ \\ 
	\end{pmatrix}
	\\
	&=
	\begin{pmatrix}
		I & Q_{12} & Q_{13} \\
		~ & Q_{22} & Q_{23} \\
		~ & Q_{32} & Q_{33}
	\end{pmatrix}.
\end{align*}
The individual matrix blocks~$Q_{ij}, i\in\{1,2,3\}, j\in\{2,3\}$ of $\inv{\mathcal{P}_{\penaltyParam,\text{mod}}}\mathcal{A}_{\penaltyParam}$ are therefore given as
\begin{align*}
Q_{12} &= -\inv{(\indexedSolid{K}_{\penaltyParam})}\trans{B}_{\penaltyParam}, \\
Q_{13} &= -\inv{(\indexedSolid{K}_{\penaltyParam})}\trans{M}, \\
Q_{22} &= I - \inv{(\indexedBeam{K}_\penaltyParam)}B_{\penaltyParam}\inv{(\indexedSolid{K}_\penaltyParam)}\trans{B}_\penaltyParam, \\
Q_{23} &= \inv{(\indexedBeam{K}_\penaltyParam)}(\trans{D}-B_\penaltyParam\inv{(\indexedSolid{K}_\penaltyParam)}\trans{M}), \\
Q_{32} &= \inv{S_\penaltyParam}(-M \inv{(\indexedSolid{K}_\penaltyParam)} \trans{B}_{\penaltyParam} + D \inv{(\indexedBeam{K}_\penaltyParam)} B_{\penaltyParam} \inv{(\indexedSolid{K}_\penaltyParam)} \trans{B}_{\penaltyParam}), \\
Q_{33} &= \inv{S_\penaltyParam}(-M \inv{(\indexedSolid{K}_\penaltyParam)} \trans{M} -D \inv{(\indexedBeam{K}_\penaltyParam)} \trans{D} + D \inv{(\indexedBeam{K}_\penaltyParam)} B_\penaltyParam \inv{(\indexedSolid{K}_\penaltyParam)} \trans{M}).
\end{align*}
The resulting asymptotic limits for changing $\penaltyParam$ based on the given factorization are stated as follows:
\begin{alignat*}{2}
\lim_{\penaltyParam\rightarrow0} \norm{Q_{12}} &{}= 0, &\qquad \lim_{\penaltyParam\rightarrow\infty} \norm{Q_{12}} &{} = \norm{-\inv{(\indexedSolid{K})}\trans{M}\inv{(M\inv{(\indexedSolid{K})}\trans{M})}D}, \\
\lim_{\penaltyParam\rightarrow0} \norm{Q_{13}} &{}= \norm{-\inv{(\indexedSolid{K})}\trans{M}}, &\qquad \lim_{\penaltyParam\rightarrow\infty} \norm{Q_{13}} &{}= 0, \\
\lim_{\penaltyParam\rightarrow0} \norm{Q_{22}} &{}= 1, &\qquad \lim_{\penaltyParam\rightarrow\infty} \norm{Q_{22}} &{} = \norm{I-\inv{(\indexedBeam{K})}\trans{D} \inv{(D\inv{(\indexedBeam{K})}\trans{D})}D}, \\
\lim_{\penaltyParam\rightarrow0} \norm{Q_{23}} &{}= \norm{\inv{(\indexedBeam{K})}\trans{D}}, &\qquad \lim_{\penaltyParam\rightarrow\infty} \norm{Q_{23}} &{}= 0, \\
\lim_{\penaltyParam\rightarrow0} \norm{Q_{32}} &{}= 0, &\qquad \lim_{\penaltyParam\rightarrow\infty} \norm{Q_{32}} &{}= \norm{\inv{(D\inv{(\indexedBeam{K})}\trans{D})}D}, \\
\lim_{\penaltyParam\rightarrow0} \norm{Q_{33}} &{}= 1, &\qquad \lim_{\penaltyParam\rightarrow\infty} \norm{Q_{33}} &{}=1.
\end{alignat*}
In contrast to the ideal augmented Lagrangian preconditioner,
choosing $\penaltyParam\rightarrow0$ or $\penaltyParam\rightarrow\infty$ does not result in $\inv{\mathcal{P}_{\penaltyParam}}\mathcal{A}_{\penaltyParam} \rightarrow I$ as individual terms
grow or shrink. This is the direct consequence of dropping $\trans{B_{\penaltyParam}}$. Especially for $\penaltyParam\rightarrow\infty$,
$Q_{32}$ does not vanish and $Q_{22}\neq I$, thus destroying the unit upper triangular structure and modifying the eigenvalue spectrum. The competing limits suggest that an effective value of $\penaltyParam$
should lie between the two asymptotic regimes to obtain a favorable eigenvalue spectrum and $\inv{\mathcal{P}_{\penaltyParam}}\mathcal{A}_{\penaltyParam} \approx I$, but we remark,
that the given analysis alone does not establish existence or uniqueness of an optimal value.

\paragraph{Individually scaled modified augmented Lagrangian preconditioner}
In a final step, we introduce a block preconditioner based on the augmented Lagrangian formulation \eqref{eq:augmented_discretized_coupled_system}
applying the modified preconditioner~\eqref{eq:modified_augmented_lagrangian_preconditioner}, but with different scaling parameters for the solid and
beam contributions. As the coupled system \eqref{eq:discretized_coupled_system} is composed of two different PDEs,
we follow the recently proposed approach from~\cite{BenziPreprint}
and rewrite the augmented system \eqref{eq:augmented_discretized_coupled_system} by introducing two independent penalty parameters~$\indexedSolid{\penaltyParam}$
and~$\indexedBeam{\penaltyParam}$ such that
\begin{equation}
\linearOperator \unknown_h = \rhs, \;\; \text{with} \;\;
\linearOperator =
\begin{pmatrix}
\indexedSolid{K} + \indexedSolid{\penaltyParam} \trans{\matSolidLm} \inv{W} \matSolidLm & - \indexedSolid{\penaltyParam} \trans{\matSolidLm} \inv{W} \matBeamLm & -\trans{\matSolidLm} \\
-\indexedBeam{\penaltyParam} \trans{\matBeamLm} \inv{W} \matSolidLm & \indexedBeam{K} + \indexedBeam{\penaltyParam} \trans{\matBeamLm} \inv{W} \matBeamLm & \trans{\matBeamLm} \\
-\matSolidLm & \matBeamLm & ~ \\ 
\end{pmatrix}
\;\; \text{and} \;\;
\rhs =
\begin{pmatrix}
\indexedSolid{\residualNonlinear} - \indexedSolid{\penaltyParam}\trans{M}\inv{W}\gap \\
\indexedBeam{\residualNonlinear} + \indexedBeam{\penaltyParam}\trans{\matBeamLm}\inv{W}\gap \\
\gap
\end{pmatrix}
\label{eq:individually_augmented_discretized_coupled_system}
\end{equation}
and therefore
\begin{equation}
\linearOperator_{\indexedSolid{\penaltyParam}, \indexedBeam{\penaltyParam}} \unknown_h = \rhs_{\indexedSolid{\penaltyParam}, \indexedBeam{\penaltyParam}}, \;\; \text{with} \;\;
\linearOperator_{\indexedSolid{\penaltyParam}, \indexedBeam{\penaltyParam}} =
\begin{pmatrix}
\indexedSolid{K}_{\indexedSolid{\penaltyParam}} & -\trans{B_{\indexedSolid{\penaltyParam}}} & -\trans{\matSolidLm} \\
-B_{\indexedBeam{\penaltyParam}} & \indexedBeam{K}_{\indexedBeam{\penaltyParam}} & \trans{\matBeamLm} \\
-\matSolidLm & \matBeamLm & ~ \\ 
\end{pmatrix}
\;\; \text{and} \;\;
\rhs_{\indexedSolid{\penaltyParam}, \indexedBeam{\penaltyParam}} =
\begin{pmatrix}
\indexedSolid{\residualNonlinear}_{\indexedSolid{\penaltyParam}} \\
\indexedBeam{\residualNonlinear}_{\indexedBeam{\penaltyParam}} \\
\gap
\end{pmatrix}
\label{eq:simple_individually_augmented_discretized_coupled_system}
\end{equation}
describes the independently row-scaled augmented system. Using the same modified augmented Lagrangian preconditioner as before applied
to \eqref{eq:simple_individually_augmented_discretized_coupled_system} yields
\begin{equation}
\mathcal{P}_{\indexedSolid{\penaltyParam}, \indexedBeam{\penaltyParam}, \text{mod}}
:=
\begin{pmatrix}
\indexedSolid{K}_{\indexedSolid{\penaltyParam}} & ~ & ~ \\
-B_{\indexedBeam{\penaltyParam}} & \indexedBeam{K}_{\indexedBeam{\penaltyParam}} & ~ \\
-\matSolidLm & \matBeamLm & S_{\indexedSolid{\penaltyParam}, \indexedBeam{\penaltyParam}} \\ 
\end{pmatrix}.
\end{equation}
We next consider the application of the individually scaled modified augmented Lagrangian preconditioner to the linear operator given
in~\eqref{eq:simple_individually_augmented_discretized_coupled_system} stated by
\begin{align*}
	\inv{\mathcal{P}_{\indexedSolid{\penaltyParam}, \indexedBeam{\penaltyParam},\text{mod}}}\mathcal{A}_{\indexedSolid{\penaltyParam}, \indexedBeam{\penaltyParam}} &= 
	\begin{pmatrix}
		I & ~ & ~ \\
		~ & I & ~ \\
		~ & ~ & \inv{S}_{\indexedSolid{\penaltyParam}, \indexedBeam{\penaltyParam}}
	\end{pmatrix}
	\begin{pmatrix}
		I & ~ & ~ \\
		~ & I & ~ \\
		\matSolidLm & -\matBeamLm  & I
	\end{pmatrix}
	\begin{pmatrix}
		I & ~ & ~ \\
		\inv{(\indexedBeam{K}_{\indexedBeam{\penaltyParam}})} B_{\indexedBeam{\penaltyParam}} & I & ~ \\
		~ & ~ & I 
	\end{pmatrix}
	\begin{pmatrix}
		\inv{(\indexedSolid{K}_{\indexedSolid{\penaltyParam}})} & ~ & ~ \\
		~ &\inv{(\indexedBeam{K}_{\indexedBeam{\penaltyParam}})} & ~ \\
		~ & ~ & I \\
	\end{pmatrix}
	\begin{pmatrix}
		\indexedSolid{K}_{\indexedSolid{\penaltyParam}} & -\trans{B_{\indexedSolid{\penaltyParam}}} & -\trans{\matSolidLm} \\
		-B_{\indexedBeam{\penaltyParam}} & \indexedBeam{K}_{\indexedBeam{\penaltyParam}} & \trans{\matBeamLm} \\
		-\matSolidLm & \matBeamLm & ~ \\ 
	\end{pmatrix}
	\\
	&=
	\begin{pmatrix}
		I & Q_{12} & Q_{13} \\
		~ & Q_{22} & Q_{23} \\
		~ & Q_{32} & Q_{33}
	\end{pmatrix},
\end{align*}
with the individual blocks defined as
\begin{align*}
Q_{12} &= -\inv{(\indexedSolid{K}_{\indexedSolid{\penaltyParam}})}\trans{B}_{\indexedSolid{\penaltyParam}}, \\
Q_{13} &= -\inv{(\indexedSolid{K}_{\indexedSolid{\penaltyParam}})}\trans{M}, \\
Q_{22} &= I - \inv{(\indexedBeam{K}_{\indexedBeam{\penaltyParam}})}B_{\indexedBeam{\penaltyParam}}\inv{(\indexedSolid{K}_{\indexedSolid{\penaltyParam}})}\trans{B}_{\indexedSolid{\penaltyParam}}, \\
Q_{23} &= \inv{(\indexedBeam{K}_{\indexedBeam{\penaltyParam}})}(\trans{D}-B_{\indexedBeam{\penaltyParam}}\inv{(\indexedSolid{K}_{\indexedSolid{\penaltyParam}})}\trans{M}), \\
Q_{32} &= \inv{S_{\indexedSolid{\penaltyParam}, \indexedBeam{\penaltyParam}}}(-M \inv{(\indexedSolid{K}_{\indexedSolid{\penaltyParam}})} \trans{B}_{\indexedSolid{\penaltyParam}} + D \inv{(\indexedBeam{K}_{\indexedBeam{\penaltyParam}})} B_{\indexedBeam{\penaltyParam}} \inv{(\indexedSolid{K}_{\indexedSolid{\penaltyParam}})} \trans{B}_{\indexedSolid{\penaltyParam}}), \\
Q_{33} &= \inv{S_{\indexedSolid{\penaltyParam}, \indexedBeam{\penaltyParam}}}(-M \inv{(\indexedSolid{K}_{\indexedSolid{\penaltyParam}})} \trans{M} -D \inv{(\indexedBeam{K}_{\indexedBeam{\penaltyParam}})} \trans{D} + D \inv{(\indexedBeam{K}_{\indexedBeam{\penaltyParam}})} B_{\indexedBeam{\penaltyParam}} \inv{(\indexedSolid{K}_{\indexedSolid{\penaltyParam}})} \trans{M}).
\end{align*}
The respective exact Schur complement is denoted by
\begin{equation*}
S_{\indexedSolid{\penaltyParam}, \indexedBeam{\penaltyParam}}:=-W\inv{(W + \indexedSolid{\penaltyParam}M\inv{(\indexedSolid{K})}\trans{M} + \indexedBeam{\penaltyParam}D\inv{(\indexedBeam{K})}\trans{D})}(M\inv{(\indexedSolid{K})}\trans{M}+D\inv{(\indexedBeam{K})}\trans{D})
\end{equation*}
with its inverse
\begin{equation*}
\inv{S_{\indexedSolid{\penaltyParam}, \indexedBeam{\penaltyParam}}}:=-\inv{(M\inv{(\indexedSolid{K})}\trans{M}+D\inv{(\indexedBeam{K})}\trans{D})}(W + \indexedSolid{\penaltyParam}M\inv{(\indexedSolid{K})}\trans{M} + \indexedBeam{\penaltyParam}D\inv{(\indexedBeam{K})}\trans{D})\inv{W}.
\end{equation*}
We now consider the asymptotic behavior of the matrix blocks under varying $\indexedSolid{\penaltyParam}$, but $\indexedBeam{\penaltyParam}$ being fixed:
\begin{alignat*}{2}
\lim_{\indexedSolid{\penaltyParam}\rightarrow0} \norm{Q_{12}} &{}= 0, &\qquad \lim_{\indexedSolid{\penaltyParam}\rightarrow\infty} \norm{Q_{12}} &{}= \norm{-\inv{(\indexedSolid{K})}\trans{M}\inv{(M\inv{(\indexedSolid{K})}\trans{M})}D}, \\
\lim_{\indexedSolid{\penaltyParam}\rightarrow0} \norm{Q_{13}} &{}= \norm{-\inv{(\indexedSolid{K})}\trans{M}}, &\qquad \lim_{\indexedSolid{\penaltyParam}\rightarrow\infty} \norm{Q_{13}} &{}= 0, \\
\lim_{\indexedSolid{\penaltyParam}\rightarrow0} \norm{Q_{22}} &{}= 1, &\qquad \lim_{\indexedSolid{\penaltyParam}\rightarrow\infty} \norm{Q_{22}} &{} = \norm{\inv{(\indexedBeam{K}_{\indexedBeam{\penaltyParam}})}\indexedBeam{K}}, \\
\lim_{\indexedSolid{\penaltyParam}\rightarrow0} \norm{Q_{23}} &{}= \norm{\inv{(\indexedBeam{K}_{\indexedBeam{\penaltyParam}})}(\trans{D}-B_{\indexedBeam{\penaltyParam}}\inv{(\indexedSolid{K})}\trans{M})}, &\qquad \lim_{\indexedSolid{\penaltyParam}\rightarrow\infty} \norm{Q_{23}} &{}= \norm{\inv{(\indexedBeam{K}_{\indexedBeam{\penaltyParam}})}\trans{D}}, \\
\lim_{\indexedSolid{\penaltyParam}\rightarrow0} \norm{Q_{32}} &{}= 0, &\qquad \lim_{\indexedSolid{\penaltyParam}\rightarrow\infty} \norm{Q_{32}} &{}= \infty, \\
\lim_{\indexedSolid{\penaltyParam}\rightarrow0} \norm{Q_{33}} &{}= 1, &\qquad \lim_{\indexedSolid{\penaltyParam}\rightarrow\infty} \norm{Q_{33}} &{}= \infty.
\end{alignat*}
For fixed $\indexedBeam{\penaltyParam}$, choosing $\indexedSolid{\penaltyParam}$ sufficiently small is favorable because the omitted block $\trans{B}_{\indexedSolid{\penaltyParam}}$ vanishes
as $\indexedSolid{\penaltyParam}\rightarrow0$. In contrast, the preconditioned operator is generically unbounded as $\indexedSolid{\penaltyParam}\rightarrow\infty$, due to $Q_{32}$ and $Q_{33}$
growing.
In a final step we investigate the cases for $\indexedSolid{\penaltyParam}\rightarrow0$ and variable $\indexedBeam{\penaltyParam}$, which only affects $Q_{23}$:
\begin{alignat*}{2}
\lim\limits_{\substack{\indexedSolid{\penaltyParam}\to 0\\ \indexedBeam{\penaltyParam}\to 0}} \norm{Q_{23}} &{}= \norm{\inv{(\indexedBeam{K})}\trans{D}},
&\qquad \lim\limits_{\substack{\indexedSolid{\penaltyParam}\to 0\\ \indexedBeam{\penaltyParam}\to \infty}}  \norm{Q_{23}} &{}= \norm{-\inv{(\indexedBeam{K})}\trans{D}\inv{(D\inv{(\indexedBeam{K})}\trans{D})}M\inv{(\indexedSolid{K})}\trans{M}}.
\end{alignat*}

As the exact application of $\inv{\mathcal{P}}_{\penaltyParam, \text{mod}}$ and $\inv{\mathcal{P}_{\indexedSolid{\penaltyParam}, \indexedBeam{\penaltyParam}, \text{mod}}}$
is not feasible in an actual computation, we use approximated versions. Following the discussion in \secref{subsec:matrix_structure} and \secref{sec:block_preconditioner},
we opt to use AMG on the solid domain and an LU-factorization on the beam and Lagrange multiplier field, respectively. The last missing  component of the block preconditioners
therefore consists of finding a suitable approximation $\hat{S}_{\penaltyParam}$ to the exact augmented Schur complement $S_{\penaltyParam}$ and
respectively $\hat{S}_{\indexedSolid{\penaltyParam}, \indexedBeam{\penaltyParam}}$ for $S_{\indexedSolid{\penaltyParam}, \indexedBeam{\penaltyParam}}$,
as its exact computation is not feasible and would result in a dense matrix.

\subsection{Choice of the augmented Schur complement}
\label{subsec:choice_of_schur_complement}

\begin{figure}
	\centering
	\begin{subfigure}[b]{0.49\textwidth}
		\includegraphics[scale=1.0]{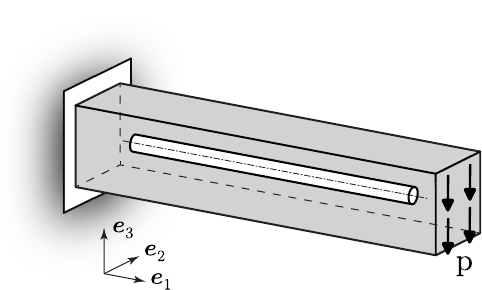}
		\caption{Geometry and boundary conditions of the model problem.}
		\label{fig:model_problem_setup}
	\end{subfigure}
	\hfill
	\begin{subfigure}[b]{0.49\textwidth}
		\includegraphics[trim={0cm 5cm 0cm 5cm}, clip, width=\textwidth]{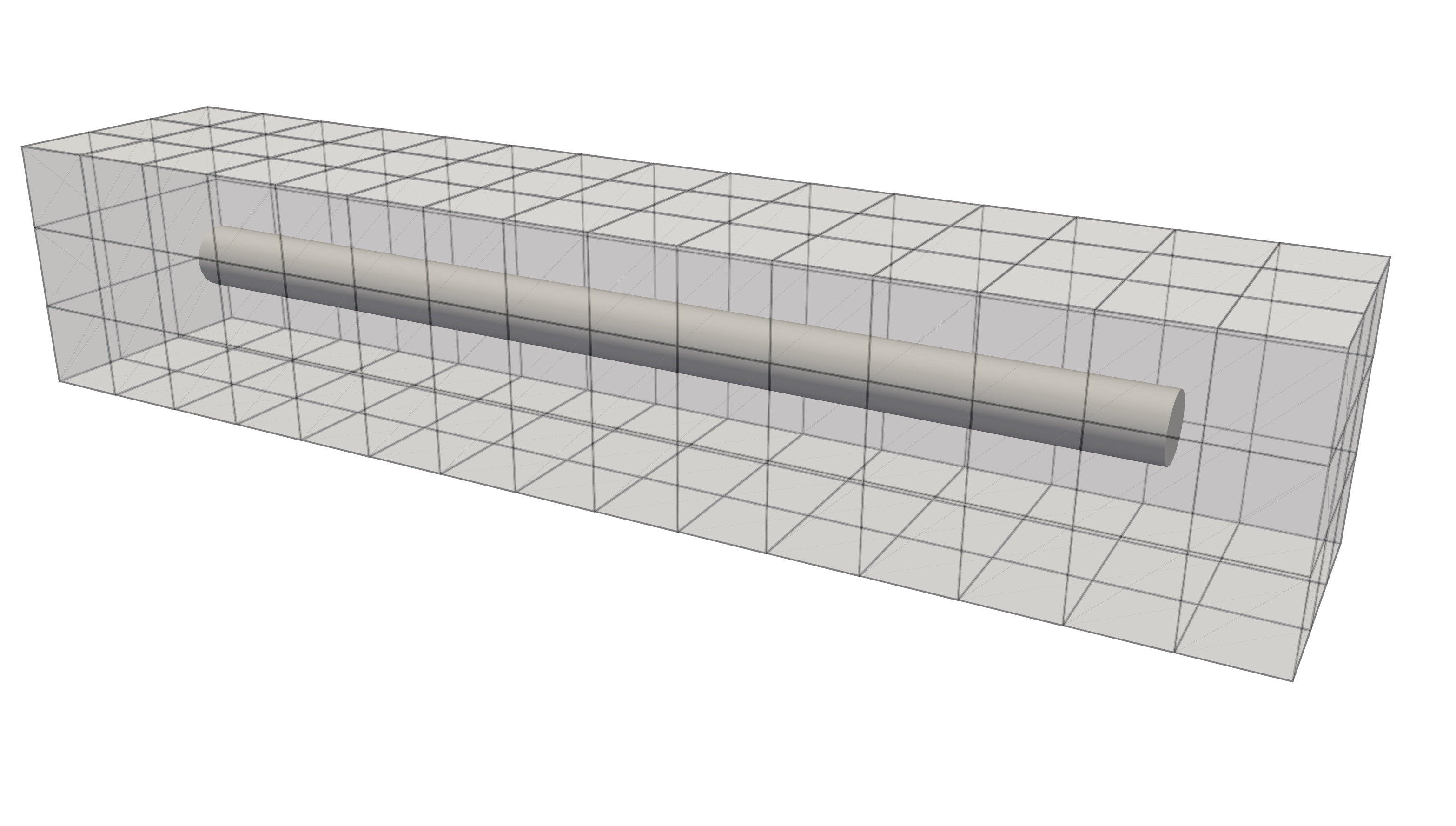}
		\caption{Visualization of the solid and beam mesh.}
		\label{fig:model_problem_mesh}
	\end{subfigure}
	\caption{Model problem consisting of a cantilever beam with one fully embedded fiber.}
	\label{fig:model_problem}
\end{figure}

In the practical block preconditioners considered in the scope of this publication, the Schur complement is no longer applied exactly, but replaced by an inexpensive
matrix whose approximation quality must be assessed separately.
Therefore, we introduce a small model problem to be used for numerical studies in the scope of this section, which is shown in \figref{fig:model_problem}. The numerical setup
consists of a cantilever beam with dimensions ${\qty{5}{\meter} \times \qty{1}{\meter} \times \qty{1}{\meter}}$ with a single fiber embedded into the center of the solid continuum.
Both fiber end-cross-sections are equally far away from the respective perpendicular solid surfaces.
The single fiber is discretized with four beam elements with a {\crosssection} radius of $\indexedBeam{\beamRadius}=\qty{0.125}{\meter}$, beam element
length $\indexedBeam{h} = \qty{1.0}{\meter}$ and a Young's modulus of $\indexedBeam{\youngs}=\qty{50}{\newton/\square\meter}$.
The solid continuum uses $\indexedSolid{\youngs}=\qty{10}{\newton/\square\meter}$ and $\nu=0.3$ as material parameters. The solid discretization consists of ${15 \times 3 \times 3}$ elements,
respectively (see \figref{fig:model_problem_mesh}). The displacements of the solid at the left side of the cantilever beam is fully constrained, while on the right side a surface
load $\mathrm{p}$ is applied to the structure (see \figref{fig:model_problem_setup}) resulting in a bending deformation. The beam is constrained by the Lagrange multiplier field,
attaching its displacement to the one from the surrounding solid.

\begin{figure}
	\includegraphics[trim={2cm 0cm 2cm 0cm}, clip, width=\textwidth]{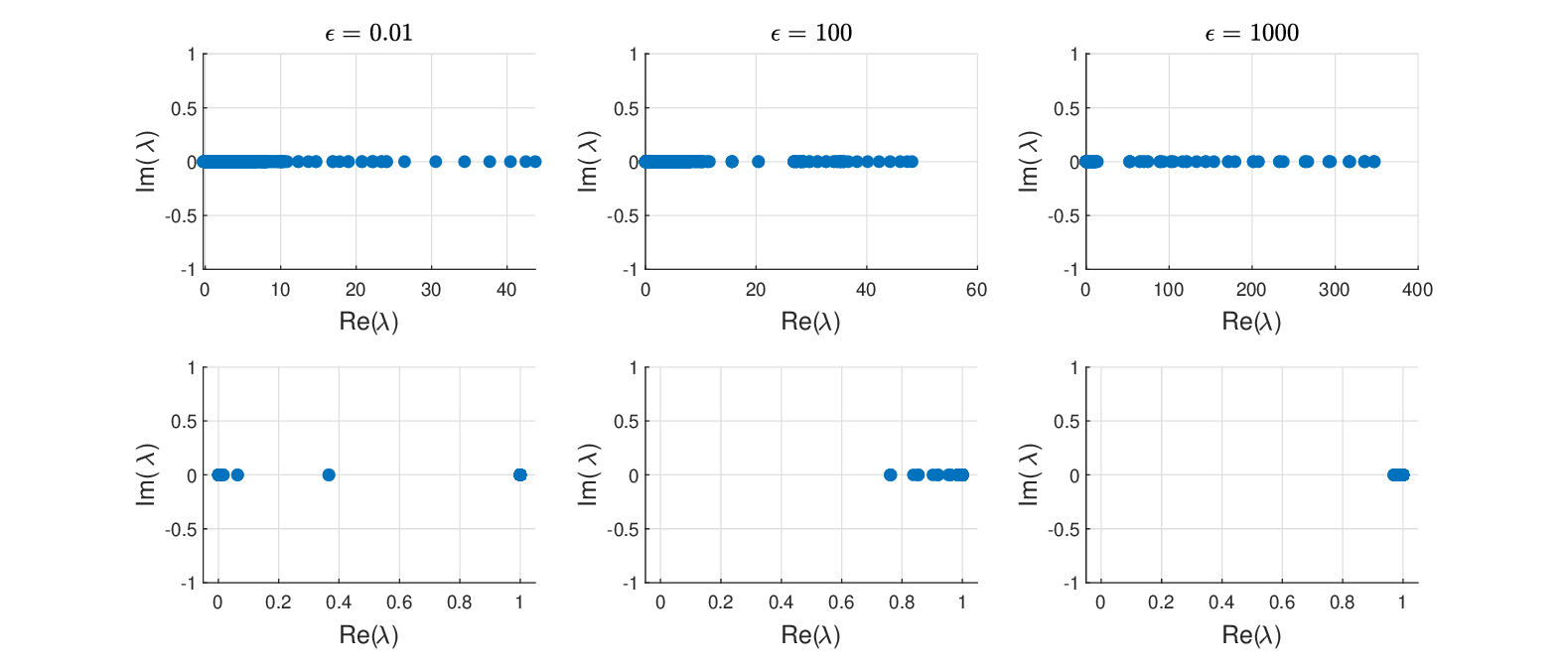}
	\caption{Comparison of the eigenvalue spectrum of the system matrix $\linearOperator_{\penaltyParam}$ (top) and the preconditioned
		operator $\inv{\mathcal{P}_{\penaltyParam}}\linearOperator_{\penaltyParam}$ (bottom) using the ideal augmented Lagrangian preconditioner
		for the {\mixeddimensional} {\beamsolid} model problem for different values of $\penaltyParam$. We note that the horizontal axes are scaled
		differently for visualization purposes.}
	\label{fig:augmented_lagrangian_preconditioner_penalty_influence}
\end{figure}

In a first step, we briefly discuss a practical choice for $W$, using the ideal augmented Lagrangian preconditioner, and present a numerical study that
shows the influence of $\penaltyParam$ on the eigenvalues of the system matrix $\linearOperator_{\penaltyParam}$ and the preconditioned
operator $\inv{\mathcal{P}_{\penaltyParam}}\linearOperator_{\penaltyParam}$. Based on the asymptotic limit of~\eqref{eq:augmented_schur_complement},
we consider $\hat{S}_{\penaltyParam} = -\frac{1}{\penaltyParam}W$ as Schur complement approximation. Hereby we specifically choose $W=\kappa$, which is
defined for each Lagrange multiplier node $j$ as given in \cite{Steinbrecher2020}, yielding
\begin{equation}
	\kappa_{jj} := \int_{\Lambda} \phi_j \; \mathrm{d}s \; I,
\end{equation}
with an identity matrix $I\in\REalSp{3\times3}$.
Following up on the analysis presented in~\secref{subsec:modified_augmented_lagrangian_block_preconditioner} and replacing~$S_\penaltyParam$
by~$\hat{S}_{\penaltyParam}$ leaves $Q_{13}$ and $Q_{23}$ unchanged, whereas the multiplier block becomes
\begin{equation*}
Q_{33} = \inv{\hat{S}_{\penaltyParam}}S_\penaltyParam = -\penaltyParam\inv{W}\inv{(\inv{S}-\penaltyParam\inv{W})} = \inv{(I-\frac{1}{\penaltyParam}\inv{S}W)}
\end{equation*}
resulting in
\begin{alignat*}{2}
\lim_{\penaltyParam\rightarrow0} \norm{Q_{33}} &{}= 0, &\qquad \lim_{\penaltyParam\rightarrow\infty} \norm{Q_{33}} &{}= 1. \\
\end{alignat*}
Despite using a simple approximation for the Schur complement, we are able to recover the properties of the exact Schur complement in the asymptotic limit~${\penaltyParam\rightarrow\infty}$
such that $\inv{\mathcal {P}_{\penaltyParam}}\mathcal{A}_{\penaltyParam}\longrightarrow I$.
\figref{fig:augmented_lagrangian_preconditioner_penalty_influence} shows the eigenvalue spectrum for different
values of the penalty parameter for the system matrix $A_{\penaltyParam}$ in the top row and $\inv{\mathcal{P}_{\penaltyParam}}\linearOperator_{\penaltyParam}$
in the bottom row. As expected, the maximum eigenvalues of $A_{\penaltyParam}$ increase and are not bounded with increasing values of $\penaltyParam$.
In contrast, for $\inv{\mathcal{P}_{\penaltyParam}}\linearOperator_{\penaltyParam}$ all eigenvalues cluster around one with $\penaltyParam\rightarrow\infty$,
as $\hat{S}_{\penaltyParam}$ is becoming spectrally closer to the true $S_{\penaltyParam}$. With the observations made in \cite{Benzi2026a, Steinbrecher2020} and
our brief numerical investigation, we consider $W = \kappa$ to be a proper choice \cite{Wathen1987}.

The choice of the penalty parameter $\penaltyParam$ (respectively $\indexedSolid{\penaltyParam}$ and $\indexedBeam{\penaltyParam}$) and the respective Schur complement approximation play a central role in the overall
performance of the preconditioners $\mathcal{P}_{\penaltyParam, \text{mod}}$ and $\mathcal{P}_{\indexedSolid{\penaltyParam}, \indexedBeam{\penaltyParam}, \text{mod}}$. As stated in \cite{Benzi2011a} and further
investigated in \cite{Benzi2011b}, $\penaltyParam$ influences the modified augmented Lagrangian preconditioner differently
as \eqref{eq:augmented_lagrangian_preconditioner}. The penalty parameter now has to be chosen
carefully, usually not being located at the asymptotic limits $\penaltyParam \rightarrow 0$ and $\penaltyParam \rightarrow \infty$. Additionally, for $\mathcal{P}_{\indexedSolid{\penaltyParam}, \indexedBeam{\penaltyParam}, \text{mod}}$
an appropriate choice of $\indexedSolid{\penaltyParam}$ and $\indexedBeam{\penaltyParam}$ is essential to obtain an efficient method~\cite{BenziPreprint}.
 In the following, we consider three types of Schur complement approximations, with variants~{\one} and {\two} being related to $\mathcal{P}_{\penaltyParam, \text{mod}}$
 and variant~{\three} to $\mathcal{P}_{\indexedSolid{\penaltyParam}, \indexedBeam{\penaltyParam}, \text{mod}}$.

\paragraph{Variant {\one}}
We consider the first case for the special case of short, independent beams inside the solid continuum, such that we
choose $\penaltyParam \ll \indexedBeam{E}$, which is rather untypical for augmented Lagrangian preconditioners, yet not uncommon for the augmented Lagrangian
formulation itself. One can see the augmentation as a pure stabilization in this case. Due to $\penaltyParam$ being small, we assume 
\begin{equation*}
\indexedSolid{K}_{\penaltyParam} \approx \indexedSolid{K} \quad , \quad \indexedBeam{K}_{\penaltyParam} \approx \indexedBeam{K} \quad , \quad
B_{\penaltyParam} = \trans{B}_{\penaltyParam} \approx 0 \quad \text{and thus} \quad S_{\penaltyParam} \approx S
\end{equation*}
to hold true.
Therefore we assume the Schur complement to be given by
\begin{equation*}
S_{\penaltyParam} = -\matSolidLm \; \inv{(\indexedSolid{K}_{\penaltyParam})} \; \trans{\matSolidLm}
- \matBeamLm \; \inv{(\indexedBeam{K}_{\penaltyParam})} \; \trans{\matBeamLm},
\end{equation*}
being very close to the original Schur complement with a small perturbation, while neglecting the additional coupling contributions given by $B_{\penaltyParam}$
and $\trans{B}_{\penaltyParam}$.
With the augmented beam sub-matrix being described by a block-diagonal sparsity pattern, its inverse can be cheaply approximated by a sparse approximate
inverse (SPAI) \cite{Grote1997}. A common choice for the solid inverse is a SIMPLE-like approximation \cite{Wiesner2021} or a block-diagonal variant \cite{Franceschini2019},
resulting in the respective approximated Schur complement given by
\begin{equation}
\hat{S}_{\penaltyParam} = -\matSolidLm \; \inv{\operatorname{diag}(\indexedSolid{K}_{\penaltyParam})} \; \trans{\matSolidLm}
- \matBeamLm \; \operatorname{SPAI}(\indexedBeam{K}_{\penaltyParam}) \; \trans{\matBeamLm}.
\label{eq:schur_complement_variant_1}
\end{equation}

\begin{figure}
\includegraphics[trim={2cm 0cm 2cm 0cm}, clip, width=\textwidth]{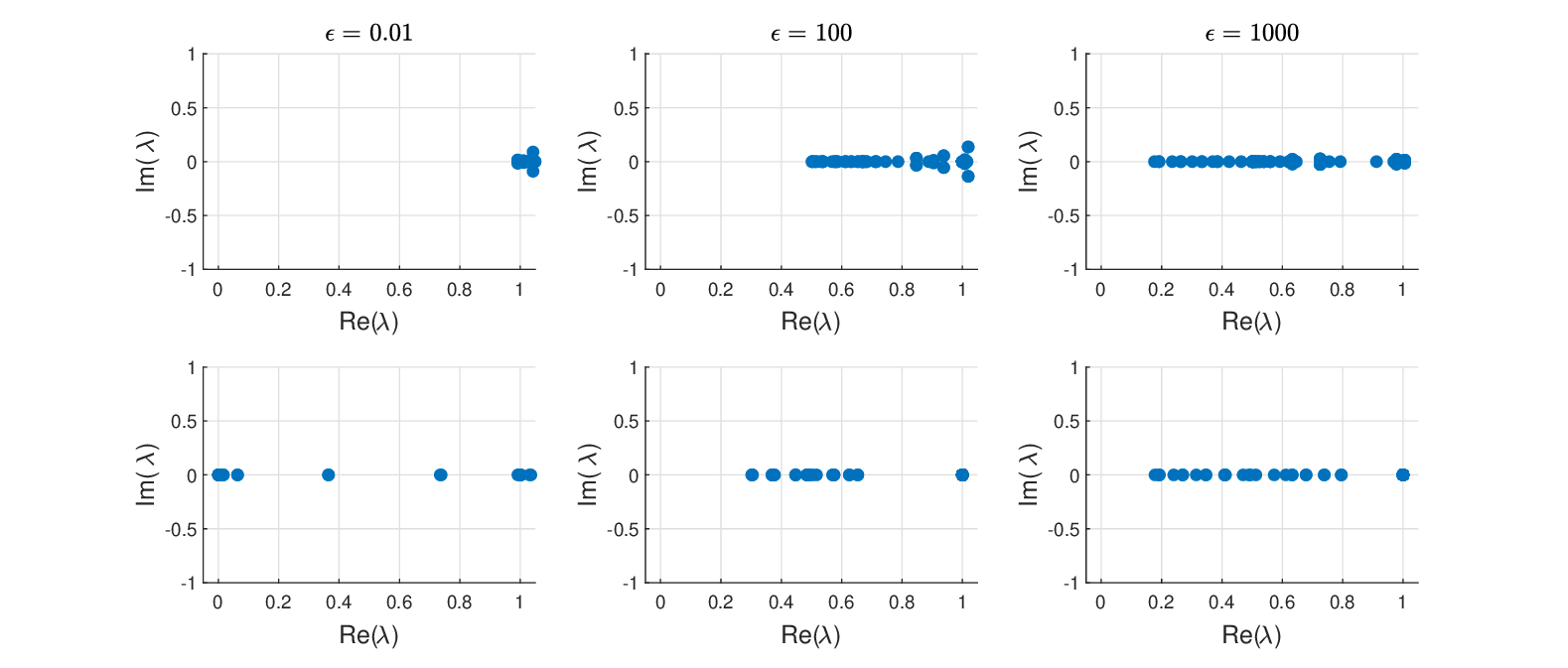}
\caption{Comparison of the eigenvalue spectrum of preconditioned operator $\inv{\mathcal{P}_{\penaltyParam, \text{mod}}}\linearOperator_{\penaltyParam}$ with Schur complement variant {\one} (top)
and the preconditioned operator $\inv{\mathcal{P}_{\penaltyParam, \text{mod}}}\linearOperator_{\penaltyParam}$ with Schur complement variant {\two} (bottom) for the {\mixeddimensional} {\beamsolid}
model problem for different values of $\penaltyParam$.}
\label{fig:modified_augmented_lagrangian_preconditioner_penalty_influence}
\end{figure}

\paragraph{Variant {\two}}
For the second case we seek to find a value for $\penaltyParam$ such that
\begin{equation*}
\hat{S}_{\penaltyParam} = -\frac{1}{\penaltyParam}  W
\end{equation*}
holds true. By setting $W = \kappa$ as highlighted before, the augmented Schur complement follows the approximation given by
\begin{equation}
\hat{S}_{\penaltyParam} = -\frac{1}{\penaltyParam} \kappa.
\label{eq:schur_complement_variant_2}
\end{equation}
For the modified augmented Lagrangian preconditioner with variant {\two}, the multiplier block also satisfies
\begin{equation*}
\lim_{\penaltyParam\rightarrow0} \norm{Q_{33}} = 0, \qquad \lim_{\penaltyParam\rightarrow\infty} \norm{Q_{33}} = 1.
\end{equation*}
Thus, as $\penaltyParam\rightarrow0$, the limiting preconditioned operator becomes singular because $Q_{33}\rightarrow0$. 
Although $Q_{33}\rightarrow I$ as $\penaltyParam\rightarrow\infty$, the full preconditioned operator does not approach the identity,
as $Q_{12}$ and $Q_{32}$ approach nonzero limits. In fact, the limiting $Q_{22}$ has eigenvalues, which are not equal to one,
hence, the overall eigenvalue spectrum does not cluster around one.
To underline our statements, we conduct a small eigenvalue analysis similar to the one presented in \secref{subsec:modified_augmented_lagrangian_block_preconditioner} with the
resulting eigenvalue spectra given in \figref{fig:modified_augmented_lagrangian_preconditioner_penalty_influence} for varying penalty parameter values.
Considering Schur complement variant {\one} in the top row, the eigenvalues of the preconditioned linear operator cluster around one and move towards
zero for increasing penalty parameter values. As the influence of $B_{\penaltyParam}$ gets stronger with bigger $\penaltyParam$, the Schur complement approximation
$\hat{S}_{\penaltyParam}$ given in \eqref{eq:schur_complement_variant_1} moves away from the actual $S_{\penaltyParam}$,
which supports our assumption. For variant {\two}, the eigenvalues are clustered around zero for small penalty parameter values. With an increasing
$\penaltyParam$ the eigenvalue spectrum is bounded away from zero, yet does not cluster around one. Increasing the penalty parameter even further
worsens the eigenvalue spectrum again, with eigenvalues moving back to zero. This underscores that the removal of $B_{\penaltyParam}^T$ violates
the properties shown in the second row of \figref{fig:augmented_lagrangian_preconditioner_penalty_influence}. We expect that there exists an optimal
value for $\penaltyParam$, such that the eigenvalues of $\inv{\mathcal{P}_{\penaltyParam, \text{mod}}}\linearOperator_{\penaltyParam}$ cluster around one. We
numerically show this behavior in \secref{subsec:penalty_parameter_choice}.

We now turn to the Schur complement approximation for $\mathcal{P}_{\indexedSolid{\penaltyParam}, \indexedBeam{\penaltyParam}, \text{mod}}$. Due to the introduction
of two independent penalty parameters and similar to~\cite{BenziPreprint},
we are able to scale the system \eqref{eq:simple_individually_augmented_discretized_coupled_system}, such that we can obtain
a favorable solution. 
Since $\trans{B_{\indexedSolid{\penaltyParam}}}$ is omitted, the asymptotic analysis motivates choosing $\indexedSolid{\penaltyParam}$
sufficiently small so that the neglected coupling contribution remains weak.
This also has the nice side effect, that $\indexedSolid{K}_{\penaltyParam} \approx \indexedSolid{K}$ for small~$\penaltyParam$ and thus standard AMG works very effectively.
In contrast, we need to choose an appropriate $\indexedBeam{\penaltyParam}$, such that the influence of $\indexedBeam{K}_{\indexedBeam{\penaltyParam}}$ is
highlighted and the eigenvalues are shifted towards one. While this leads to a severe ill-conditioning of the beam sub-problem, its influence on the actual preconditioner implementation
is, in our case, negligible as we solve the beam sub-problem with a direct method.

\paragraph{Variant {\three}}
For the third case, we set
\begin{equation*}
\hat{S}_{\indexedSolid{\penaltyParam}, \indexedBeam{\penaltyParam}} = -\frac{1}{\indexedBeam{\penaltyParam}}  W
\end{equation*}
and choose to use $ \indexedBeam{\penaltyParam}$ as the overall problem is dominated by the beam penalty contribution.
By setting $W = \kappa$ as highlighted before, the augmented Schur complement follows the approximation given by
\begin{equation}
\hat{S}_{\indexedSolid{\penaltyParam}, \indexedBeam{\penaltyParam}} = -\frac{1}{\indexedBeam{\penaltyParam}} \kappa.
\label{eq:schur_complement_variant_3}
\end{equation}
Considering the respective Schur complement approximation for the individually scaled case we obtain the following
asymptotic limits for changing $\indexedSolid{\penaltyParam}$ and fixed $\indexedBeam{\penaltyParam}$:
\begin{alignat*}{2}
\lim_{\indexedSolid{\penaltyParam}\rightarrow0} \norm{Q_{32}} &{}= 0, &\qquad \lim_{\indexedSolid{\penaltyParam}\rightarrow\infty} \norm{Q_{32}} &{}= \norm{\indexedBeam{\penaltyParam}\inv{W}D\inv{(\indexedBeam{K}_{\indexedBeam{\penaltyParam}})}\indexedBeam{K}}, \\
\lim_{\indexedSolid{\penaltyParam}\rightarrow0} \norm{Q_{33}} &{}= \norm{\indexedBeam{\penaltyParam}\inv{(W+\indexedBeam{\penaltyParam}D\inv{(\indexedBeam{K})}\trans{D})}(M\inv{(\indexedSolid{K})}\trans{M}+D\inv{(\indexedBeam{K})}\trans{D})}, &\qquad \lim_{\indexedSolid{\penaltyParam}\rightarrow\infty} \norm{Q_{33}} &{}= \norm{\indexedBeam{\penaltyParam}\inv{W}D\inv{(\indexedBeam{K}_{\indexedBeam{\penaltyParam}})}\trans{D}}.
\end{alignat*}
Using $\indexedSolid{\penaltyParam}\rightarrow0$ is favorable as $Q_{32}$ vanishes and we are able to recover a block upper-triangular operator. In contrast to the exact Schur complement, $Q_{33}$ does
not approach the identity, but a more complicated limit depending on $\indexedBeam{\penaltyParam}$. Thus, we finally consider the special case of $\indexedSolid{\penaltyParam}\rightarrow0$ under
varying $\indexedBeam{\penaltyParam}$ resulting in
\begin{alignat*}{2}
\lim\limits_{\substack{\indexedSolid{\penaltyParam}\to 0\\ \indexedBeam{\penaltyParam}\to 0}} \norm{Q_{33}} &{}= 0, 
&\qquad \lim\limits_{\substack{\indexedSolid{\penaltyParam}\to 0\\ \indexedBeam{\penaltyParam}\to \infty}}  \norm{Q_{33}} &{}= \norm{I + \inv{(D\inv{(\indexedBeam{K})}\trans{D})} M\inv{(\indexedSolid{K})}\trans{M}}.
\end{alignat*}
Similar to the cases before, there exists an optimal value for $\indexedBeam{\penaltyParam}$ away from the asymptotic limits.

\begin{figure}
\includegraphics[trim={2cm 0cm 2cm 0cm}, clip, width=\textwidth]{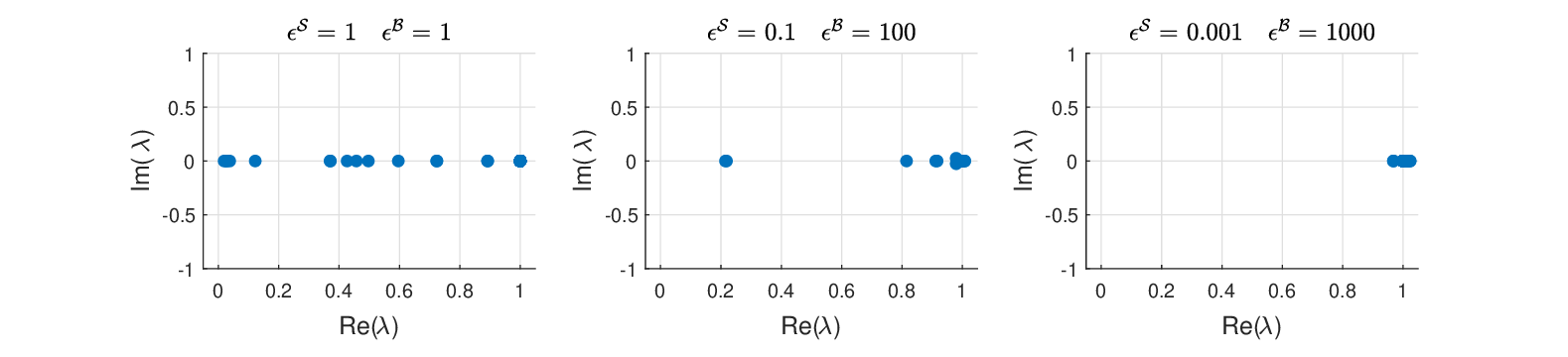}
\caption{Comparison of the eigenvalue spectrum of preconditioned operator
$\inv{\mathcal{P}_{\indexedSolid{\penaltyParam}, \indexedBeam{\penaltyParam}, \text{mod}}}\linearOperator_{\indexedSolid{\penaltyParam}, \indexedBeam{\penaltyParam}}$ with
Schur complement variant {\three} for the {\mixeddimensional} {\beamsolid} model problem for different values of $\indexedSolid{\penaltyParam}$ and $\indexedBeam{\penaltyParam}$.}
\label{fig:modified_individually_augmented_lagrangian_preconditioner_penalty_influence}
\end{figure}

As shown in \figref{fig:modified_individually_augmented_lagrangian_preconditioner_penalty_influence}, the eigenvalues nicely cluster around one for $\indexedSolid{\penaltyParam} \rightarrow 0$
and growing $\indexedBeam{\penaltyParam}$. Yet we remark that there are a few outliers, which we do not show explicitly. For an increasing beam penalty parameter, we expect an over-penalization
of the system, thus shifting eigenvalues above one. This assumption is backed by the asymptotic limit given by $\indexedBeam{\penaltyParam}\rightarrow\infty$. We study this effect for different
combinations of $\indexedSolid{\penaltyParam}$ and $\indexedBeam{\penaltyParam}$ in more detail in \secref{subsec:penalty_parameter_choice}.

While there exist occasions, where it makes sense to use an approximation of the full augmented Schur complement \eqref{eq:augmented_schur_complement} \cite{He2018},
we omit this variant in the scope of this publication, but mention it here for the sake of completeness.

\subsection{Comparison to existing methods in literature}
\label{sec:comparison_with_literature}

Having discussed all details of the proposed preconditioner, we now position it relative to representative block preconditioners from the literature. In particular,
we compare the presented modified augmented Lagrangian methods with operator preconditioning methods for {\saddlepoint} systems~\cite{Baerland2019,Dimola2024,Kuchta2019,Kuchta2016}
and with penalty-based block factorization strategies~\cite{Budisa2024, Cerroni2019, Firmbach2024a}.

In contrast to our approach, the preconditioners from~\cite{Baerland2019, Dimola2024, Kuchta2019, Kuchta2016} are tailored to {\saddlepoint} problems
stemming from the {\mixeddimensional} coupling of scalar transport equations and are based on the concept of operator preconditioning, resulting in block
diagonal preconditioners. The arising Schur complement is approximated by a spectrally matching fractional Laplacian, which constitutes
the main difference. This idea is not directly transferable to the present vector-valued coupling of 3D elasticity and 1D {\torsionfree} {\KirchhoffLove} beams
without a new operator preconditioning analysis for the beam block and the mixed-dimensional coupling operators. We therefore regard these methods as
relevant reference points, but not as out-of-the-box competitors for the present problem class. Still, the application of operator preconditioning
to the present problem type is an attractive topic for future research.

For the penalty-based case, the most relevant method is the approximate block factorization preconditioner developed in~\cite{Firmbach2024a}, which addresses
exactly the same beam-solid coupling setting. In contrast to the present augmented Lagrangian formulation, that approach solves the penalty-regularized system~\eqref{eq:penalty_discretized_coupled_system}
and relies on an approximate factorization of the resulting $2\times2$ block structure. It therefore provides the closest predecessor and is considered as the
primary baseline for assessing the benefit of the modified augmented Lagrangian preconditioners later in \secref{subsec:hybrid_composite_plate}.
The multigrid method proposed in~\cite{Budisa2024} is also designed for penalty-regularized coupled problems, however, its application to domains with different
degrees of freedom and physics has not yet been explored. Likewise, the exact block preconditioning approach presented in~\cite{Cerroni2019} has not been investigated
in the given setting too.

\section{Numerical examples}
\label{sec:numerical_examples}

To evaluate the proposed block preconditioners with the corresponding Schur complement variants, we carry out a set of experiments designed to quantify the effect of the modeling
parameters, assess parallel scalability and demonstrate the relevance of the proposed methods for engineering applications. In a first step we examine the influence of the penalty
parameter on the overall convergence behavior of the preconditioned linear solver based on the discussion presented in \secref{sec:block_preconditioner}. The second
numerical experiment is designed to investigate the dependency of the preconditioners on varying model parameters as introduced in \secref{subsec:modeling_parameters}.
Afterwards, we assess weak and strong scalability of the proposed methods. In a last step we show the applicability of the block preconditioners to an engineering application
consisting of a ﬁber-reinforced hybrid composite and compare the newly proposed approaches to the commonly used incomplete LU (ILU) factorization preconditioner and the
penalty-based approximate block factorization preconditioner from~\cite{Firmbach2024a}.

\begin{figure}
\centering
\begin{subfigure}{0.32\textwidth}
\includegraphics[trim={40cm 1cm 40cm 1cm}, clip, width=\textwidth]{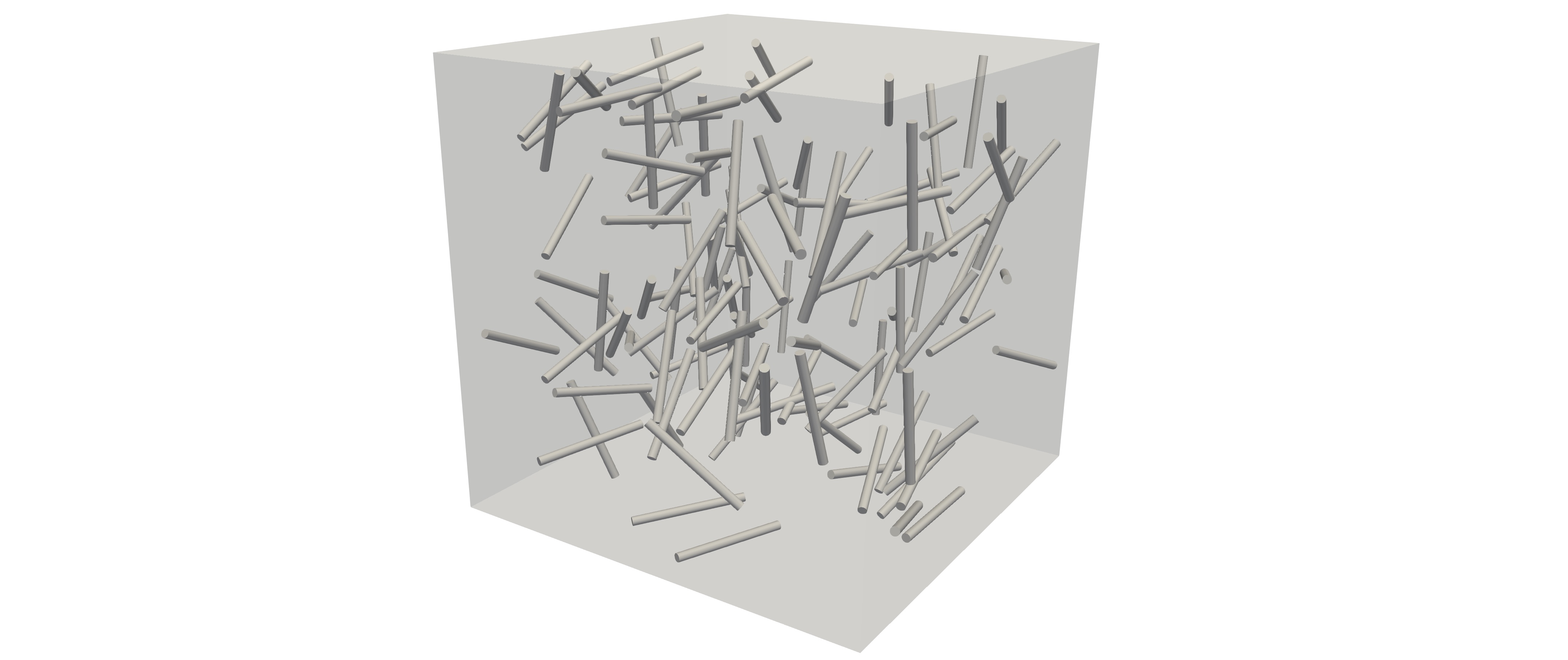}
\caption{$\mathcal{\volume}=1\%$}
\end{subfigure}
\hfill
\begin{subfigure}{0.32\textwidth}
\includegraphics[trim={40cm 1cm 40cm 1cm}, clip, width=\textwidth]{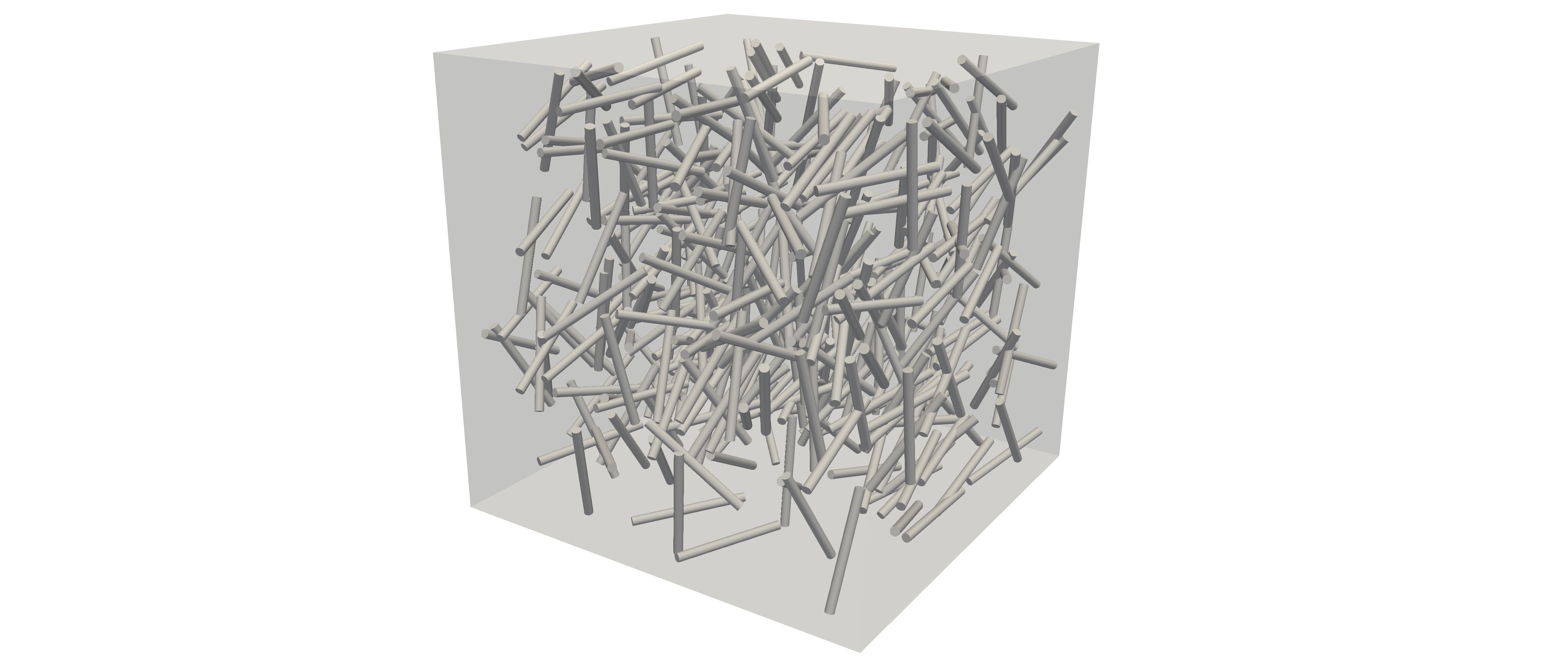}
\caption{$\mathcal{\volume}=3\%$}
\end{subfigure}
\hfill
\begin{subfigure}{0.32\textwidth}
\includegraphics[trim={40cm 1cm 40cm 1cm}, clip, width=\textwidth]{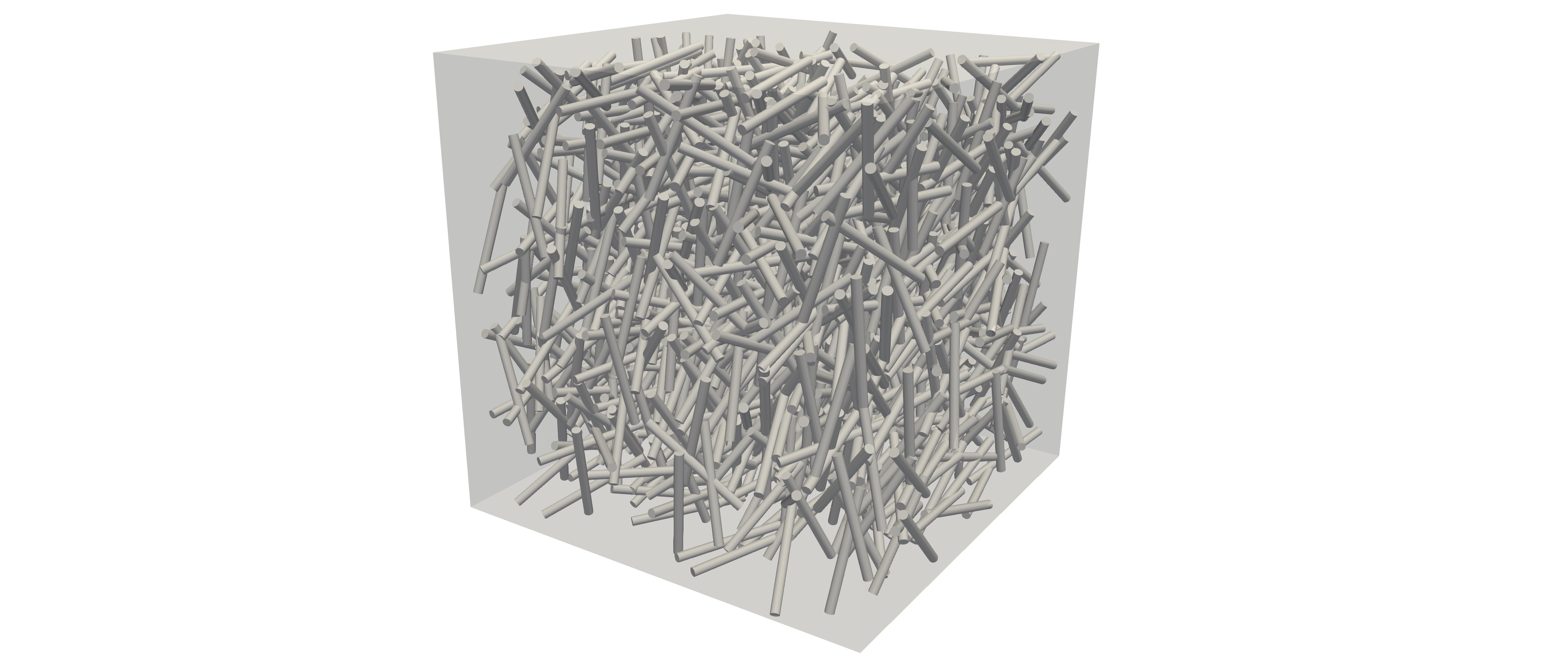}
\caption{$\mathcal{\volume}=10\%$}
\end{subfigure}
\caption{Visualization of the numerical test setup for different {\beamtosolid} volume ratios for a constant beam {\crosssection} radius.}
\label{fig:numerical_test_setup}
\end{figure}

The general numerical setup for the first three experiments in \secref{subsec:penalty_parameter_choice}--\ref{subsec:scalability} is inspired by
the test cases given in \cite{Lauff2025b, Lauff2025a} and the ones introduced in \cite{Firmbach2024a}.
It mimics a generic short fiber-reinforced material modeled by a representative volume element (RVE). The domain of the solid continuum is given by a unit cube
with fixed volume $\indexedSolid{\volume}=\qty{1}{\cubic\meter}$, which is filled with randomly oriented fibers of equal length $l=\qty{0.25}{\meter}$.
An illustration of the geometry for different {\beamtosolid} volume ratios is given in \figref{fig:numerical_test_setup}. The solid body is clamped
at the bottom face setting the displacement in all directions to $\indexedSolid{u}=0$, while the top surface is displaced in vertical direction by $\qty{0.01}{\meter}$,
which is equal to $1\%$ of the cube's edge length.
In the following, we consider the solid mesh size $\indexedSolid{h}$, the beam mesh size $\indexedBeam{h}$, the beam Young's modulus $\indexedBeam{\youngs}$ and the beam
{\crosssection} radius $\indexedBeam{R}$ to be varying parameters. Therefore, also the stiffness contrast~$\mathcal{E}$ and volume ratio~$\mathcal{\volume}$
vary depending on the chosen values. Based on the findings in \cite{Steinbrecher2020}, we consider {\beamtosolid} mesh size ratios
of $\mathcal{H} \geq 2.5$ to avoid an overconstraining by the discrete Lagrange multipliers. As outlined in \secref{sec:equations}, the solid is modeled
as {\StVenantKirchhoff} material with the Young's modulus being fixed to $\indexedSolid{\youngs}=\qty{1}{\newton/\square\meter}$ and the Poisson's ratio set to~$\poisson=0.3$.
For this numerical study, the fibers are represented by a single {\torsionfree} {\KirchhoffLove} beam finite element each. This is justified since we consider short, straight fibers,
which are well modeled by one element only~\cite{Firmbach2024a, Steinbrecher2020} due to the high approximation quality of the {\torsionfree} {\KirchhoffLove} beam formulation
making use of cubic Hermite polynomials for the interpolation of the centerline displacements~\cite{Meier2019a}. In practice, the formulation therefore allows even relatively
long fibers to be approximated using only a few beam elements.
We employ the discretization scheme considered above, using continuous linear {\Lagrangean} finite elements for
the solid and the Lagrange multiplier field, and cubic Hermite polynomials for the beams. All geometries and meshes are generated using {\beamme} \cite{BeamMe}.

We use {\trilinos} \cite{Heroux2005a, Mayr2025a} and {\baci} \cite{4C} for the simulation of the numerical experiments.
We run both codes in a distributed-memory fashion building upon the Message Passing Interface ({\mpi}) communication model.
GMRES from {\belos} \cite{Bavier2012a} is used to solve the arising linear systems. The iterative solver is assumed to be
converged, when the residual is reduced by a factor of $10^8$. The modified augmented Lagrangian block preconditioners are implemented using the block
preconditioning package {\teko} \cite{Cyr2016a}. We apply the algebraic multigrid implementation from {\muelu}~\cite{BergerVergiat2023a}, specifically 
smoothed aggregation (SA-AMG) and plain aggregation (PA-AMG). We solve the coarsest level of the multigrid hierarchy with the distributed version
of {\superlu} \cite{Li2003a} as direct solver. The coarse level is reached if $6500$ or fewer unknowns remain. On each level, a pre- and post-smoothing is applied with
Chebychev polynomials of $3$rd order. Coarse levels of the multigrid hierarchy are repartitioned by a coordinate-based approach~\cite{Deveci2015} implemented in~{\zoltan2}
to ensure a proper workload per MPI process.
Direct sub-block solves are performed by~{\amesos}~\cite{Bavier2012a}, again using the distributed version of~{\superlu}.
All simulations are run on CPU nodes of our in-house cluster (one CPU node features 2x Intel Skylake CPUs with 24 cores each).

\subsection{Choice of the Schur complement and the penalty parameter}
\label{subsec:penalty_parameter_choice}

\begin{figure}[t]
\centering
\includegraphics[width=\textwidth]{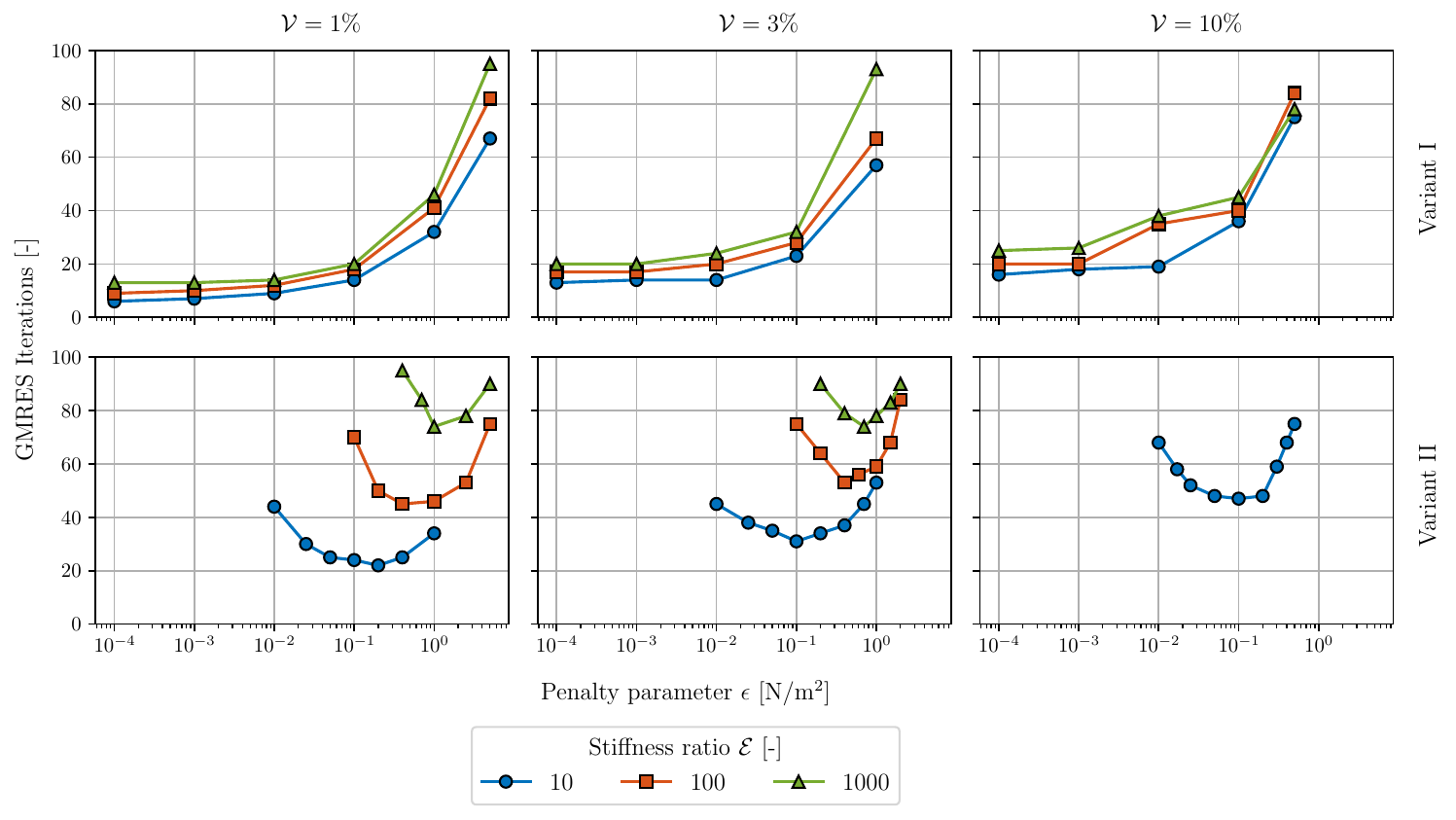}
\caption{Iteration count of the linear solver based on Schur complement variant {\one} (top) and variant {\two} (bottom) for a changing penalty parameter.
In addition the results are shown for different stiffness ratios and a varying {\beamtosolid} volume ratio of $\mathcal{\volume}=1\%$ (left),
$\mathcal{\volume}=3\%$ (middle) and $\mathcal{\volume}=10\%$ (right).}
\label{fig:penalty_parameter_influence}
\end{figure}

\begin{figure}[t]
\centering
\includegraphics[width=\textwidth]{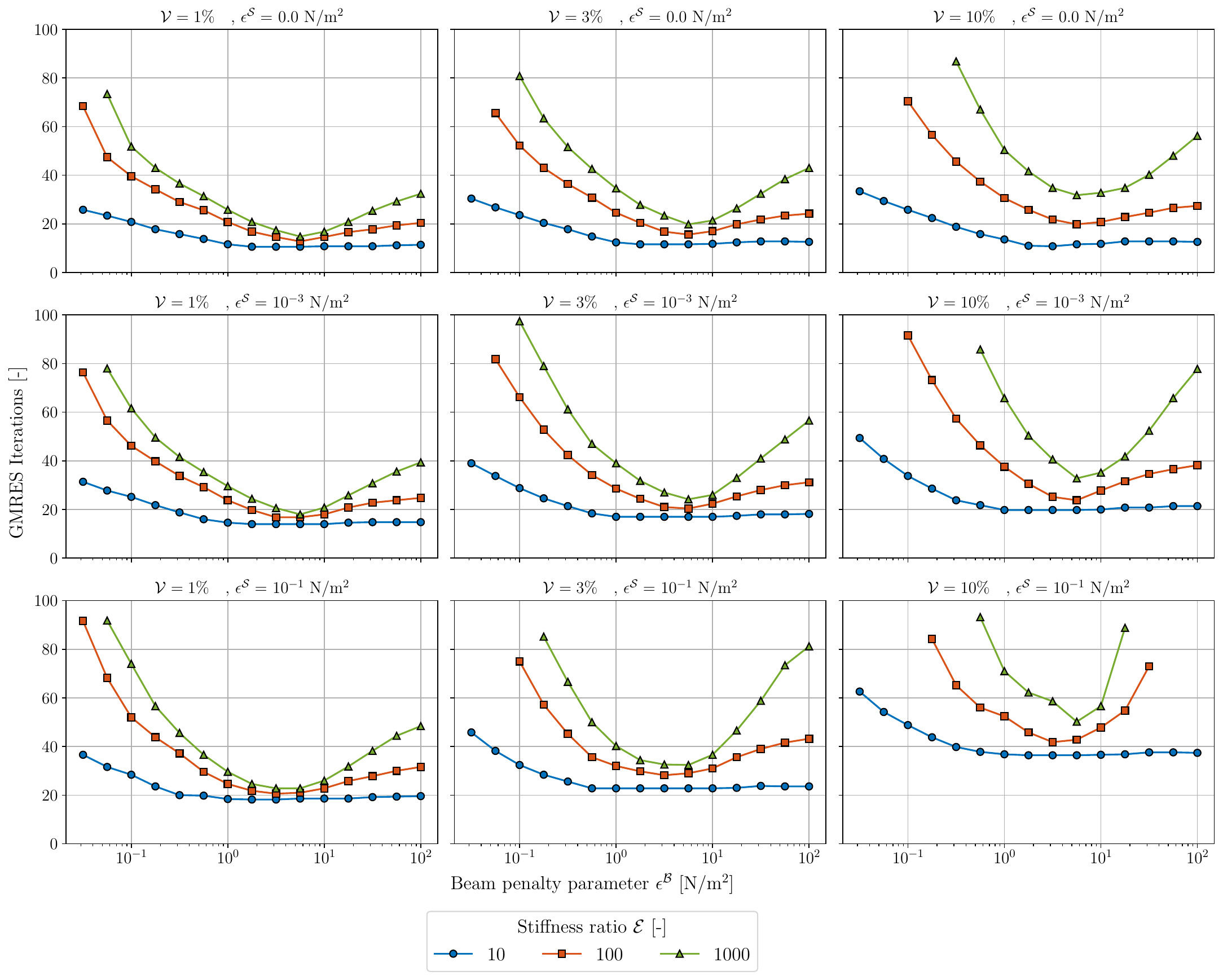}
\caption{Iteration count of the linear solver based on Schur complement variant {\three} for changing solid and beam penalty parameters. 
In addition the results are shown for different stiffness ratios and a varying {\beamtosolid} volume ratio of $\mathcal{\volume}=1\%$ (left),
$\mathcal{\volume}=3\%$ (middle) and $\mathcal{\volume}=10\%$ (right).}
\label{fig:penalty_parameter_influence_2}
\end{figure}

The choice of the penalty parameter $\penaltyParam$ (respectively $\indexedSolid{\penaltyParam}$ and $\indexedBeam{\penaltyParam}$ for variant~{\three} of the Schur complement)
has a major influence on the overall performance of the proposed block preconditioners, as it steers the approximation quality of the Schur complement variants.
In a first study, we examine how the number of iterations of the linear solver changes due to a
varying~$\penaltyParam \in [\qty{e-4}{\newton/\square\meter}, \qty{10}{\newton/\square\meter}]$
(additionally $\indexedSolid{\penaltyParam}\in \{\qty{e-1}{\newton/\square\meter}, \qty{e-3}{\newton/\square\meter}, \qty{0.0}{\newton/\square\meter}\}$ and $\indexedBeam{\penaltyParam}\in [\qty{e-2}{\newton/\square\meter}, \qty{e2}{\newton/\square\meter}]$),
whereas the penalty-regularized formulation from~\cite{Steinbrecher2020} proposes a fixed value $\penaltyParam\approx\indexedBeam{E}$.
We utilize the three Schur complement approximations from \secref{subsec:choice_of_schur_complement}
given by variant~{\one} \eqref{eq:schur_complement_variant_1}, variant~{\two} \eqref{eq:schur_complement_variant_2} and variant~{\three} \eqref{eq:schur_complement_variant_3}.
The dependency of the number of linear iterations on the value of $\penaltyParam$ is shown in \figref{fig:penalty_parameter_influence} for variants~{\one}
and~{\two}, with the first row corresponding to variant {\one}, while the second row represents variant {\two}. The respective behavior for variant~{{\three}}
is visualized in \figref{fig:penalty_parameter_influence_2} for varying $\indexedSolid{\penaltyParam}$ and $\indexedBeam{\penaltyParam}$.
Experiments with a constant stiffness ratio $\mathcal{E} \in \{10, 100, 1000\}$
are distinguished by different colors. Each column corresponds to a different {\beamtosolid} volume ratio~$\mathcal{\volume}\in\{1\%, 3\%, 10\%\}$.
We assume a constant mesh size ratio of $\mathcal{H} = 7.5$, which turns out to be a good choice considering the results given in \secref{subsec:parameter_robustness}
and a constant beam {\crosssection} radius of $\indexedBeam{\beamRadius} =\qty{0.01}{\meter}$.

For Schur complement variant {\one}, we introduced the assumption, that the respective approximation is valid for $\penaltyParam \ll \indexedBeam{E}$ with its
quality degrading for $\penaltyParam \rightarrow \indexedBeam{E}$. This behavior can exactly be observed in the first row of \figref{fig:penalty_parameter_influence}
throughout all columns. For small values of $\penaltyParam$, the iteration count stays low for different stiffness ratios and volume ratios, only increasing
slightly for higher values of $\mathcal{E}$ and $\mathcal{\volume}$. Shifting the penalty parameter closer to $\indexedBeam{E}$ results in a degradation of the
approximation quality of the Schur complement and therefore results in unbounded values for the number of iterations. We conclude that for sufficiently
small values of $\penaltyParam$ far away from $\indexedBeam{E}$, variant {\one} is a valid and robust approximation for the Schur complement.

The observed behavior changes considerably for variant~{\two}. Similar to the observations made in~\cite{Benzi2011b}, there exists a value
for $\penaltyParam$ that minimizes the iteration count, which can be seen in the second row of \figref{fig:penalty_parameter_influence}.
In addition, the Schur complement approximation appears to depend on the  stiffness ratio, resulting in a potentially unbounded iteration count.
For $\mathcal{\volume} = 10\%$, only the characteristic behavior of $\mathcal{E}=10$ is shown, as higher material contrasts exceed the allowed iteration
count and are therefore omitted. The value of $\penaltyParam$ that minimizes the iteration count is also not identical for different
stiffness ratios. The optimal penalty parameter for variant {\two} should be chosen higher for larger values of $\mathcal{E}$. Increasing the stiffness contrast
reduces the parameter range that leads to fast convergence, thereby complicating the identification of the optimal penalty parameter value.
For application cases in our range of consideration, which feature a medium material contrast, a valid choice is~$\penaltyParam\in[0.001\indexedBeam{E}, 0.01\indexedBeam{E}]$.
	
Regarding variant {\three}, the behavior is again slightly different as illustrated in \figref{fig:penalty_parameter_influence_2}. Each diagram plots the
number of linear iterations over a changing value of the beam penalty parameter for a unique combination of a fixed {\beamtosolid} volume ratio and solid penalty parameter.
Again, there exists a parameter combination of $\indexedSolid{\penaltyParam}$ and $\indexedBeam{\penaltyParam}$, which minimizes the iteration count and
in contrast to variant {\two} results in a limited dependency to the stiffness and volume ratio. For a low stiffness ratio, an increasing volume ratio affects the number of linear iterations taken
only slightly if $\indexedSolid{\penaltyParam}$ is chosen sufficiently small. This can be observed when comparing the iteration results of $\indexedSolid{\penaltyParam}=\qty{e-1}{\newton/\square\meter}$
and $\indexedSolid{\penaltyParam}=\qty{0.0}{\newton/\square\meter}$ for a constant value of $\mathcal{E}=10$. For the first parameter combination, an
increasing volume ratio also leads to an increase in iterations, while for the second one the iteration count remains constant. For larger material contrasts,
this behavior marginally degrades, with the choice
of $\indexedBeam{\penaltyParam}$ becoming more important, as the curves of linear iterations for a constant stiffness ratio now feature a distinct minimum.
A general observation is, that for $\indexedSolid{\penaltyParam} \rightarrow 0$ the number of linear iterations decreases steadily, while an appropriate choice of
the beam penalty parameter reduces the influence on the modeling parameters for $\mathcal{\volume}<10\%$ for the given model problem.
	
The overall results for variant~{\two} are very similar to the findings from \cite{Benzi2011b, Franceschini2022a} showing the effect of $\penaltyParam$ related to
problems from fluid and contact mechanics. While Schur complement variant {\one} is robust {\wrt} different stiffness and volume ratios,
the respective penalty parameter has to be chosen sufficiently small and far away from $\indexedBeam{E}$. In addition, the calculation
of the sparse approximate inverse of the beam matrix sub-block as well as two triple matrix products to form the Schur complement present a computational complexity, which
should not be underestimated. In contrast, Schur complement variant {\two} is represented by a simple diagonal and, thus, is computationally cheap.
Yet, it lacks independence from the stiffness and volume ratio and needs a matching $\penaltyParam$ to be effective.
Variant~{\three} combines the best of the two other Schur complement approaches, also being represented by a diagonal and able to partly regain independence
to the modeling parameters and low iteration counts by steering $\indexedSolid{\penaltyParam}$ and $\indexedBeam{\penaltyParam}$ correctly.
The presented results are mostly in line with the behavior observed in~\cite{BenziPreprint}.
In the following numerical experiments, we further highlight all three variants discussing the need for parameter robustness or performance.

\subsection{Robustness under varying model parameters}
\label{subsec:parameter_robustness}

\begin{figure}[t]
\centering
\includegraphics[width=\textwidth]{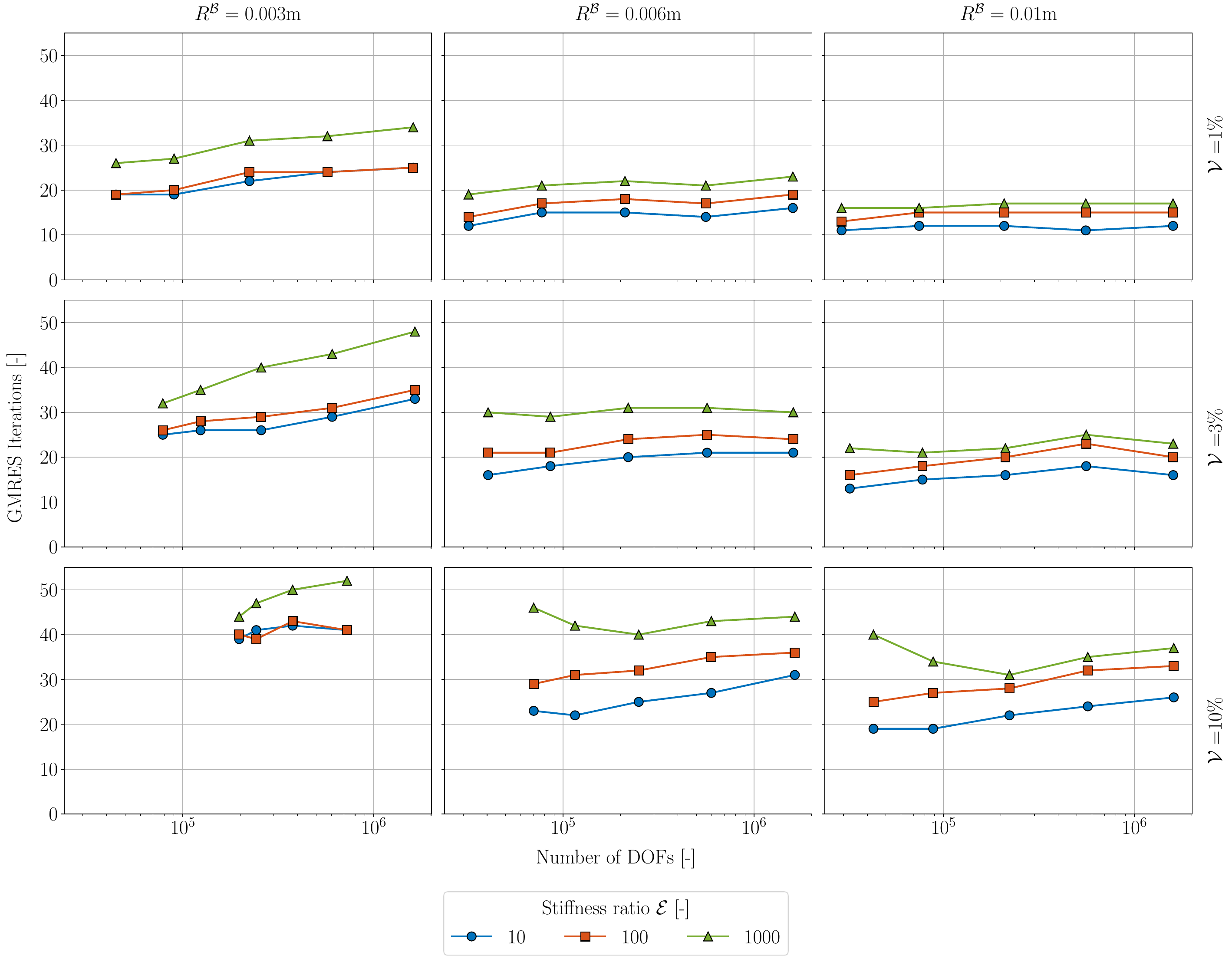}
\caption{Visualization of the iteration count based on different solid mesh sizes for different stiffness ratios,
a varying {\beamtosolid} volume ratio of $\mathcal{\volume}=1\%$ (top), $\mathcal{\volume}=3\%$ (middle) and $\mathcal{\volume}=10\%$ (bottom)
and changing beam {\crosssection} radius $\indexedBeam{\beamRadius} = \qty{0.003}{\meter}$ (left), $\indexedBeam{\beamRadius} = \qty{0.006}{\meter}$
(middle) and $\indexedBeam{\beamRadius} = \qty{0.010}{\meter}$ (right) for Schur complement variant {\one}.}
\label{fig:parameter_robustness}
\end{figure}

Now, we study the robustness of the block preconditioners {\wrt} model parameter changes and how these
influence the number of iterations of the linear solver. Based on the initial results obtained in \secref{subsec:penalty_parameter_choice},
in particular the challenges in identifying an optimal penalty parameter for Schur complement variant~{\two}
we only use variant~{\one} with a constant penalty parameter $\penaltyParam=\qty{0.01}{\newton/\square\meter}$
and variant~{\three} with fixed $\indexedSolid{\penaltyParam}=\qty{0.0}{\newton/\square\meter}$ and $\indexedBeam{\penaltyParam}=\qty{5.0}{\newton/\square\meter}$.
For the parameter study, we consider varying values of the {\beamtosolid} volume ratio $\mathcal{\volume} \in \{1\%, 3\%, 10\%\}$, the stiffness
ratio $\mathcal{E} \in \{10, 100, 1000\}$ and the beam {\crosssection} radius $\indexedBeam{\beamRadius} = \{\qty{0.003}{\meter}, \qty{0.006}{\meter}, \qty{0.010}{\meter}\}$.
We also study changing values of the mesh size ratio~$\mathcal{H}$ by maintaining a constant beam mesh size and varying the solid mesh size.
Smaller solid element sizes are generated by uniform mesh refinement of the base configuration.
The results are visualized in \figsref{fig:parameter_robustness}{fig:parameter_robustness_2} for variants~{\one} and~{\three}, respectively,
showing the number of linear iterations plotted over an increasing number of degrees of freedom (DOFs). Each row of the plot shows different values of the volume
ratio with $\mathcal{\volume}=1\%$ (top), $\mathcal{\volume}=3\%$ (middle) and $\mathcal{\volume}=10\%$ (bottom). The columns are
related to the beam {\crosssection} radius with $\indexedBeam{\beamRadius}=\qty{0.003}{\meter}$ (left), $\indexedBeam{\beamRadius}=\qty{0.006}{\meter}$ (middle)
and $\indexedBeam{\beamRadius}=\qty{0.01}{\meter}$ (right).
As outlined in Section~\ref{subsec:modeling_parameters}, all parameter combinations considered in the following remain within the intended validity regime of the mixed-dimensional model. The parameter variations
are chosen to represent practically relevant changes in quantities such as the beam {\crosssection} radius, beam volume ratio, and beam-to-solid stiffness ratio. Accordingly, the results are interpreted primarily 
with respect to the influence of these application-relevant parameters on solver performance. The cases combining the finest solid discretization with a beam {\crosssection} radius of $\qty{0.010}{\meter}$ are close
to the limit of the underlying modeling assumption but remain within its intended regime.

In the following, we consider the parameter combination $\mathcal{\volume}=1\%$ and $\indexedBeam{\beamRadius}=\qty{0.01}{\meter}$ in the upper right corner
of \figsref{fig:parameter_robustness}{fig:parameter_robustness_2} as baseline. For this case, both preconditioners show a limited dependency on the modeling
parameters with decreasing  solid mesh size, as the iteration count stays perfectly constant for variant~{\one} and marginally increases for variant~{\three}. 

\begin{figure}[t]
\centering
\includegraphics[width=\textwidth]{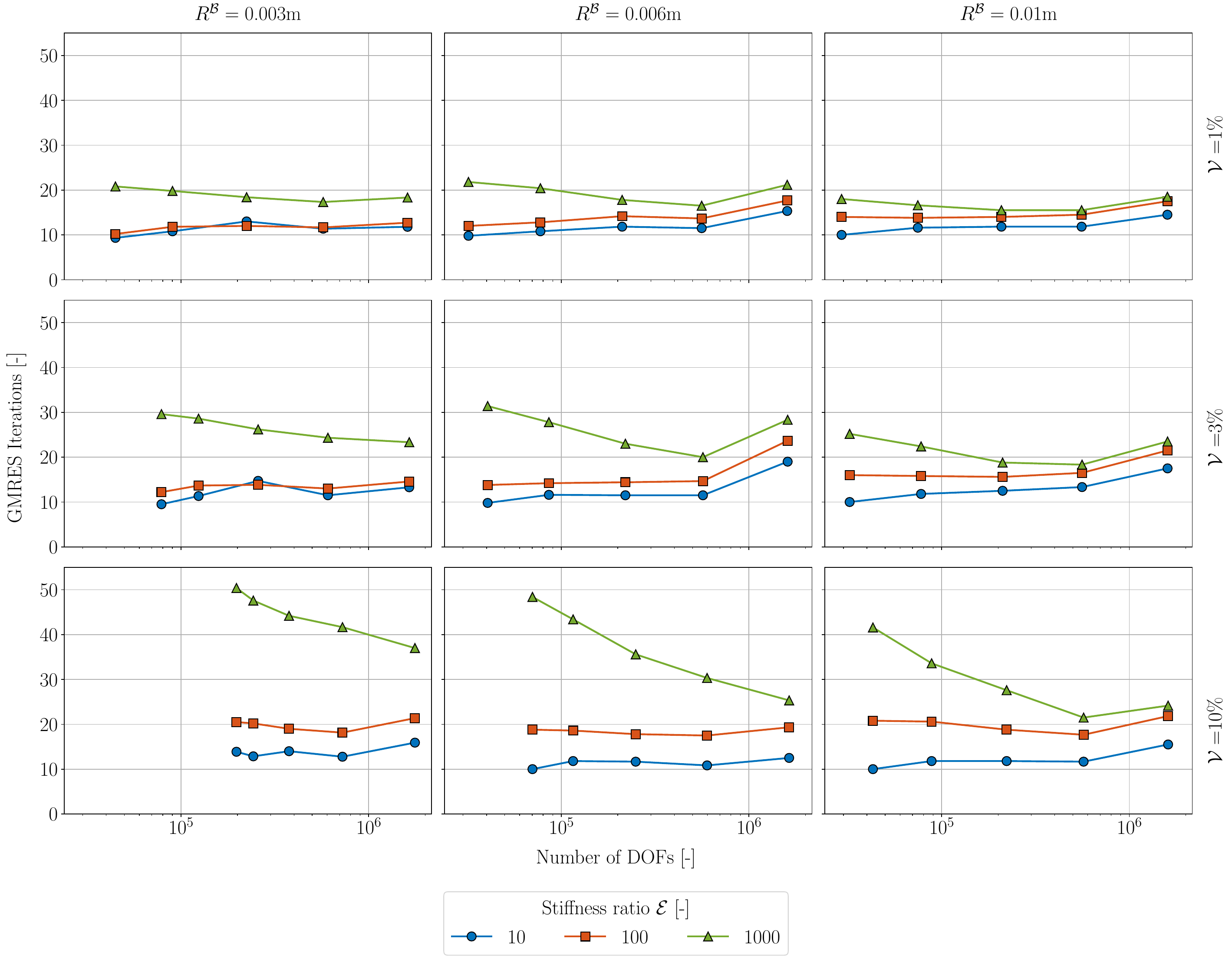}
\caption{Visualization of the iteration count based on different solid mesh sizes for different stiffness ratios, a varying {\beamtosolid} volume ratio
	of $\mathcal{\volume}=1\%$ (top), $\mathcal{\volume}=3\%$ (middle) and $\mathcal{\volume}=10\%$ (bottom)
and changing beam {\crosssection} radius $\indexedBeam{\beamRadius} = \qty{0.003}{\meter}$ (left), $\indexedBeam{\beamRadius} = \qty{0.006}{\meter}$
(middle) and $\indexedBeam{\beamRadius} = \qty{0.010}{\meter}$ (right) for Schur complement variant {\three}.}
\label{fig:parameter_robustness_2}
\end{figure}

We first discuss the behavior of the block preconditioner using Schur complement variant~{\one} when the beam {\crosssection} radius is reduced. Hereby, the block preconditioner
loses some of its parameter independence showing a weak relation of the iteration count {\wrt} the stiffness ratio in case of a decreasing solid mesh size. In addition, the number of iterations
reacts more sensitive to increasing stiffness ratios with more iterations necessary to find a solution to the linear problem. Reducing the {\crosssection} radius while keeping $\mathcal{\volume}$
constant effectively results in more fibers and, thus, more beam elements interacting with the solid continuum. It is observed in \cite{Dimola2024}, that the {\crosssection}
radius of the one-dimensional domain is a relevant parameter for parameter robustness, specifically related to the discretization approach itself. In our case, $\indexedBeam{\beamRadius}$
has a similar influence at first glance, but due to constant {\beamtosolid} volume ratios enforced for the robustness study the parameter is tightly connected to the amount of
interaction partners present.
%Therefore, we remark that the beam {\crosssection} radius needs further investigation in an isolated setting, whereas the present study is tailored more
%towards user parameters relevant for applications.

In a second step, we increase $\mathcal{\volume}$ to $3\%$, which also shows a dependency of the iteration
count to the stiffness ratio with decreasing solid mesh size. In addition, for large solid element mesh sizes the iteration count seems to decrease first, while
increasing again for decreasing values of $\indexedSolid{h}$. This seems to be related to an overconstraining of the coupling conditions by the Lagrange multiplier
field and is not discussed further in the scope of this publication, as we mainly deal with the construction of block preconditioners and do not specifically
consider an in-depth analysis of the discretization and respective formulation. For the parameter
combination $\mathcal{\volume}=10\%$ and $\beamRadius=\qty{0.003}{\meter}$,
the numerical solver breaks down due to extremely populated matrix sub-blocks arising from the multitude of coupling contributions, which marks the limit of the coupling approach
and Schur complement variant~{\one} itself.
While showing near-independence in parameters for low values of $\mathcal{\volume}$ and $\beamRadius=\qty{0.01}{\meter}$, the iteration count is showing an increasing
dependency on mesh refinement for growing stiffness ratios. In addition, this effect is amplified for examples with small beam radii and large {\beamtosolid}
volume ratios. In general, the iteration count increases modestly for a decreasing $\indexedBeam{\beamRadius}$ and increasing $\mathcal{\volume}$.

Now we discuss the results for the block preconditioner using Schur complement variant {\three}. Again, we consider a reduction of the beam {\crosssection} radius first
keeping $\mathcal{\volume}$ fixed and thus looking at each row of \figref{fig:parameter_robustness_2} independently. Hereby, the preconditioner shows to be robust with
a decreasing mesh size as the number of iterations as well as the overall iteration behavior remains similar. This holds true for all stiffness ratios considered. Only for the highest
material contrast, an increase in linear iterations can be observed, which amplifies for increasing values of $\mathcal{\volume}$. Especially for $\mathcal{\volume}=10\%$
and $\mathcal{E}=1000$, this effect is pronounced as the iteration count reduces drastically with smaller mesh size.

In a next step we consider fixed beam radii and vary the volume ratio, thus considering each column separately at a time. For an increasing volume ratio, the number of
linear iterations for the lowest stiffness contrast~$\mathcal{E}=10$ remains constant for each beam {\crosssection} radius. For $\mathcal{E}=100$ the growth in iterations is still small, yet
the highest material contrast~$\mathcal{E}=1000$ again shows to be sensitive to parameter changes.
In contrast to Schur complement variant~{\one}, we are mostly able to maintain independence from the modeling parameters across different beam {\crosssection} radii and volume ratios
for low to medium material contrast values. For high stiffness ratios, the iteration count reduces with decreasing mesh size, which is the opposite behavior as for variant~{\one},
where the number of iterations increases.

The overall iteration count for both block preconditioners stays between $10$ and $50$ iterations for the given parameter setup, which is still in an acceptable range for
our applications. It should also be highlighted that the preconditioners are still parameter robust for a variety of combinations.

\subsection{Strong and weak scalability}
\label{subsec:scalability}

In the following, we study the parallel efficiency of the proposed block preconditioners by performing a strong and weak scaling study. We conduct the
numerical experiment with a fixed parameter configuration based on the findings from \secref{subsec:penalty_parameter_choice} and
\secref{subsec:parameter_robustness}. To offer enough computational work and a robust setup, we consider a {\beamtosolid} volume ratio of~$\mathcal{\volume}=3\%$,
a stiffness ratio of~$\mathcal{E}=10$ and a mesh size ratio of $\mathcal{H} = 7.5$, while the beam {\crosssection} radius is set to~$\indexedBeam{\beamRadius}=\qty{0.01}{\meter}$.
The respective penalty parameter for each Schur complement is given by $\penaltyParam=\qty{0.01}{\newton/\square\meter}$ for variant~{\one},
$\penaltyParam=\qty{0.15}{\newton/\square\meter}$ for variant~{\two} and $\indexedSolid{\penaltyParam}=\qty{0.0}{\newton/\square\meter}$,
$\indexedBeam{\penaltyParam}=\qty{5.0}{\newton/\square\meter}$ for variant~{\three}.
We adapt the cube length of the solid domain for each problem size such that the overall geometric ratios are preserved during weak scaling. The number of DOFs
of each discretization for the different problem sizes is shown in \tabref{tab:system_sizes}. Each problem scale starts at an initial processor count $P$ featuring
around~$n^{\text{total}}_{DOF}/P \approx \num{90000}$ DOFs per processor, which is halved several times for the strong scaling setup, until we reach the strong
scaling limit or the maximum number of processors~$P=\num{512}$ in our setup. Hereby,~$n^{\text{total}}_{DOF}$ describes the total problem size. All timings
are obtained by averaging the respective numbers from $\num{20}$ individual simulation runs.

\begin{table}
\caption{Number of DOFs $n_{DOF}$ of each part of the discretization for different problem sizes used for the scaling study.}
\centering
\begin{tabular}{c c c c c c}
\hline
ID & $\indexedSolid{n}_{DOF}$ & $\indexedBeam{n}_{DOF}$ & $n^{\lambda}_{DOF}$ & $n^{\text{total}}_{DOF}$ \\
\hline
$1$ & $\num{89373}$    & $\num{4584}$    & $\num{2292}$   & $\num{96249}$    \\
$2$ & $\num{346053}$   & $\num{18336}$   & $\num{9168}$   & $\num{373557}$   \\
$3$ & $\num{1361613}$  & $\num{73344}$   & $\num{36672}$  & $\num{1471629}$  \\
$4$ & $\num{5401533}$  & $\num{293364}$  & $\num{146682}$ & $\num{5841579}$  \\
%$5$ & $\num{21516573}$ & $\num{1173420}$ & $\num{586710}$ & $\num{23276703}$ \\
\hline
\end{tabular}
\label{tab:system_sizes}
\end{table}

\begin{figure}
\centering
\begin{subfigure}{\textwidth}
\includegraphics[width=\textwidth]{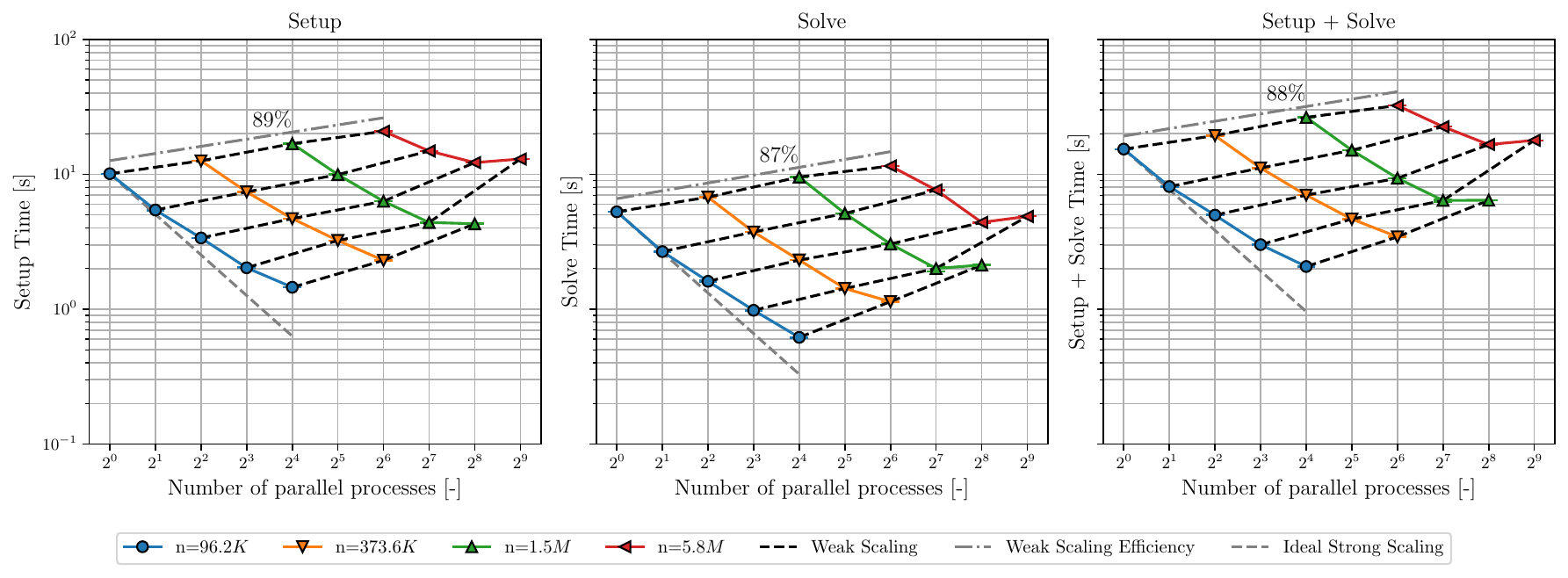}
\caption{Schur complement variant {\one} scaling results for the short fiber composite RVE for different problem sizes.}
\label{fig:weak_and_strong_scalability_I}
\end{subfigure}
\begin{subfigure}{\textwidth}
\includegraphics[width=\textwidth]{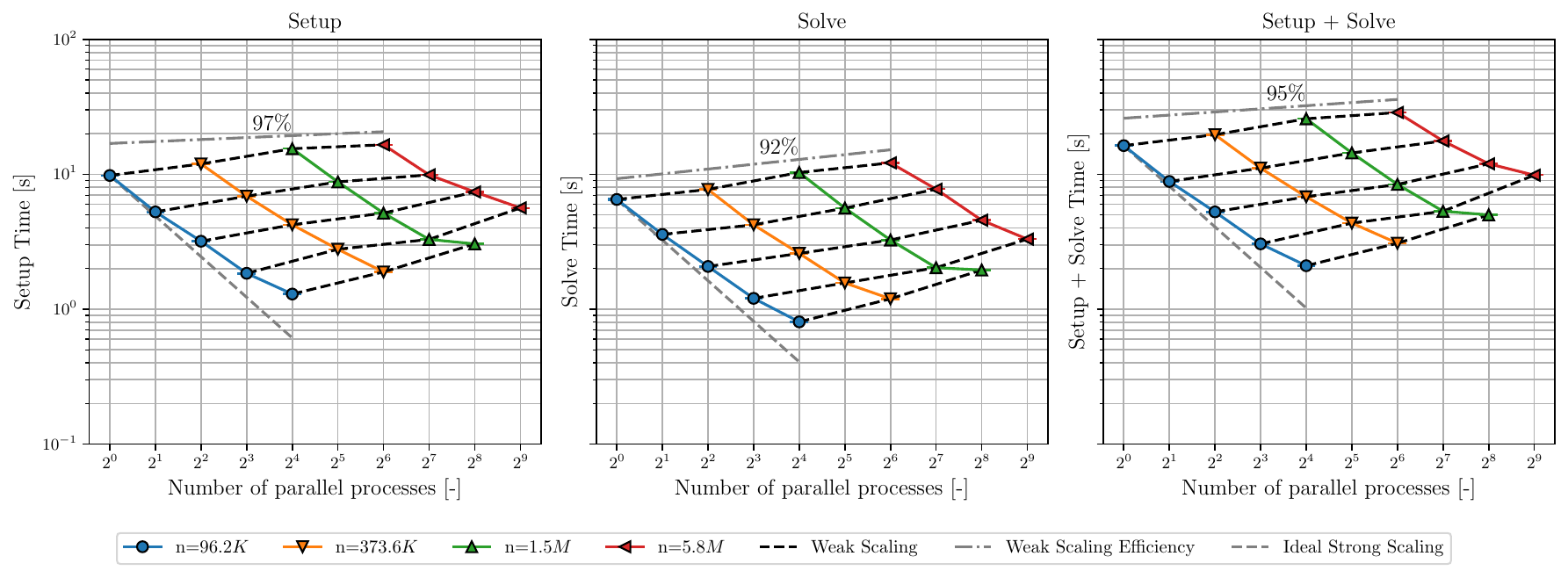}
\caption{Schur complement variant {\two} scaling results for the short fiber composite RVE for different problem sizes.}
\label{fig:weak_and_strong_scalability_II}
\end{subfigure}
\begin{subfigure}{\textwidth}
\includegraphics[width=\textwidth]{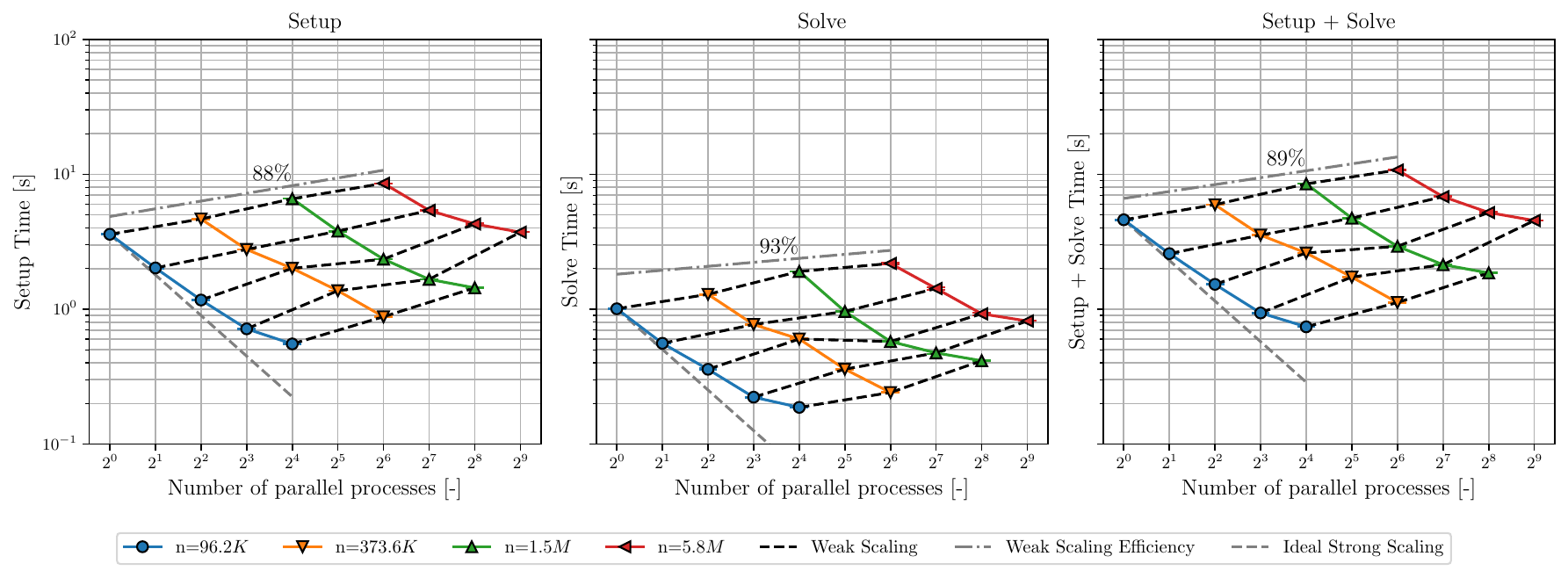}
\caption{Schur complement variant {\three} scaling results for the short fiber composite RVE for different problem sizes.}
\label{fig:weak_and_strong_scalability_III}
\end{subfigure}
\caption{Weak and strong scaling results for Schur complement variant~{\one} (top), variant~{\two} (middle) and variant~{\three} (bottom).
Ideal strong scaling is shown for reference, along with a line corresponding to the estimated weak scaling efficiency over the problem scales.
Ideal weak scaling is a horizontal line.}
\label{fig:weak_and_strong_scalability}
\end{figure}

As initial verification of our test setup, we check the iteration count of the linear solver for each problem scale. For variant~{\one} and {\two}
the number of iterations stays in between $\num{30}$ and $\num{38}$ for all different processor counts, while for variant~{\three} the values
range from $\num{15}$ to $\num{20}$. The actual scaling results are given as timings and are summarized in \figref{fig:weak_and_strong_scalability_I}
for Schur complement variant~{\one}, \figref{fig:weak_and_strong_scalability_II} for variant~{\two} and \figref{fig:weak_and_strong_scalability_III} for variant~{\three}.
We additionally compute the standard deviation for each data point and plot it as error bars. Owing to the near-identical timings for the generation of
one data point over $\num{20}$ runs, the deviation is negligible and therefore the actual error bar is hard to visualize, yet we state that the standard
deviation is at most $0.1229$ for all data points.
We differentiate between the setup, the solve and the combined setup + solve phase.
Lines of the same color represent strong scaling for a constant problem size, with the bright gray line representing ideal strong scaling. The black
dashed lines illustrate weak scaling over a constant problem size per {\mpi} process.

Regarding strong scalability, we are able to achieve good scaling results for the solve and setup phase of the first three problem scales until performance
degrades when reaching a load per {\mpi} process of~$n^{\text{total}}_{DOF}/P \approx \num{12000}$. This holds true for all variants, whereas variant~{\one}
shows slightly worse strong scaling. For bigger problem sizes and higher processor counts, this behavior is more pronounced. The problem with ID~$4$
(see \tabref{tab:system_sizes}) already hits the strong scaling limit at~$n^{\text{total}}_{DOF}/P \approx \num{24000}$, with both solve and setup timings
rising again for smaller values of~$n^{\text{total}}_{DOF}/P$.
%Using more processors for the biggest problem size has nearly no effect on timings.
%Despite the solve time being reduced, the setup time stays constant.
For variant {\two}, the effect is not as dramatic considering the same values for~$n^{\text{total}}_{DOF}/P$.
While strong scaling also slows down for higher processor counts, the setup and solve times are still decreasing.
This behavior also holds true for the block preconditioner using Schur complement variant {\three}.
The difference in the strong scaling efficiency comes from the different Schur complement approaches themselves.
For Schur complement variant {\two} and {\three}, a simple inversion of a diagonal matrix is sufficient, while for variant {\one} we employ a direct solver
for preconditioning the respective sub-block. As we go up to~$n^{\lambda}_{DOF}=\num{146682}$ for the Lagrange multiplier block ({\cf}~\tabref{tab:system_sizes}), 
the compute time of the direct method dominates the whole setup phase due to limited exploitation of parallelism and higher order of computational complexity
in comparison to {\eg} AMG, especially for an increasing number of processors.

The results for weak scalability show a similar picture. For variant {\one}, the setup phase is dominated by the factorization of the Lagrange multiplier sub-block,
resulting in a reduced weak scalability to large problem sizes and processor counts. We achieve efficiencies up to $89\%$ for the setup phase and acceptable $87\%$ for the solve
phase. The overall block preconditioner reaches a weak scaling efficiency of around $88\%$. For Schur complement variant {\two}, the setup phase shows very good
weak scalability up to $97\%$ and the solve phase reaches $92\%$. The combined setup and solve phase achieves a weak scaling efficiency of around $95\%$.
For variant~{\three}, we achieve similar solve efficiencies of approximately $93\%$, yet the efficiency of the setup phase degrades to $88\%$. As we
set~$\indexedSolid{\penaltyParam}=\qty{0.0}{\newton/\square\meter}$ for variant~{\three}, the number of {\nonzeros} of the solid sub-matrix is greatly reduced as the
penalty contribution is neglected. This results in less computational work per {\mpi} process compared to variant~{\two} and results in a less favorable parallel
scaling. The overall solve phase reaches a parallel efficiency of $89\%$.

We conclude that variant {\two} shows the best strong and weak scalability. Variant~{\three} slightly outperforms variant~{\one} in terms of the reported parallel efficiencies.
The overall timings suggest that the block preconditioner using Schur complement variant~{\three} enables the fastest solution process, being considerably faster than
the other two for the given problem setup. We remark that the scalability results reported here are representative for the short-fiber regime with block diagonal beam
sub-problems. For longer or more continuously connected fibers, where the beam block loses its block diagonal structure and the coupling matrices feature more {\nonzeros},
the cost of the direct sub-solves and the overall setup phase may scale less favorably.

\subsection{Application: Hybrid composite plate}
\label{subsec:hybrid_composite_plate}

\begin{figure}
\centering
\begin{subfigure}[b]{0.49\textwidth}
\includegraphics[width=\textwidth]{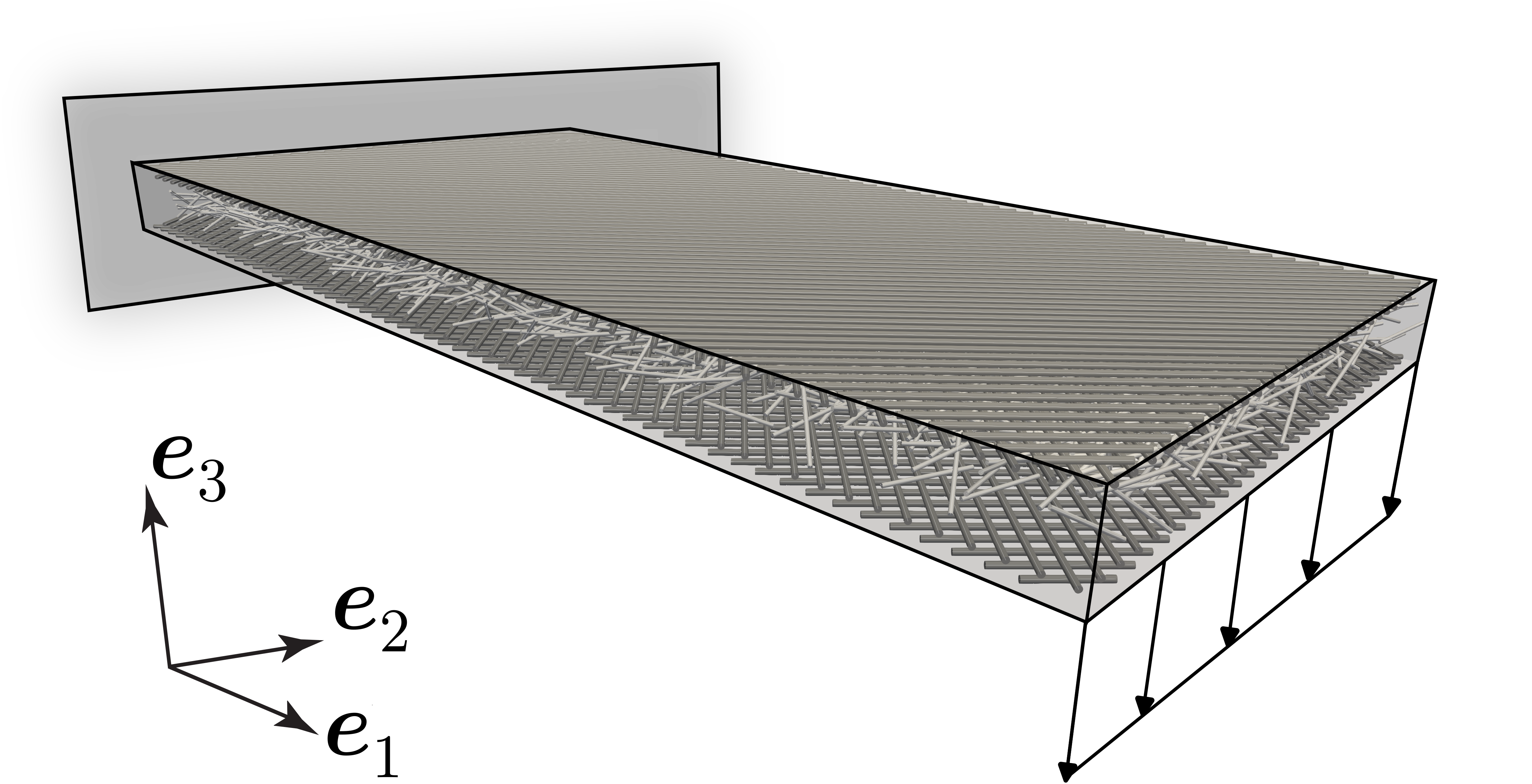}
\caption{Load and boundary setup of the hybrid composite plate model.}
\label{fig:hybrid_composite_model}
\end{subfigure}
\hfill
\begin{subfigure}[b]{0.49\textwidth}
\includegraphics[width=\textwidth]{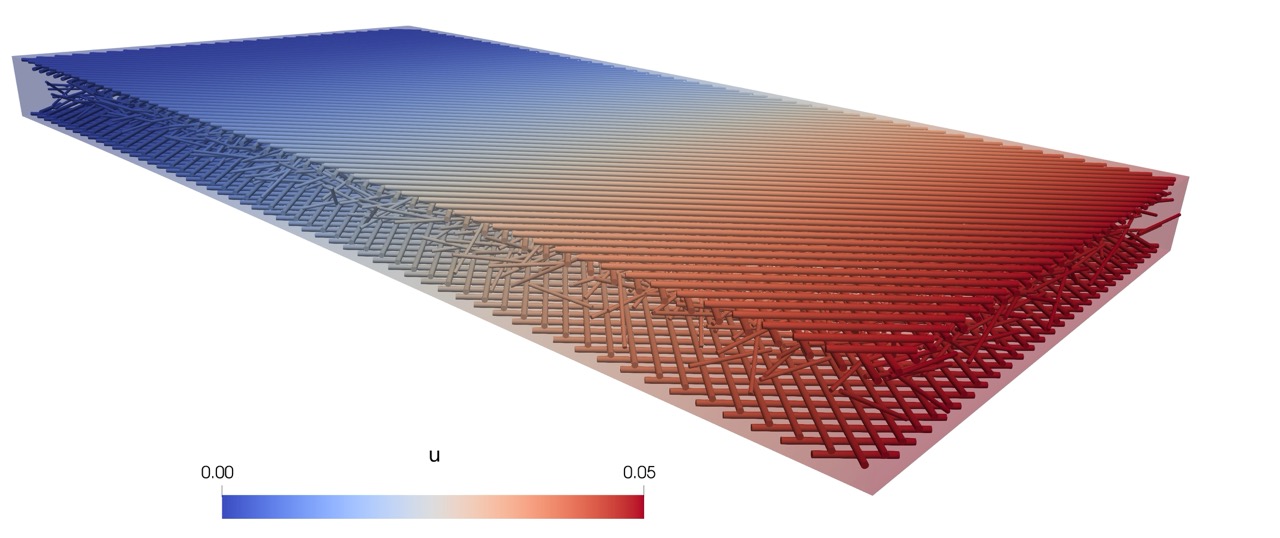}
\caption{Displacement field due to a prescribed deformation at the free end.}
\label{fig:hybrid_composite_model_displacement}
\end{subfigure}
\\
\begin{subfigure}{\textwidth}
\includegraphics[width=\textwidth]{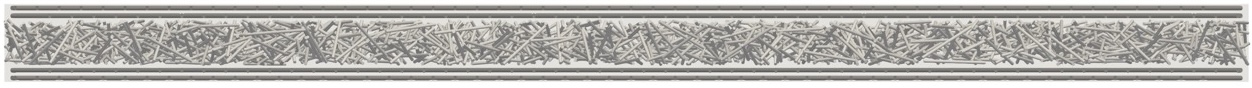}
\caption{Cross-section view with the solid being shown transparent, the upper and lower fiber layers (dark gray) and the short fiber core (light gray).}
\label{fig:hybrid_composite_model_cross_section}
\end{subfigure}
\caption{A hybrid composite plate with a prescribed deformation at its free end.}
\end{figure}

To assess the performance of the presented block preconditioners under an application-oriented configuration, we consider their application to a hybrid composite
plate (inspired by the example shown in \cite{Lauff2025a}) with different layers of fiber orientations and distributions. The problem setup is
shown in \figref{fig:hybrid_composite_model} with a visualization of the {\crosssection} given in \figref{fig:hybrid_composite_model_cross_section}.
We consider the problem setup to resemble a sandwich plate with the dimensions~$\qty{2.0}{\meter}\times\qty{4.0}{\meter}\times\qty{0.25}{\meter}$. The
top and bottom sheets of the structure consist of two continuous fiber layers each rotated by~$\qty{45}{\degree}$ and $\qty{-45}{\degree}$,
respectively. The individual fibers are discretized adaptively based on their length with short fibers being represented by only one beam element,
while longer ones contain up to $\num{11}$ elements. The beam {\crosssection} radius is set to~$\indexedBeam{\beamRadius}=\qty{0.0075}{\meter}$ and the
Young's modulus to~$\indexedBeam{\youngs}=\qty{210}{\newton/\square\meter}$.  The inner part of the structure consists of a
composite core with short fibers of constant length $l=\qty{0.25}{\meter}$. Each fiber is discretized by one {\torsionfree}
{\KirchhoffLove} beam element with~$\indexedBeam{\beamRadius}=\qty{0.005}{\meter}$ and~$\indexedBeam{\youngs}=\qty{70}{\newton/\square\meter}$.
The solid part of the domain features the same material properties throughout the whole continuum: a {\StVenantKirchhoff} material with Young's
modulus~$\indexedSolid{\youngs}=\qty{30}{\newton/\square\meter}$ and Poisson's ratio~$\poisson=0.3$. It is discretized by first-order hexahedral
finite elements. The left side of the structure is fully clamped, the right side is displaced by~$\qty{0.05}{\meter}$ in negative~$e_3$-direction forcing
the plate to bend (see \figref{fig:hybrid_composite_model_displacement}), while all other sides are subject to natural boundary conditions.
We consider a hierarchy of three different meshes for the solid domain, which are created by uniform mesh refinement of the base configuration.
The numbers of DOFs for the beam and the Lagrange multiplier field remain constant.

We study GMRES as linear solver with different preconditioners, namely a naive approach with a textbook ILU preconditioner, the approximate block factorization
preconditioner from~\cite{Firmbach2024a} as well as the proposed augmented Lagrangian block preconditioners with Schur complement variants {\one}, {\two} and {\three}
from~\secref{sec:block_preconditioner}. The factorization-based preconditioner reads
\begin{equation*}
	\mathcal{P}_{\text{fact}}
	=
	\begin{pmatrix}
		\hat{S} & -\trans{B}_{\penaltyParam} \\
		~ &  \indexedBeam{K}_{\penaltyParam} \\
	\end{pmatrix}
	\begin{pmatrix}
		I & ~ \\
		-\inv{(\indexedBeam{K}_{\penaltyParam})}B_{\penaltyParam} & I \\
	\end{pmatrix}
\end{equation*}
with $\hat{S} = \indexedSolid{K}_{\penaltyParam} - B_{\penaltyParam} \operatorname{SPAI}(\indexedBeam{K}_{\penaltyParam}) \trans{B}_{\penaltyParam}$.
While the block factorization preconditioner is not directly depending on a specific choice of the penalty parameter, we use $\penaltyParam = \qty{210.0}{\newton/\square\meter}$
to fulfill the constraints of the penalty-regularized system~\eqref{eq:penalty_discretized_coupled_system} properly.
For variant~{\one}, the penalty parameter is set to~$\penaltyParam = \qty{0.01}{\newton/\square\meter}$, while we use~$\penaltyParam = \qty{0.2}{\newton/\square\meter}$
for variant~{\two}. As the interaction of the continuous fiber layers with the surrounding solid is non-local, the resulting coupling matrices~$D$ and~$M$
as well as the contributions due to the augmented Lagrangian formulation feature proportionally many {\nonzero} entries. Instead of SA-AMG, we therefore
employ PA-AMG to reduce the fill-in on coarse levels of the augmented solid sub-problem for both preconditioner variants. A more detailed insight to this
effect and how it influences AMG is given in~\cite{Cerroni2019,Thomas2019} for example, while a more general overview on the solution process of these
types of systems is provided in~\cite{Benzi2024}. For Schur complement variant~{\three}, we consider three different parameter combinations. Option {\three}a
considers~$\indexedSolid{\penaltyParam}=\qty{e-3}{\newton/\square\meter}$ , while variant~{\three}b and~{\three}c set the penalty parameter related to the
solid contribution to zero. Both make use of PA-AMG, while the last option~{\three}c uses SA-AMG as setting~$\indexedSolid{\penaltyParam}=\qty{0}{\newton/\square\meter}$
avoids excessive fill-in due from penalty contributions on coarse multigrid levels. All different combinations of variant~{\three} use~$\indexedBeam{\penaltyParam}=\qty{1.0}{\newton/\square\meter}$.
For the beam and Schur complement sub-blocks, we use an LU factorization. In the case of a diagonal Schur complement approximation, we directly invert the matrix sub-block.
An overview of all different parameter combinations for the modified Lagrangian preconditioner cases is provided in Table~\ref{tab:hybrid_composite_variant_setups}.

\begin{table}
	\centering
	\caption{Block-LU and modified augmented Lagrangian preconditioner setup for the different Schur complement variants {\one}, {\two} and {\three} based on the penalty parameter
		and numerical methods applied to approximate the solid, beam and Schur complement contributions, respectively.}
	\label{tab:hybrid_composite_variant_setups}
	\begin{tabular}{c c c c c c}
		\hline
		Variant & \multicolumn{2}{c}{$\penaltyParam$  [$\unit{\newton/\square\meter}$]} & $\indexedSolid{\hat{\mathcal{P}}}$ & $\indexedBeam{\hat{\mathcal{P}}}$ & $\hat{\mathcal{P}}^{\lambda}$ \\
		\hline
		Block-LU & \multicolumn{2}{c}{$210.0$} & PA-AMG & LU & - \\
		\hline
		{\one} & \multicolumn{2}{c}{$0.01$} & PA-AMG & LU & LU \\
		{\two} & \multicolumn{2}{c}{$0.2$} & PA-AMG & LU & - \\
		\hline
		~ & $\indexedSolid{\penaltyParam}$ [$\unit{\newton/\square\meter}$] & $\indexedBeam{\penaltyParam}$ [$\unit{\newton/\square\meter}$] & ~ & ~ & ~ \\
		\hline
		{\three}a & $10^{-3}$ & $1.0$ & PA-AMG & LU & - \\
		{\three}b & $0.0$ & $1.0$ &  PA-AMG & LU & - \\
		{\three}c & $0.0$ & $1.0$ &  SA-AMG & LU & - \\
		\hline
	\end{tabular}
\end{table}

The linear solver is assumed to be converged, when the residual is reduced by a factor of~$10^{6}$. The corresponding tolerance is lower than in the previous numerical examples,
but this is justified for the application setting, where such a residual reduction is usually sufficient to obtain a good approximation of the linear system within the nonlinear solution
process and, in addition, allows comparibility to the application example results from~\cite{Firmbach2024a}. The solution methods and their iteration counts and timings are
summarized in \tabref{tab:hybrid_composite_results}. The timings are broken down into the setup time for each individual sub-block, namely the multigrid method
on the solid field~$\tSetup^{\indexSolid}$, the direct method used for the beam contribution~$\tSetup^{\indexBeam}$ and the different variants for the Schur complement~$\tSetup^{\lambda}$.
In addition, we consider the overall setup time for the block preconditioner~$\tSetup$, the solve time of the linear method~$\tSolve$ and finally the
total time spent for solving the linear system~$\tTotal$. All timings are obtained as averages of ten individual simulation runs. We also consider distributed
sparse direct solvers, yet none of them is able to solve the linear system in feasible time, thus we do not report results for those methods.

\begingroup
\setlength{\tabcolsep}{0.3em}
\begin{table}[h]
\centering
\caption{Comparison of averaged linear solver timings per nonlinear iteration for a hybrid composite sandwich plate
(Block-LU -- a GMRES solver using an approximate block factorization preconditioner,
Variant {\one}, {\two} and {\three} --- a GMRES solver using the block preconditioners proposed in \secref{sec:block_preconditioner}
 with the respective Schur complement variant). The best value in each category is highlighted in bold font.}
\label{tab:hybrid_composite_results}
\begin{tabular}{c c c c c c c c c c c c c c}
\hline
$\nproc$ & $n^{\indexSolid}_{DOF}$ & $n^{\indexBeam}_{DOF}$ & $n^{\lambda}_{DOF}$ & $n^{total}_{DOF}$ & Method & $\#$iter & \multicolumn{6}{c}{CPU time [$\unit{\second}$]} \\
~ & ~ & ~ & ~ & ~ & ~ & ~ &  $\tSetup^{\indexSolid}$ &  $\tSetup^{\indexBeam}$ & $\tSetup^{\lambda}$ & $\tSetup$ & $\tSolve$ & $\tTotal$ \\
\hline
\multirow{6}{*}{\num{8}} & \multirow{6}{*}{\num{430353}} & \multirow{6}{*}{\num{43344}} & \multirow{6}{*}{\num{21672}} & \multirow{6}{*}{\num{495369}} & Block-LU  & 74 & 3.638 & 0.112 & - & 21.153 & 62.473 & 83.625 \\
~ & ~ & ~ & ~ & ~ & Variant {\one}  & 47 & 0.902 & \textbf{0.104} & 5.160 & 7.255 & 13.827 & 21.082 \\
~ & ~ & ~ & ~ & ~ & Variant {\two}     & 81 & 0.949 & \textbf{0.104} & -  & 1.052 & 22.477 & 23.532 \\
~ & ~ & ~ & ~ & ~ & Variant {\three}a & 79 & 0.779 & 0.115 & -  & 0.912 & 19.735 & 20.647 \\
~ & ~ & ~ & ~ & ~ & Variant {\three}b & 23 &  \textbf{0.247} &0.107 & - & \textbf{0.375} & 1.5631 & \textbf{1.9383} \\
~ & ~ & ~ & ~ & ~ & Variant {\three}c & \textbf{20} & 0.859 & 0.111 & -  & 0.977 & \textbf{1.5017} & 2.4783 \\
\hline
\multirow{6}{*}{\num{64}} & \multirow{6}{*}{\num{3255903}} & \multirow{6}{*}{\num{43344}} & \multirow{6}{*}{\num{21672}} & \multirow{6}{*}{\num{3320919}} & Block-LU  & 72 & 7.962 & 0.116 & - & 37.104 & 131.414 & 168.518 \\
~ & ~ & ~ & ~ & ~ & Variant {\one}  & 37 & 0.745 & 0.188 & 0.945 & 3.690 & 13.967 & 17.662 \\
~ & ~ & ~ & ~ & ~ & Variant {\two}      & 82 & 0.781 & 0.198 & - & 0.967 & 19.829 & 20.796 \\
~ & ~ & ~ & ~ & ~ & Variant {\three}a & 75 & 0.770 & 0.189 & - & 0.954 & 11.505 & 12.459 \\
~ & ~ & ~ & ~ & ~ & Variant {\three}b & 37 & \textbf{0.550} & 0.184 & - & \textbf{0.745} & 4.391 & \textbf{5.136} \\
~ & ~ & ~ & ~ & ~ & Variant {\three}c & \textbf{21} & 3.037  & \textbf{0.182} & - & 3.2243 & \textbf{3.0494} & 6.274 \\
\hline
\multirow{6}{*}{\num{512}} & \multirow{6}{*}{\num{25308603}} & \multirow{6}{*}{\num{43344}} & \multirow{6}{*}{\num{21672}} & \multirow{6}{*}{\num{25373619}} & Block-LU  & 61 & 8.731 & 0.149 & - &  32.224 & 119.987 & 152.213 \\
~ & ~ & ~ & ~ & ~ & Variant {\one}  & 33 & 1.754 & 0.434 & 0.383 &  4.132 & 13.427 & 17.559 \\
~ & ~ & ~ & ~ & ~ & Variant {\two} & 82 & 1.770 & 0.470 & -     &  2.061 & 32.953 & 35.014 \\
~ & ~ & ~ & ~ & ~ & Variant {\three}a & 55  & 1.579 & 0.388 & - &  1.944 & 12.750 & 14.694 \\
~ & ~ & ~ & ~ & ~ & Variant {\three}b & 54 & \textbf{0.875} & 0.342 & - & \textbf{1.258} & 8.926 & 10.184 \\
~ & ~ & ~ & ~ & ~ & Variant {\three}c & \textbf{23} & 3.397 & \textbf{0.314} & - & 3.943 & \textbf{4.201} & \textbf{8.144}  \\
\hline
\end{tabular}
\end{table}
\endgroup

We consider the naive preconditioning approach first, which does not result
in convergence for any of the three meshes, emphasizing the importance of block preconditioning for this type of problem. 
In a next step, we consider the approximate block factorization preconditioner from \cite{Firmbach2024a} to serve as baseline for the following discussion.
The preconditioner achieves iteration counts of $\approx 70$ for the first two mesh configurations and $61$ for the finest mesh.
The proposed block preconditioner with Schur complement variant~{\two} performs slightly worse, showing a constant iteration count of~$\approx 80$ for all different mesh refinements.
Still, the number of iterations taken to find a solution for both approaches is rather high compared to variant~{\one}, which shows a decrease in the iteration count for an increasing
problem size, reducing from $47$ to $33$. This behavior is most likely connected to the ratio of solid and beam interaction pairs related to the overall
number of solid elements, which determines the density of the Lagrange multiplier matrix sub-block. With an increasing number of solid elements and
a constant number of beam elements, the sub-matrix is getting sparser, helping to find better approximations of the Schur complement. Variant~{\three}a
shows a similar characteristic with the iteration count decreasing from $79$ to $55$. In contrast, the parameter combinations related to {{\three}b} lead
to an increase in iterations from $23$ to $54$. Only variant~{{\three}c} shows a low and constant number of iterations of~$\approx 20$ over all three mesh
configurations. Overall, variant~{\three} outperforms variant~{\two} and the block-LU preconditioner, since setting $\indexedSolid{\penaltyParam}\rightarrow 0$
eliminates the approximation error introduced by the dropped off-diagonal block $\trans{B}_{\penaltyParam}$, which in turn leads to lower iteration counts. Moreover,
choosing~$\indexedSolid{\penaltyParam}=\qty{0}{\newton/\square\meter}$ removes the additional augmentation of the solid block, making SA-AMG
affordable for the respective sub-problem because excessive fill-in on coarse levels is avoided, thereby enabling proper scalability.

Further, we compare the timings of the individual preconditioners used for each of the sub-blocks~$\tSetup^{\indexSolid}$, $\tSetup^{\indexBeam}$
as well as~$\tSetup^{\lambda}$.  As we employ PA-AMG for the matrix block representing the solid stiffness, we can expect reasonable scalability
to bigger problem sizes. While roughly double the time is necessary between the base problem and the finest refinement for the multigrid setup, the
method still stays below~$\qty{2}{\second}$, while the problem size increases by a factor of~$\approx 60\times$. The setup cost of the block-LU
preconditioner stands out being considerably higher compared to the other preconditioning variants. This relates to the use of the augmented solid Schur
complement for the setup, which features substantially more {\nonzeros} than the augmented solid sub-matrix considered in all other methods.
The setup time of the SA-AMG method lies in between the benchmark preconditioner and the other variants, increasing
about a factor of four from~$\approx \qty{0.9}{\second}$ to~$\approx \qty{3.4}{\second}$ between the smallest and the biggest problem. 
Due to the beam sub-block being represented by a block-diagonal sparsity pattern, the direct factorization
works fast with a slight increase in the setup time mainly due to communication overhead due to the increased number of parallel processes.
The setup time of the factorization for the Lagrange multiplier block drastically decreases from the first to the second mesh refinement step. As the
amount of DOFs stays constant and the overall density of the matrix block decreases, reduced timings are to be expected. The timings between
variant~{\one}, {\two} and {\three}a are similar as the individual preconditioners act on the same matrix sub-blocks. For {\three}b, one can expect
a faster setup time for the PA-AMG method, as the missing penalty contribution results in a lower bandwidth and, thus, a reduced time. Another difference
is the construction of the Schur complement block, which for variant~{\two} and {\three} reduces to the inversion of a diagonal matrix and therefore we
neglect time measurements for this case. Next, we consider the total setup time $\tSetup$, with all variants showing a different behavior over all mesh
refinements. Variant~{\one} results in an expensive construction procedure for the block preconditioner due to the more expensive Schur complement
approximation requiring two triple matrix products and the SPAI calculation. While variant~{\two} shows
a slightly increasing trend, variant~{\one} gives a decrease in timings being dominated by the factorization of the Lagrange multiplier block.
The setup time of variant {\three}a behaves very similar to variant {\two} as both use the same methods and operate on the same sparsity patterns.
The block preconditioner  related to variant {\three}b achieves the overall lowest setup time, as the augmented Lagrangian terms of the solid contribution are omitted, which represent
the major bottleneck for the preconditioner setup due to an increased bandwidth of the sparsity pattern. The setup cost of variant~{\three}c is comparable
to the one of variant~{\one}, as SA-AMG is more expensive to construct, with both being the overall most costly methods in regard to the setup
of the modified augmented Lagrangian preconditioners presented. Still, all methods are able to beat the approximate block factorization preconditioner.

For the timings of the solve phase $\tSolve$, the behavior switches, with variant {\one} showing faster solve times due to a lower iteration count compared
to variant~{\two}. While variant~{\two} shows a constant iteration count over all problem sizes, the solve time still increases. While the augmentation adds
proportionally more {\nonzero} values to the matrix for each refinement step, the matrix-vector products inside the Krylov method therefore also get more
expensive and do not scale as efficiently. Variant~{\three}a performs as good and in some cases even better than variant~{\one}. Overall, the solve timings
related to variant~{\three}b and variant~{\three}c are the lowest. Lastly, we discuss the total time spent in the linear solver. For the base problem, the total
time $\tTotal$ for variants~{\one}, {\two} and {\three}a is nearly identical, with {\three}b and {\three}c both outperforming the other preconditioners
by~$\approx 10\times$. This changes for the other meshes with variant~{\one} being~$\approx 2\times$ faster than variant~{\two} for the biggest problem.
Especially in the largest mesh refinement case, a clear ordering of the methods becomes visible with variant~{\two} performing the worst followed by
variant~{\one}. The variants~{\three}a, {\three}b and {\three}c perform consistently better in exactly that order. The best method is~$\approx 4\times$
faster than the worst. Again, all methods are able to beat the benchmark case.

We conclude, that most preconditioning methods show a bounded iteration count over all different problem sizes. While variant {\one} is robust under mesh refinement,
its implementation comes with more effort and complexity, which also shows up in the timings. Variant {\two} is simple to construct, yet still robust
to a decreasing mesh size with the high iteration count as major drawback. Variant {\three} leaves more room to steer the Schur complement
approximation, thus all of its parameter combinations outperform the other variants in terms of timings. We consider the block preconditioner using variant~{\three}c
to perform the best for this problem type as it shows nearly constant iteration counts over all three mesh configurations and also provides almost always the best
time-to-solution. Compared with the block-LU preconditioner, the advantages of modified augmented Lagrangian preconditioning for the {\saddlepoint} problem are clear
and highlight its importance for the presented application case. This result, however, was to be expected as the approximate block factorization preconditioner is based
on the penalty-regularized system, and operates on matrices with much higher bandwidth and thus considerably more {\nonzeros}.

\section{Conclusion and outlook}
\label{sec:conclusion_and_outlook}

In this work, we proposed and analyzed modified augmented Lagrangian block preconditioners for the
{\mixeddimensional} {\beamsolid} coupling of a three-dimensional solid with embedded one-dimensional
{\torsionfree} {\KirchhoffLove} beams and constraint enforcement through Lagrange multipliers. Starting from the
underlying variational formulation and its finite element discretization, we studied the structure of the
resulting linear system, the presence of a pure Neumann sub-problem on the beam sub-block,
and the role of physically relevant modeling parameters.

To overcome the given limitations, we reformulated the discrete coupled problem in an augmented
Lagrangian fashion. This formulation preserves the exact enforcement of the coupling constraints while
adding a mild regularization that also improves solvability in the case of singular sub-blocks. Based on this augmented system,
we derived an ideal augmented Lagrangian block preconditioner and then constructed more practical,
block-triangular ``modified'' variants that allow for separate and efficient treatment of the solid, beam
and Schur complement blocks. We analyze the asymptotic limits of the preconditioned operator using
the presented block preconditioners and discuss practical approximations of the appearing inverses.
The solid sub-problem is handled by AMG, while the beam and Schur
complement blocks can be treated by sparse direct methods or are inverted directly in case of diagonal matrices.
Central to the effectiveness and performance of the approach is
an appropriate Schur complement approximation and the choice of the penalty parameter. We introduced three
Schur complement variants: one that approximates the Schur complement of the original system using simple inverses of
the solid and beam operators, a second one that exploits a scaling matrix, and a third one that introduces
different penalty parameters for the solid and beam sub-problem to improve the approximation with that scaling
matrix. This scaling matrix has been chosen as a mass-like matrix at the coupling interface of the Lagrange
multipliers to retain spectral equivalence of the Schur complement approximation.

The numerical experiments suggest that the proposed modified augmented Lagrangian block preconditioner, together with
suitable choices of the Schur complement approximation and penalty parameter, exhibits only a mild dependence on a broad
range of modeling parameters relevant for short-fiber-reinforced materials. For this tailored approximation and given parameter
combinations, the method shows iteration counts with a moderate mesh dependence over the admissible refinement
range for moderate stiffness contrasts and beam-to-solid volume ratios. It maintains acceptable iteration numbers even in more
challenging regimes characterized by small beam radii, large stiffness
ratios, and high beam-to-solid volume ratios. At the same time, the results also indicate that small beam radii combined with high
stiffness ratios remain extremely challenging for the proposed method. Moreover, the present study is restricted to short-fiber
configurations, the behavior for long or continuous fibers is not yet investigated. The influence of the penalty parameter on
convergence has been studied systematically for all three Schur complement variants, revealing the expected trade-off between
performance and independence from model parameters. Overall, the numerical results indicate that suitable choices of the Schur
complement approximation and penalty parameter can reduce the sensitivity of the iteration counts to simultaneous variations in
the considered modeling parameter range.

The parallel scalability studies show that the proposed preconditioners are suitable for large-scale parallel simulations
of short-fiber {\mixeddimensional} {\beamsolid} coupling scenarios. For representative RVE problems with several million
degrees of freedom, we observe favorable weak and strong scaling on distributed-memory architectures, with good efficiency
of both the AMG-based solid sub-solver and the block preconditioners as a whole, especially for Schur complement approximation
variant {\two} and {\three}. For longer or more continuously connected fibers, the beam and coupling blocks feature more {\nonzeros},
so the setup cost and direct sub-solves are expected to scale less favorably, which is left for future work.

Finally, we studied an application-oriented test case where the proposed coupling and preconditioning strategy
has been applied to a hybrid composite plate with layered fiber reinforcement and a short fiber-reinforced core.
The problem features moderate stiffness contrasts and heterogeneous length scales, posing a challenging
{\mixeddimensional} and multi-material setting. The results demonstrate that not only textbook preconditioners do not yield convergence,
but that the newly proposed modified augmented Lagrangian block
preconditioners enable convergence and remain robust and scalable in the given configuration for an appropriate choice of the Schur
complement approximation and penalty parameter. Compared to a block factorization preconditioner based on the
penalty-regularized system, the presented method is competitive regarding iteration counts and is overall considerably faster.

Further work will extend the proposed coupling scheme and block preconditioner to be usable with a Simo-Reissner beam formulation.
One difficulty in this regard is the introduction of additional rotational DOFs in the beam domain, which enables the modeling of more complex
interaction scenarios with initially curved fibers. Treating the coupling of both the solid and beam domain in a purely positional way without
any additional modifications is not sufficient anymore, as the rotations are not constrained properly, thus leaving components of the rigid
body modes unconstrained. Application areas of interest are {\mixeddimensional} {\beamsolid} volume and surface coupling. Addressing
these challenges will further broaden the applicability of the proposed methods and enable the efficient simulation of a wider range of
fiber-reinforced structures and {\mixeddimensional} {\beamsolid} interaction problems.

\section*{Acknowledgements}

The work described in this contribution has been funded by the \emph{Deutsche Forschungsgemeinschaft (DFG, German Research Foundation)} % -- project number 528397555.
within the project ``Stable discretization methods and scalable solvers for embedded fiber\slash{}solid coupling'' (project number 528397555)
as well as \emph{dtec.bw - Digitalization and Technology Research Center of the Bundeswehr}
under the project ``hpc.bw - Competence Platform for High Performance Computing''. dtec.bw is funded by the European Union – NextGenerationEU.
%The authors gratefully acknowledge the computing resources provided by the Data Science \& Computing Lab at the University of the Bundeswehr Munich.

\section*{Data Availability Statement}

Building blocks of the modified augmented Lagrangian preconditioner developed and applied in this study are openly
available in {\trilinos} at \url{https://github.com/trilinos/Trilinos}~\cite{Heroux2005a,Mayr2025a}. The implementation
of the {\mixeddimensional} {\beamsolid} coupling is available in 4C at \url{https://github.com/4C-multiphysics/4C}~\cite{4C}.
All other data that support the findings of this study are available from the corresponding author
upon reasonable request.

\section*{Declaration of Competing Interest}

The authors declare that they have no known competing financial interests or personal relationships
that could have appeared to influence the work reported in this paper.

\section*{Declaration of generative AI and AI-assisted technologies in the writing process}

During the preparation of this work, the authors used ChatGPT-5.5 Instant to fix writing errors and to improve the
readability and language in some parts of the manuscript. After using these tools, the authors reviewed and
edited the content and take full responsibility for the content of the published article.

%%%%%%%%%%%%%%%%%%%%%%%%%%%%%%%%%%%%%%%%
\section*{CRediT authorship contribution statement}

% Possible contributions: https://authorservices.wiley.com/author-resources/Journal-Authors/open-access/credit.html

\textbf{Max Firmbach:} Writing – original draft, Writing – review \& editing, Visualization, Validation, Software, Methodology,
Investigation, Formal analysis, Data curation, Conceptualization;
\textbf{Ivo Steinbrecher:} Writing – review \& editing, Software, Conceptualization;
\textbf{Alexander Popp:} Writing – review \& editing, Supervision, Funding acquisition;
\textbf{Matthias Mayr:} Writing – review \& editing, Writing – original draft, Supervision, Project administration, Methodology, Funding acquisition, Conceptualization.

All authors read and approved the final manuscript.

\bibliographystyle{abbrv}
\bibliography{bibliography}

\end{document}

%% file: defines.tex
\newcommand{\secref}[1]{Section~\ref{#1}} % reference to 'section #1'
\newcommand{\Secref}[1]{Section~\ref{#1}} % reference to 'Section #1'
\newcommand{\figref}[1]{Figure~\ref{#1}} % reference to 'figure #1'
\newcommand{\figsref}[2]{Figures~\ref{#1} and~\ref{#2}} % reference to 'figures #1 and #2'
\newcommand{\tabref}[1]{Table~\ref{#1}} % reference to 'table #1'
\newcommand{\tensor}[1]{\boldsymbol{\mathrm{#1}}}
\newcommand{\mat}[1]{\ensuremath{\mathbf{#1}}} %second order matrix
\newcommand{\mr}[1]{\ensuremath{\mathrm{#1}}} % write upright letters in math mode

\newcommand{\indexBeam}{\mathcal{B}}

\newcommand{\indexSolid}{\mathcal{S}}

\newcommand{\indexedDomain}[2]{#1^{#2}}

\newcommand{\indexedBeam}[1]{\indexedDomain{#1}{\indexBeam}}

\newcommand{\indexedSolid}[1]{\indexedDomain{#1}{\indexSolid}}

\newcommand{\inv}[1]{#1^{-1}} % inverse
\newcommand{\trans}[1]{#1^{\mr{T}}} % transpose
\newcommand{\RealSp}{\mathbb{R}} % vector space with real-numbered vectors
\newcommand{\REalSp}[1]{\RealSp^{#1}} % vector space with real-numbered vectors of dimension #1

\newcommand{\norm}[1]{\left\|#1\right\|} % any norm of a vector or matrix
\newcommand{\matDiv}{\mathrm{Div}} % divergence symbol in material configuration

\newcommand{\dom}{\ensuremath{\Omega}} % domain symbol
\newcommand{\volume}{V}
\newcommand{\beamRadius}{R}

\newcommand{\lm}{\lambda} % Lagrange multiplier
\newcommand{\weakForm}{\delta\mathscr{W}} % generic symbol denoting a weak form
\newcommand{\variation}{\delta} % prefix to denote a variation

\newcommand{\youngs}{E} % Young's modulus
\newcommand{\poisson}{\nu} % Young's modulus

\newcommand{\gap}{g} % gap function
\newcommand{\penaltyParam}{\epsilon} % penalty parameter
\newcommand{\nproc}{n^\mr{proc}} % number of processors/cores
\newcommand{\ndim}{d} % number of dimensions
\newcommand{\tSetup}{T_{\mathrm{setup}}}
\newcommand{\tSolve}{T_{\mathrm{solve}}}
\newcommand{\tTotal}{T_{\mathrm{total}}}

\newcommand{\linearOperator}{\mat{\mathcal{A}}} % linear operator / matrix in linear equations
\newcommand{\unknown}{u} % unknown solution vector in linear equations
\newcommand{\rhs}{f} % right-hand side vector in linear equations

\newcommand{\residualNonlinear}{f}

\newcommand{\matBeamLm}{D}
\newcommand{\matSolidLm}{M}

\newcommand{\beamsolid}{beam\hyp{}solid}
\newcommand{\beamtosolid}{beam\hyp{}to\hyp{}solid}

\newcommand{\fibersolid}{fiber\hyp{}solid}

\newcommand{\mixeddimensional}{mixed\hyp{}dimensional}

\newcommand{\nonzero}{non\hyp{}zero}
\newcommand{\nonzeros}{{\nonzero}s}

\newcommand{\torsionfree}{torsion\hyp{}free}
\newcommand{\saddlepoint}{saddle\hyp{}point}
\newcommand{\crosssection}{cross\hyp{}section}

\newcommand{\Lagrangian}{Lagrangean}
\newcommand{\Lagrangean}{Lagrangean}

\newcommand{\GreenLagrange}{Green--Lagrange}

\newcommand{\PiolaKirchhoff}{Piola--Kirchhoff}
\newcommand{\KirchhoffLove}{Kirchhoff--Love}

\newcommand{\SimoReissner}{Simo--Reissner}
\newcommand{\StVenantKirchhoff}{St.-Venant--Kirchhoff}

\newcommand{\SoftwarePackage}[1]{\textsc{#1}} % special font type for software packages
\newcommand{\baci}{\SoftwarePackage{4C}}
\newcommand{\beamme}{\SoftwarePackage{BeamMe}}
\newcommand{\muelu}{\SoftwarePackage{MueLu}}
\newcommand{\superlu}{\SoftwarePackage{SuperLU}}
\newcommand{\trilinos}{\SoftwarePackage{Trilinos}}
\newcommand{\zoltan}{\SoftwarePackage{Zoltan}}
\newcommand{\belos}{\SoftwarePackage{Belos}}
\newcommand{\teko}{\SoftwarePackage{Teko}}
\newcommand{\amesos}{\SoftwarePackage{Amesos2}}
\newcommand{\mpi}{\SoftwarePackage{MPI}}

\newcommand{\ie}{i.e.,}

\newcommand{\eg}{e.g.,}

\newcommand{\cf}{cf.}
\newcommand{\wrt}{w.r.t.}

\newcommand{\one}{I}
\newcommand{\two}{II}
\newcommand{\three}{III}